\pdfoutput=1

\documentclass[11pt,a4paper]{article}

\usepackage[normalem]{ulem}
\usepackage{jheppub}
\usepackage{amsmath}
\usepackage{amssymb}
\usepackage{appendix}
\usepackage{multirow}
\usepackage{rotating}
\usepackage{diagbox}
\usepackage{float}
\usepackage{xcolor}
\usepackage{cancel}
\usepackage{tabularx}
\usepackage{color}

\def\stat{({\rm stat})}
\def\syst{({\rm syst})}

\newcommand{\bea}{\begin{eqnarray}}
\newcommand{\eea}{\end{eqnarray}}

\usepackage{color}
\definecolor{niceblue}{rgb}{0,0,1}
\definecolor{nicered}{rgb}{0.7,0.1,0.1}
\definecolor{nicegreen}{rgb}{0.1,0.5,.1}

\usepackage{graphicx}
\usepackage{subcaption}
\graphicspath{{figures/}{./}}
\title{Optimised observables and new physics prospects in the penguin-mediated decays $B_{d(s)}\to K^{(*)0}\phi$}

\author[a]{Aritra Biswas,}
\author[b]{S\'ebastien Descotes-Genon,}
\author[a]{Joaquim Matias,}
\author[b,c]{Gilberto Tetlalmatzi-Xolocotzi}

\affiliation[a]{Universitat Aut\`onoma de Barcelona, 08193 Bellaterra, Barcelona,\\
Institut de F\'{i}sica d'Altes Energies (IFAE), The Barcelona Institute of Science and Technology, Campus UAB, 08193 Bellaterra (Barcelona)}
\affiliation[b]{Universit\'e Paris-Saclay, CNRS/IN2P3, IJCLab, 91405 Orsay, France}
\affiliation[c]{Theoretische Physik 1, Center for Particle Physics Siegen (CPPS), Universit\"at Siegen, Walter-Flex-Str. 3, 57068 Siegen, Germany}

\abstract{We study the penguin-mediated $\bar{B}_{d(s)}\to\bar{K}^{*0}(K^{*0})\phi$ transitions, proposing a new optimised observable $L_{K^*\phi}$ from the ratio of longitudinal branching ratios of these decays, with limited hadronic uncertainties and enhanced sensitivity to New Physics. This observable exhibits a deviation at the $1.48\, \sigma$ level between its experimental value and its SM determination within QCD factorisation. This result can be accommodated together with the significant deviations found for the $K^{(*)}\bar{K}^{(*)}$ modes in our earlier works if New Physics affects either the QCD penguin operator $Q_4$ or the chromomagnetic dipole operator $Q_{8g}$ for both $b\to d$ and $b\to s$ transitions. The allowed range for the Wilson coefficients $C_{4s,8gs}$ is narrower compared to $C_{4d,8gd}$ since the $b\to s$ transition channel $\bar{B}_d\to\bar{K}^{*0}\phi$ is in better agreement with the SM. However, if we add the measured branching ratio for the $\bar{B}_{d}\to \bar{K}^{0}\phi$ to our analysis, the simultaneous explanation of all the experimental data for the $K^{(*)}\bar{K}^{(*)}$ and the $K^*(\bar{K}^{(*)})\phi$ channels in terms of New Physics in the Wilson coefficients $C_{4d,s}$ or $C_{8gd,s}$ only, is not possible. This set of observables can be explained more easily if we assume New Physics in both  $f=d,s$ sectors, either in $(C_{4f},C_{6f})$ or $(C_{6f},C_{8gf})$.  The addition of the branching ratios of the charged modes $B^-\to K^{(*)-}\phi$ to the above mix of observables results in a reduction of the parameter space for the $(C_{4f},C_{6f})$ scenario, while discarding the $(C_{6f},C_{8gf})$ scenario completely. This can be traced to the discrepancy of more than 1 $\sigma$ between the experimental measurements of the branching ratios of the $\bar{B}_{d}(B^-)\to\bar{K}^0(K^-)\phi$ transitions. This should provide a strong incentive for the LHCb/Belle II experiments to measure the branching ratios of $\bar{B}_{d}\to\bar{K}^{(*)0}\phi$, $\bar{B}_s\to K^{*0}\phi$ and particularly $B^-\to K^-\phi$ to confirm or dismiss the conclusions hinted at by present data.}

\emailAdd{abiswas@ifae.es}
\emailAdd{sebastien.descotes-genon@ijclab.in2p3.fr}
\emailAdd{matias@ifae.es}
\emailAdd{gtx@physik.uni-siegen.de}

\arxivnumber{}

\begin{document}
%\pacs{}
\begin{flushright}

SI-HEP-2024-06\\
P3H-24-020

\end{flushright}

%%%%%%%%%%%%%%%%%%%%%%%%%%%%%%%%%
%%%%%%%%%%%%%%%%%%%%%%%%%%%%%%%%%
\maketitle
%%%%%%%%%%%
\section{Introduction}\label{intro}

In recent years we have been witnessing persistent deviations between the theoretical prediction and the experimental measurements of some $B$-meson observables related to $b\to s \mu^+\mu^-$ and $b \to c \tau \nu$ transitions. For instance, the  significant deviation in the optimised angular observable $P_5^\prime$~\cite{ Descotes-Genon:2012isb}~\footnote{The sensitivity of this observable to hadronic uncertainties, either from form factors or from non-local charm-loop contributions has been extensively discussed in the recent literature. Interestingly the recent analysis presented by the LHCb collaboration~\cite{LHCb:2023gel,LHCb:2023gpo}, using all data from $B\to K^*(\to K\pi)\mu\mu$ and a very general parametrisation for the charm-loop contribution, concludes that the anomaly in $P_5^\prime$ requires a  Wilson coefficient ${\cal C}_{9\mu}$ deviating from the SM value at the 2$\sigma$ level, a result in remarkable agreement with the global fit in Ref.~\cite{Alguero:2023jeh} including all channels and using a conservative theory estimate for the non-local charm-loop contributions.}, the deficit observed in a large set of $b \to s \mu\mu$ branching ratios (denoted by $\mathcal{B}$ in the following) and in the Lepton Flavour Universality (LFU) ratios $R_{D, D^*}$ comparing $B\to D(^*)\tau \nu$ with decays involving lighter leptons. Unfortunately, the LFU observables $R_{K,K^*}$ comparing $b\to se^+e^-$ and $b\to s\mu^+\mu^-$ transitions have disappeared from this list, due to the difficulties encountered by the LHCb collaboration with the PID of electrons~\cite{LHCb:2022vje,LHCb:2022qnv}. It will thus be necessary to wait for Belle II and CMS measurements to assess  the size of the violation of LFU (if any) in these modes~\footnote{Interestingly, CMS has provided a first measurement of this quantity albeit with huge asymmetric errors: $R_K=0.78^{+0.47}_{-0.23}$~\cite{CMS:2023klk}.}.

In any case, if the previous deviations in semileptonic $B$ decays increase and are confirmed to stem from New Physics (NP), it is expected that other tensions in rare $b\to s$ or even $b \to d$ transitions may also show up, for instance in rare hadronic decays. These tensions can be unveiled through optimised observables with an enhanced sensitivity to NP because of their reduced sensitivity to the hadronic uncertainties affecting non-leptonic $B$ decays.
We pursued this idea in a recent article~\cite{Biswas:2023pyw}, where we studied the anatomy of optimised observables involving $U$-spin connected branching ratios of pseudoscalar penguin-mediated decays (${\bar B}_s \to K^{(*)0}\bar{K}^{(*)0}$ versus $B_d \to K^{(*)0}\bar{K}^{(*)0}$). This study followed a previous article focusing on the longitudinal polarisation of the vector modes ${\bar B}_{d,s}\to K^{*0} \bar{K}^{*0}$~\cite{Alguero:2020xca} which showed an unexpectedly large breaking of $U$-spin symmetry among longitudinal polarisations\footnote{In Ref.~\cite{Gronau:2013mda},
the test relation $-\frac{A_s^{CP} \tau(B_d) {\cal B}_s}{A_d^{CP} \tau(B_s) {\cal B}_d}=1$ was proposed for pion and kaon charged modes to assess
the size of $U$-spin breaking. Even though a similar test could in principle be derived for the neutral kaon modes considered in Ref.~\cite{Biswas:2023pyw}, we cannot exploit it given that some of the CP-asymmetries needed are not measured yet. Checking this relation provides an interesting motivation for experimentalists to measure these CP-asymmetries.}. We analysed this issue 
by exploiting the fact that these penguin-mediated decays have a peculiar hadronic structure in terms of QCD contributions. Taking advantage of the correlations among hadronic uncertainties, we were able to construct new optimised observables called $L_{K^*K^*}$ (longitudinally polarised vector ratio) and $L_{KK}$ (pseudoscalar ratio), and predict their values more accurately than the branching ratios themselves. The SM predictions of these optimised observables within QCD factorisation (QCDF)
turned out to exhibit tensions of 2.6$\sigma$ and 2.4$\sigma$ respectively with experimental measurements. Interestingly, the tensions in both observables could be easily accommodated by common NP contributions to the Wilson coefficient of the chromomagnetic operator and/or one of the QCD penguin operators (see Ref.~\cite{Lizana:2023kei} for a possible explanation with a model based on a scalar leptoquark $S_1$).

Our aim in the present article is to explore to which extent this picture that emerged from the decays ${\bar B}_{d,s}\to\bar{K}^{*0}K^{*0}$ and ${\bar B}_{d,s}\to\bar{K}^0K^0$  holds for other similar penguin mediated rare decays. We will explore the $B \to VV$ modes ${\bar B}_d\to \bar{K}^{*0}\phi$ and ${\bar B}_s\to{K}^{*0} \phi$. All these modes are 
penguin mediated, dominated by the  QCD penguin amplitude and include a penguin-annihilation contribution, with branching ratios in a range between 10$^{-6}$ to 10$^{-5}$. One primary difference with the $K^{(*)} \bar{K}^{(*)}$ modes considered in Refs.~\cite{Alguero:2020xca,Biswas:2023pyw} is the fact that in the ${\bar B}_{d(s)} \to\bar{K}^{*0}(K^{*0})\phi$ decays, the $\bar{B}_d$ decay corresponds to a $b\to s $ transition and the $\bar{B}_s$ decay to a $b \to d $ one, opposite to the case of $\bar{B}_{d,s} \to K^{(*)} \bar{K}^{(*)}$ modes where the light flavour of the transition and the flavour of the spectator quark coincides.
Another important difference is the presence of two types of topologies entering the decays ${\bar B}_{d(s)} \to\bar{K}^{*}(K^{*})\phi$, with different sensitivities to Wilson coefficients as compared to the $K^{(*)} K^{(*)}$ case. Moreover, the two decays ${\bar B}_{d(s)} \to\bar{K}^{*}(K^*)\phi$ are not equal in the $U$-spin limit~\footnote{The amplitudes of the two decays ${\bar B}_{d(s)} \to\bar{K}^{*0}(K^{*0})\phi$ are different linear combinations of the same two reduced amplitudes, as can be seen in Sec.~IV.D of Ref.~\cite{Soni:2006vi}. These modes are not $U$-spin partners, but they share the same dominant hadronic structure that is the main property required to select the modes used to build optimised observables.}.

\begin{table}[t]
\begin{center}

\tabcolsep=0.31cm\begin{tabular}{|c|c|}
\hline\multicolumn{2}{|c|}{$\bar{B}_d\rightarrow\bar{K}^{*0}\phi$ }  \\ 
\hline
Longitudinal polarization fraction ($f_L$)&
Total Branching Ratio ($\mathcal{B}$)\\
\hline
$0.497\pm0.017$ \cite{BaBar:2008lan,Belle:2013vat,LHCb:2014xzf}&$(1.00\pm0.05)\times 10^{-5}$\cite{CLEO:2001ium,BaBar:2008lan,Belle:2013vat}\\
\hline
\end{tabular}

\vskip 1pt

\tabcolsep=0.5cm\begin{tabular}{|c|c|}
\hline\multicolumn{2}{|c|}{$\bar{B}_s\rightarrow K^{*0}\phi$ }  \\ 
\hline
Longitudinal polarization fraction ($f_L$)&
Total Branching Ratio ($\mathcal{B}$)\\
\hline
$0.51\pm 0.17$~\cite{LHCb:2013nlu}&$(1.14\pm 0.30)\times 10^{-6}$ \cite{LHCb:2013nlu}\\
\hline
\end{tabular}

\vskip 1pt

\tabcolsep=0.43cm\begin{tabular}{|c|c|}
\hline\multicolumn{2}{|c|}{$B^-\rightarrow K^{*-}\phi$ }  \\ 
\hline
Longitudinal polarization fraction ($f_L$)&
Total Branching Ratio ($\mathcal{B}$)\\
\hline
$0.50\pm 0.05$~\cite{BaBar:2007bpi, Belle:2005lvd}&$(1.00\pm 0.20)\times 10^{-5}$ \cite{BaBar:2007bpi, Belle:2003ike}\\
\hline
\end{tabular}
\caption{Measured branching ratios (all polarisations included) and polarisation fractions for $K^*\phi$ final states.} 
\label{tab:exp_inputs_Kstphi}
\end{center}
\end{table}

%\begin{table}[t]
%\begin{center}

%\tabcolsep=1.25cm\begin{tabular}{|c|}
%\hline
%$f_L(B^-\rightarrow K^{*-}\phi)$\\
%\hline
%$0.50\pm0.05$ \cite{BaBar:2007bpi, Belle:2005lvd}\\
%\hline
%\end{tabular}

%\vskip 1pt

%\tabcolsep=0.5cm\begin{tabular}{|c|c|}
%\hline\multicolumn{2}{|c|}{Measured branching ratios}  \\ 
%\hline
%$\mathcal{B}(B^-\rightarrow K^{*-}\phi)_{\rm all \ pol}$&
%$\mathcal{B}(B^-\rightarrow K^{-}\phi)$\\
%\hline
%$(1.00\pm 0.20)\times 10^{-5}$\cite{BaBar:2007bpi, Belle:2003ike}&$(8.8^{+0.7}_{-0.6})\times %10^{-6}$ \cite{BaBar:2012iuj,CDF:2005apk,Belle:2004drb,CLEO:2001ium}\\
%\hline
%\end{tabular}
%\caption{Measured branching ratios (all polarisations included for the VV mode), polarisation fraction for $B^-\to K^{*-}\phi$ and branching ratio for final $B^-\to K^-\phi$ transition.} 
%\label{tab:exp_inputs}
%\end{center}
%\end{table}

On the experimental side, it should be highlighted that there is a very large difference between the measured longitudinal polarisation fraction of both ${\bar B}_{d,s}\to K^{*0}\bar{K}^{*0}$, namely $f_L^{B_d \, {\rm exp}}=0.73\pm0.05$~\cite{LHCb:2019bnl,BaBar:2007wwj,Alguero:2020xca} and $f_L^{B_s \, {\rm exp}}=0.240\pm0.040$~\cite{LHCb:2019bnl}, whereas the corresponding SM predictions are rather close $f_L^{B_s \, {\rm QCDF}}=0.72^{+0.16}_{-0.21}$ and $f_L^{B_d \, {\rm QCDF}}=0.69^{+0.16}_{-0.20}$~\cite{Beneke:2006hg} as expected from $U$-spin symmetry. In contrast, as can be seen in Table~\ref{tab:exp_inputs_Kstphi},  for the modes ${\bar B}_{d,s} \to \bar{K}^{*0} \phi$   both experimental numbers for the longitudinal polarisation fraction $f_L^{B_d \, {\rm exp}}=0.497\pm 0.017$ and $f_L^{B_s \, {\rm exp}}=0.51\pm0.17$ are almost equal and in agreement with the QCDF prediction within $1~\sigma$~\cite{Beneke:2006hg,Bartsch:2008ps} albeit with large uncertainties. Such a low longitudinal polarisation fraction close to $50\%$ is not expected in general in the heavy-quark limit, but, it can be accommodated in QCDF due to, for instance, large annihilation contributions to the negative-helicity penguin amplitude or enhanced electromagnetic contributions to the transverse amplitude that can lead to significant transverse polarisations~\cite{Beneke:2006hg}. In this sense the observed value of the longitudinal polarisation fractions in $K^*\phi$ is not necessarily in disagreement with general QCD expectations (and could be attributed to a specific pattern of hadronic contributions), contrary to the $K^*\bar{K}^*$ case (where the data do not follow general $U$-spin expectations). 

Since there might be large hadronic uncertainties making the transverse amplitudes difficult to assess theoretically, we will focus on optimised observables that do not depend on the transverse amplitudes and moreover can benefit from specific cancellations of hadronic uncertainties. 
However, a complete study will also entail individual branching ratios, specially when the corresponding optimized observable is not yet measured (focusing on longitudinally polarised states in the case of vector vector final states), even though the uncertainties for their theoretical predictions are much larger.

The structure of the article is as follows. In section~\ref{sec:twoapproaches}, we recall the two main approaches (flavour symmetries and factorisation) considered to tackle nonleptonic $B$-decays. In section~\ref{sec:theory}, we discuss the theoretical framework needed to construct the $L_{K^*\phi}$ optimised observable. Section~\ref{sec:LKstphi_est} presents the theoretical and experimental determinations of the new observable, followed by a short discussion regarding the modelling of the annihilation contributions as they may have a large impact on hadronic uncertainties. 
In section~\ref{sec:LKstphi_dis}, we present an Effective Field Theory analysis of NP contributions, discussing not only $L_{K^*\phi}$ and the corresponding branching ratios  but also constraints from our previous article from $K^{*0}\bar{K}^{*0}$ and $K^0\bar{K}^0$~\cite{Biswas:2023pyw}. 
Section \ref{sec:PV_VP} explores the impact
of including the constraints coming from branching ratios of measured mixed modes ($B \to VP$ with $V=K^*,\phi$ and $P=K$) on the results of the EFT analyses presented in the previous section . In section~\ref{sec:allmodes} 
we study the consistency among all constraints which requires us to generalise the one-operator scenario of the EFT analysis of section~\ref{sec:LKstphi_dis} to two-operator scenarios. We then discuss the impact of the constraints from the charged $b\to s$ modes $B^-\to K^{(*)-}\phi$.
We conclude in section~\ref{sec:conclusion}. Finally, details on the input parameters used and the operator structure of the EFT are provided in the appendices.

\section{Theoretical approaches to non-leptonic $B$ decays and NP searches} \label{sec:twoapproaches}

Non-leptonic $B$ decays are difficult to assess theoretically due to the hadronic uncertainties coming from long-distance QCD contributions. Over the decades, two main categories of approaches have been considered.

A first approach, inherited from the early studies of the theory of strong interactions but still very powerful consists of flavour symmetries. They can be used to relate hadronic matrix elements of different modes, often in order to extract information on weak phases or NP contributions. Isospin symmetry was used, for instance, in the context of determining the CKM angle $\alpha$ from measurements of branching ratios and CP asymmetries~\cite{Gronau:1990ka,Lipkin:1991st,Gronau:2001ff}. $U$-spin has been used for various $B_d$- and $B_s$-decays in order to extract hadronic matrix elements and potential NP contributions~\cite{Fleischer:2002zv,London:2004uj,
London:2004ej,Jung:2009pb,Fleischer:2010ib,Fleischer:2016jbf,Bhattacharya:2022akr}. More general studies have been performed in the context of general $SU(3)$ analyses~\cite{Zeppenfeld:1980ex,Savage:1989ub,Gronau:1994rj,Gronau:1995hm,Gronau:1995hn}, with applications to various modes~\cite{Deshpande:2000jp,Chiang:2004nm,London:2004uj,Cheng:2014rfa,Hsiao:2015iiu,Beaudry:2017gtw,Fleischer:2017vrb,Fleischer:2018bld}. If these symmetries are very useful guides to analyse the data, 
the accuracy of these approaches, in particular regarding NP searches, depends on the amount of flavour symmetry breaking induced by the difference of light-quark masses and charges, which can depend on the modes considered. In general, too large values of flavour breaking are considered as a good hint of NP.
For instance, in Ref.~\cite{Berthiaume:2023kmp} a global fit to a large set of modes (in this case $B\to PP$) under the assumption of $SU(3)$ symmetry is performed to determine how well the SM may explain a large set of data. In this particular case, the fit under the hypothesis of exact flavour symmetry shows a discrepancy of $3.6\sigma$ with respect to the SM. Solving this discrepancy would require a 1000$\%$ breaking of the $SU(3)$ symmetry, which is way beyond what is expected in the SM, suggesting an NP source for this pattern of deviation.

A second, more recent approach consists in performing a $1/m_b$ expansion of the matrix elements in order to exploit the hierarchy of scales between the light-meson masses, the typical QCD scale and the $b$-quark mass. At the leading-order of this expansion, it is possible to perform a separation (or factorisation) between short and long distances, so that the hadronic matrix elements can be expressed in terms of nonperturbative universal quantities (form factors, light-cone distribution amplitudes) and perturbative process-dependent quantities (hard-scattering kernels). The seminal papers \cite{Beneke:2000ry,Beneke:2001ev,Beneke:2003zv} of this QCD factorisation framework illustrated how this idea could be related to earlier works on factorisation~\cite{Buras:1998us,Buras:1998ra}, and showed the breakdown of this factorisation at higher orders in $1/m_b$ for some contributions. This was later understood in the framework of the Soft-Collinear Effective Theory~\cite{Bauer:2000yr,Beneke:2000wa,Bauer:2001cu,Bauer:2001yt,Bauer:2002nz,Beneke:2002ph,Beneke:2002ni,Bauer:2002uv,Bauer:2002aj,Bauer:2004tj,Bauer:2005wb,Bauer:2005kd} providing an explanation of factorisation by separating soft, collinear and hard degrees of freedom in the relevant hadronic matrix elements and providing further elements on the structure of $1/m_b$ corrections in this picture. It has been used in several instances to discuss the possible presence of NP, for instance in Ref.~\cite{Bordone:2020gao}. This requires some additional modelling for these higher-order contributions. For this reason, analysis of non-leptonic $B$ decays within QCD factorisation have often used flavour symmetries to constrain long-distance contributions, in particular for global fits (see for instance Refs.~\cite{Cheng:2011qh,Huber:2021cgk}).

This highlights the interest of combining both approaches. We follow the approach first considered in Refs.~\cite{Descotes-Genon:2006spp,Descotes-Genon:2007iri} combining flavour symmetries together with the minimal input from QCD factorisation, mainly an infrared-free finite quantity, in order to obtain more precise predictions than based on flavour symmetries alone. We will now discuss how to exploit this combined approach in the context of $B_{d,s}\to \phi K^{(*)}$ decays.

\section{QCDF framework and structure of the observable $L_{K^*\phi}$} \label{sec:theory}
In this section we present a new optimised observable that we will call $L_{K^*\phi}$. It will be constructed from the ratio of (longitudinally polarised) branching ratios of the modes $\bar{B}_d \to \bar{K}^*\phi$ and $\bar{B}_s \to K^* \phi$, following the methodology used in Refs.~\cite{Alguero:2020xca,Biswas:2023pyw}. We will discuss its theoretical structure, paying particular attention to the differences with the $L_{K^*K^*}$ and $L_{KK}$ observables introduced in Refs.~\cite{Alguero:2020xca,Biswas:2023pyw}.

\begin{figure}[ht]
\subfloat[]{\label{fig:Bd_CA}\includegraphics[width=0.45\textwidth,height=0.30\textwidth]{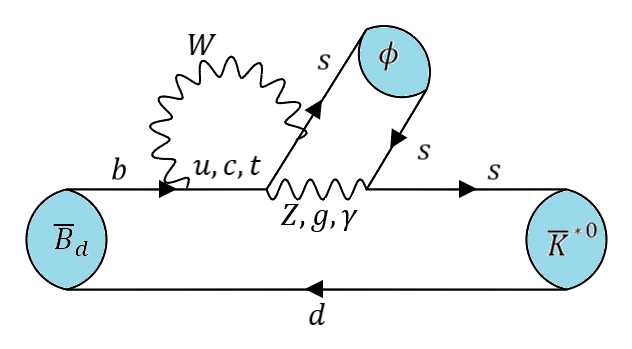}}~~~~~~
\subfloat[]{\label{fig:Bd_CS}\includegraphics[width=0.45\textwidth,height=0.35\textwidth]{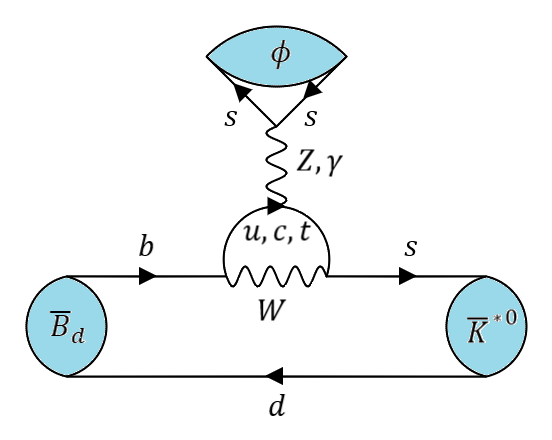}}\par 
\subfloat[]{\label{fig:Bs_CA}\includegraphics[width=0.45\textwidth,height=0.30\textwidth]{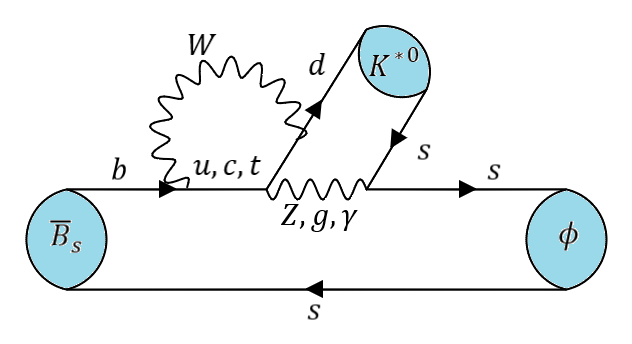}}~~~~~~
\subfloat[]{\label{fig:Bs_CS}\includegraphics[width=0.45\textwidth,height=0.35\textwidth]{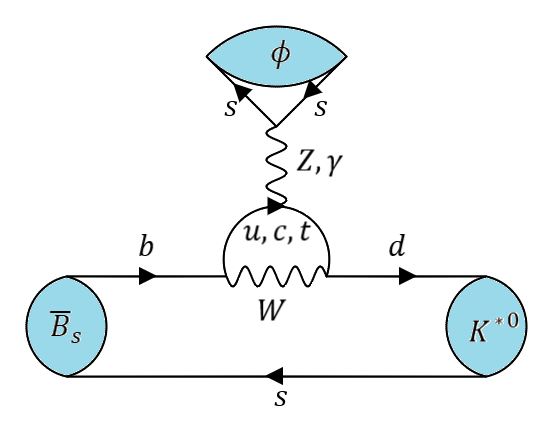}}
\caption{The lowest-order (in $\alpha_s$) topologies contributing to the $\bar{B_d}\to\bar{K}^{*0}\phi$ (top row) and $\bar{B}_s\to K^{*0}\phi$ (bottom row) transitions. Additional diagrams involving QCD penguins  are not displayed, but can be drawn from
Figs.~\ref{fig:Bd_CS} and~\ref{fig:Bs_CS}, replacing the $Z,\gamma$ line by a gluon line and adding a second gluon exchange between quark lines to ensure the correct colour flow.
}
\label{fig:topologies}
\end{figure}

\subsection{The optimised observable $L_{K^*\phi}$ }

Our goal is to define observables with a low sensitivity to hadronic uncertainties for these modes. 
As discussed in Refs.~\cite{Beneke:2006hg,Kagan:2004ia,Kagan:2004uw,Bartsch:2008ps} only the longitudinal component of such non-leptonic $B$ decays into two light vector mesons
can be reliably computed in QCDF,
as the longitudinally polarised final states are dominant  in the heavy-quark limit.
We thus choose quantities that involve only longitudinally polarised vector-vector final states. 

In agreement with Eq.~(3.6) of Ref.~\cite{Alguero:2020xca}, we define a new observable involving $K^{*0}(\bar{K}^{*0})\phi$ final states as:
\begin{equation}\label{eq:LKstarphi}
L_{K^*\phi}=\rho(m_{K^{*0}},m_{\phi})\frac{{\cal B}({\bar{B}_d \to {\bar K^{*0}}} \phi)}{{\cal B}({\bar{B}_s \to K^{*0} \phi)}}\frac{ f_L^{B_d}}{ f_L^{B_s}}=\frac{|A_0^s|^2+ |\bar A_0^s|^2}{|A_0^d|^2+ |\bar A_0^d|^2}\,,
\end{equation}
where  $\rho(m_1,m_2)$ stands for the ratio of phase-space factors 
\begin{equation} \label{eq:phasespace}
\rho(m_1,m_2)=\frac{\tau_{B_s}}{
    \tau_{B_d}}\frac{m_{B_d}^3}{m_{B_s}^3}\frac{\sqrt{(m_{B_s}^2-(m_1+m_2)^2)(m_{B_s}^2-(m_1-m_2)^2)}}{\sqrt{(m_{B_d}^2-(m_1+m_2)^2)(m_{B_d}^2-(m_1-m_2)^2)}}\,.
\end{equation}
In accordance with Refs.~\cite{Alguero:2020xca, Biswas:2023pyw}, $L_{K^*\phi}$ is constructed as a ratio with the $b\to s$ transition in the numerator versus the $b\to d$ in the denominator. But in contrast with the  $K^{(*)}K^{(*)}$ cases, the $b \to s$ ($b \to d$) transition entails a $\bar{B}_d \to \bar{K}^{*0}\phi$ ($\bar{B}_s \to K^{*0} \phi$) decay.

\begin{table}
\begin{center}
\begin{tabular}{|c|c|c|c|}
\hline
& $T\times 10^{7}$ & $P\times 10^{7}$ & $\Delta\times 10^{8}$ \\\hline
${\cal B}(\bar{B}_s \to K^{*0}\phi)$ & $3.40^{+1.24}_{-1.21}-4.07^{+2.00}_{-1.24}i$  &$3.18^{+1.23}_{-1.21}-5.11^{+2.01}_{-1.27} i$ & $2.15^{+0.67}_{-0.66}+1.04^{+0.76}_{-0.74} i$ 
\\
${\cal B}(\bar{B}_d \to\bar{K}^{*0}\phi)$ & $3.35^{+1.04}_{-1.02}-4.02^{+1.68}_{-1.08} i$  & $3.22^{+1.04}_{-1.02}-5.03^{+1.70}_{-1.31} i$ & $1.25^{+0.61}_{-0.59}+1.03^{+0.84}_{-0.83} i$ 
\\
\hline
\end{tabular}
\end{center}
\caption{$T$ and $P$ hadronic matrix elements as well as their difference $\Delta$ predicted in the SM using QCDF, with the inputs in App.~\ref{app:inputs}. We recall that we consider only the longitudinal polarisation in $K^{*0}(\bar{K}^{*0})\phi$.}
\label{tab:T_P_delta}
\end{table}

In the above expression, $A_0^q$ represents the amplitude corresponding to a $\bar{b}\to\bar{q}$ decay leading to longitudinally polarised vector mesons and $f_L^{B_{d,s}}$ are the longitudinal polarisation fractions of the $B_d$ and $B_s$ decays respectively. Following Refs.~\cite{Alguero:2020xca,Biswas:2023pyw},
\begin{eqnarray}\label{eq:PTdefinition}
 \bar{A}_0^q &=& \bar{A}_f = \lambda_u^{(q)} T_q + \lambda_c^{(q)} P_q    =\lambda_u^{(q)} \Delta_q - \lambda_t^{(q)} P_q\,,\\ 
 A_0^q &=& A_{\bar f}=(\lambda_u^{(q)})^* T_q + (\lambda_c^{(q)})^* P_q    =(\lambda_u^{(q)})^* \Delta_q - (\lambda_t^{(q)})^* P_q\,, 
\end{eqnarray}
where the CKM factor is defined by $\lambda_U^{(q)}=V_{Ub} V_{Uq}^*$ while $T_q$ (``tree'') and $P_q$ (``penguin'') refer to the hadronic matrix elements accompanying $\lambda_u^{(q)}$
and $\lambda_c^{(q)}$, respectively.
The amplitude $\bar{A}_f$ is defined as $\bar{A}_f = A(\bar{B}_{d(s)} \to \bar{K}^{*0}(K^{*0}) \phi)$ whereas its CP-conjugate counterpart
$A_{\bar f}$ can be obtained as $A_{\bar f} = \eta_f A(B_{d(s)} \to K^{*0} (\bar{K}^{*0}) \phi)$ where $\eta_f$ is the CP-parity of the final state, given for $j = 0, ||, \perp$ respectively as 1, 1, -1.

Despite the usual tree and penguin terminology that is used for the hadronic matrix elements, we stress that
both $T_q$ and $P_q$ receive only penguin contributions in the case of the penguin-mediated decays considered here and in Ref.~\cite{Alguero:2020xca}, as can be also seen from  Fig.~\ref{fig:topologies}. However, the modes considered here also receive contributions from additional penguin topologies compared to the modes discussed in Refs.~\cite{Alguero:2020xca, Biswas:2023pyw}.

As discussed extensively in Refs.~\cite{Descotes-Genon:2006spp,Alguero:2020xca,Biswas:2023pyw}, the penguin-mediated decays exhibit interesting features from the point of view of hadronic uncertainties. Indeed, unlike $T_q$ and $P_q$, the quantity $\Delta_q=T_q-P_q$ is free from the endpoint divergences arising in power suppressed long-distance contributions (hard scattering, weak annihilation).  Eq.~(\ref{eq:LKstarphi}) can be recast in terms of the divergence-free difference $\Delta_q$ as:
\begin{equation}\label{eq:Lgeneraldiscussion}
L_{K^*\phi}=\frac{|A_0^s|^2+ |\bar A_0^s|^2}{|A_0^d|^2+ |\bar A_0^d|^2}
=\kappa \left|\frac{P_s}{P_d}\right|^2 
 \left[\frac{1+\left|\alpha^s\right|^2\left|\frac{\Delta_s}{P_s}\right|^2
 + 2 {\rm Re} \left( \frac{ \Delta_s}{P_s}\right) {\rm Re}(\alpha^s) 
 }{1+\left|\alpha^d\right|^2\left|\frac{\Delta_d}{P_d}\right|^2
  + 2 {\rm Re} \left( \frac{ \Delta_d}{P_d}\right) {\rm Re}(\alpha^d)} \right]\,,
\end{equation}
where the CKM factors read 
\begin{eqnarray}\label{eq:CKM}
\kappa&=&\left|\frac{\lambda^s_u+\lambda^s_c}{\lambda^sd_u+\lambda^d_c} \right|^2=22.91^{+0.48}_{-0.47}, \nonumber\\
\alpha^d&=&\frac{\lambda^d_u}{\lambda^d_u+\lambda^d_c}=-0.0135^{+0.0123}_{-0.0124} +0.4176^{+0.0123}_{-0.0124}i, \nonumber\\
\alpha^s&=&\frac{\lambda^s_u}{\lambda^s_u+\lambda^s_c}=0.0086^{+0.0004}_{-0.0004}-0.0182^{+0.0006}_{-0.0006}i.
\end{eqnarray}
suggesting a dominance of the ratio $P_s/P_d$ that we will discuss in the following.

\subsection{Anatomy of the hadronic amplitudes}
The amplitudes involved in the $L_{K^*\phi}$ observable are given in QCDF notation by~\cite{Bartsch:2008ps}:
\begin{eqnarray}
\label{eq:QCDFstructure}
A(\bar{B}_s \to K^{*0}\phi)&=&\sum_{p=u,c}\lambda_p^{(d)} \left[\left(a_4^p -\frac{1}{2}a_{10}^p \right)A_{\phi K^*} + \left(a_3+a_5-\frac{1}{2} (a_7^p+a_9^p)\right) A_{K^*\phi}\right]\nonumber\\
&&+ \Bigl(\lambda^{(d)}_u + \lambda^{(d)}_c\Bigl) \Bigl(b_3 -\frac{1}{3}b^{\rm EW}_3\Bigl) B_{K^{*}\phi},\\
A(\bar{B}_d \to {\bar K}^{*0}\phi)&=&\sum_{p=u,c}\lambda_p^{(s)}
\left(a_4^p+a_3+a_5-\frac{1}{2} (a_7^p+a_9^p+a_{10}^p)\right) A_{K^*\phi}\nonumber\\
&&+ \Bigl(\lambda^{(s)}_u + \lambda^{(s)}_c\Bigl) \Bigl(b_3 -\frac{1}{3}b^{\rm EW}_3\Bigl) B_{K^{*}\phi} \label{eq:QCDFstructure2}\end{eqnarray}
out of which the expressions for $P_q$ and $T_q$ can be read directly from Eq.(\ref{eq:PTdefinition}).
The normalisation factors are given as
\begin{equation}
A_{V_1V_2} = \frac{iG_F}{\sqrt{2}} m_B^2 f_{V_2} A_0^{B\to V_1}(0),\quad B_{V_1 V_2} =\frac{i G_F}{\sqrt{2}}f_B f_{V_1} f_{V_2}
\end{equation}
where $f_{V}$ is the decay constant of the corresponding meson $V$ and $A_0^{B\to V_1}$ is one of the $B\to V_1$ form factors (so that the $V_1$ meson picks up the spectator quark of the initial $B$ meson). The order of the mesons ($K^*\phi$ or $\phi K^*$) in the subscript of the normalisation factors $A_{V_1V_2}$ has to be respected also for the evaluation of the accompanying coefficients 
$a_i(V_1V_2)$. These $a_i(V_1V_2)$ coefficients are convolutions of hard-scattering kernels and light-cone distribution amplitudes which can be found in Refs.~\cite{Beneke:2006hg, Bartsch:2008ps}. As expected in the case of decays mediated by penguin topologies, the dominant $a_4$ amplitude contains contributions from QCD penguins and the chromomagnetic dipole  operator, $a_3^p$ and $a_5^p$ from QCD penguins and $a_{10}^p$, $a_7^p$ and $a_9^p$ from electroweak penguin operators and the electromagnetic dipole operator. The terms $b_3$ and $b^{\rm EW}_3$ address the annihilation contributions that will be discussed in more detail in section~\ref{sec:ann}.

We assume ideal mixing between the $\omega$ and $\phi$ mesons, as supported phenomenologically, so that $\omega$ $=$ $\omega_q$ $\sim$ $(u\bar{u}+d\bar{d})/\sqrt{2}$ and $\phi$ $=$ $\phi_s$ $=$ $s\bar{s}$. The more general form away from ideal mixing can be obtained from Eq.~(122) (for $\bar{B}^0\to \bar{K}^{*0}\phi$) and Eq.~(134) (for $\bar{B}_s\to K^{*0}\phi$) of Ref.~\cite{Beneke:2003zv}.

A few additional comments are in order, in relation with Fig.~\ref{fig:topologies}:
\begin{itemize}
    \item  Both $\bar{B}_d\to\bar{K}^{*0}\phi$ and $\bar{B}_s\to K^{*0}\phi$ receive contributions from the same two topologies: a first one associated to the amplitudes $a_4^p -\frac{1}{2}a_{10}^p$ and a second one associated with $a_3+a_5-\frac{1}{2} (a_7^p+a_9^p)$. This is in clear contrast with the $\bar{B}_{d(s)}\to {K^{*}\bar{K}^{*}}$ modes contributing to $L_{K^{*}\bar{K}^{*}}$ where only the former (dominant) topology contributes.
    \item The presence of this ``additional'' penguin topology in the case of the $K^*\phi$ final states stems from the structure of the $\phi$ meson. The $\phi$ being essentially an $s\bar{s}$ state, it can be produced from the splitting of a $Z$, a photon or a gluon (with the exchange of a second gluon) as can be seen from Fig.~\ref{fig:topologies}. The same additional topology would contribute to decays where $\phi$ is replaced by  $\omega$ or $\rho^0$.
    \item For $\bar{B}_d\to\bar{K}^{*0}\phi$, both topologies 
    correspond to the $\bar{K}^{*0}$ meson picking up the spectator quark (the $M_1$ meson in the notation of Ref.~\cite{Beneke:2006hg}). On the other hand, the spectator quark is picked up by different final-state mesons in the two topologies corresponding to the $\bar{B}_s\to K^{*0}\phi$ decay. 
    This can also be clearly seen in Eq.~(\ref{eq:QCDFstructure}) featuring two different contributions $A_{K^{*}\phi}$, $A_{\phi K^{*}}$  in $\bar{B}_s\to \bar{K}^{*0}\phi$,
    but only one contribution $A_{K^*\phi}$ for $\bar{B}_d\to K^{*0}\phi$.
\end{itemize}

\section{SM prediction and experimental determination of $L_{K^*\phi}$}\label{sec:LKstphi_est}

\subsection{Longitudinal polarisations and branching ratios} \label{sec:longpolandbrKstphi}

The experimental measurements for the longitudinal polarisations and branching ratios (all polarisations included) are provided in Table~\ref{tab:exp_inputs_Kstphi}. The $\bar{B}_d\rightarrow\bar{K}^{0*}\phi$ branching ratio summing over all polarisations is an average of a 2021 CLEO~\cite{CLEO:2001ium}, 2008 Babar~\cite{BaBar:2008lan} and 2013 Belle~\cite{Belle:2013vat} measurements with good mutual compatibility. The longitudinal polarisation has been measured by Babar and Belle in the same articles, but also by LHCb~\cite{LHCb:2014xzf}. One may notice that the LHCb collaboration has not provided a measurement for the $\bar{B}_d\rightarrow\bar{K}^{0*}\phi$ branching ratio itself. On the other hand, the $\bar{B}_s\rightarrow  K^{0*}\phi$ branching ratio and polarisation have been measured only by LHCb~\cite{LHCb:2013nlu}.

\begin{table}[t]
\begin{center}
\tabcolsep=2.33cm\begin{tabular}{|c|c|}
\hline\multicolumn{2}{|c|}{Longitudinal
$\mathcal{B}(\bar{B}_d\rightarrow\bar{K}^{*0}\phi)\times10^{-6}$ }  \\ 
\hline
SM (QCDF)& Experiment\\
\hline
$4.53^{+2.19}_{-1.79}$&$4.96^{+0.31}_{-0.30}$\\
\hline
\end{tabular}

\vskip 1pt

\tabcolsep=2.33cm\begin{tabular}{|c|c|}
\hline\multicolumn{2}{|c|}{Longitudinal
$\mathcal{B}(\bar{B}_s\rightarrow K^{*0}\phi )\times 10^{-7}$ }  \\ 
\hline
SM (QCDF)& Experiment\\
\hline
$2.19^{+1.05}_{-0.94}$&$5.56^{+2.78}_{-2.27}$\\
\hline

\end{tabular}
\vskip 1pt

\tabcolsep=2.33cm\begin{tabular}{|c|c|}
\hline\multicolumn{2}{|c|}{ Longitudinal
$\mathcal{B}(B^-\rightarrow K^{*-}\phi )\times 10^{-6}$}  \\ 
\hline
SM (QCDF)& Experiment\\
\hline
$4.94^{+2.34}_{-1.91}$&$4.96^{+1.16}_{-1.08}$\\
\hline

\end{tabular}

\caption{Branching ratios for longitudinally polarised $K^*\phi$ final states. Both experimental measurements and theoretical values in SM (based on QCDF) are provided.} 
\label{tab:BR_est}
\end{center}
\end{table}

The $f_L$ measurements for the $K^*\phi$ final state exhibit a rather different pattern compared to those for the $K^*\bar{K}^*$ final state discussed for instance in Ref.~\cite{Alguero:2020xca, Biswas:2023pyw}. $f_L(\bar{B_d}\to\bar{K}^{*0}\phi)$ and $f_L(\bar{B_s}\to K^{*0}\phi)$ are very close, but with a surprisingly low value of  $\sim0.5$. As discussed in Ref.~\cite{Beneke:2006hg} such a rather low value is not expected within the framework of QCDF since the transverse polarisations are $1/m_B$ suppressed compared to the longitudinal ones, so that $f_L$ is expected to be close to 1. Different explanations have been proposed to address the discrepancies observed in the hierarchy of polarisations for $K^*\phi$ final states: some weak-annihilation contributions leading to significant transverse polarisation, an enhanced transverse flavour-singlet QCD penguin amplitude caused by significant spectator-scattering, or a large contribution to the electroweak penguin amplitude from the electromagnetic dipole operator~\cite{Beneke:2006hg}.
We recall that such interesting ideas are unable to explain the very low measurement of $f_L(\bar{B}_s\to K^{*0}\bar{K}^{*0}) = 0.240\pm 0.040$~\cite{LHCb:2019bnl}, and its deviation from $f_L(\bar{B}_d\to K^{*0}\bar{K}^{*0})$ contrary to $U$-spin expectations. As a matter of fact, Ref.~\cite{Beneke:2006hg} predicts QCDF central values for $f_L(\bar{B}_s\to K^{*0}\bar{K}^{*0})$ ranging from $0.67$ to $0.72$, rather close to $f_L(\bar{B}_d\to K^{*0}\bar{K}^{*0})$, as expected from $U$-spin, in stark disagreement with the experimental measurements.

We may use QCDF to predict the observables involving only longitudinally polarised $K^*\phi$ final states within the SM, following the same methodology as in Refs.~\cite{Alguero:2020xca, Biswas:2023pyw}. The required inputs for QCDF are given in appendix~\ref{app:inputs}. In practice, we follow a numerical procedure where each input parameter is assumed to be a Gaussian random variable with the corresponding mean given by the central value and the uncertainty representing the corresponding standard deviation~\footnote{The only exception are the parameters used to address the divergences in annihilation ($X_A$) and hard spectator scattering  ($X_H$), as well as  the energy scale ($\mu$). For $X_{A,H}$ we use the prescription provided in Ref.~\cite{Beneke:2003zv}, assuming them to be universal for $\bar{B}_{s,d}$ decays and parametrizing them as $X_{H,A}=(1+\rho_{H,A}e^{i\phi_{H,A}})~\ln(\frac{m_b}{\Lambda_{h}})$ with both $\rho_{H,A}$ and $\phi_{H,A}$ being varied according to uniform distributions within $[0,1]$ and $[0,2\pi]$ respectively. This is discussed in further detail in section~\ref{sec:ann}. For $\mu$, we also consider a uniform distribution within the range $[m_b/2,2 m_b]$.}. We generate $10^5$ points from a multinormal distribution constructed from all of these Gaussian input random variables taken together and compute a value for the observables (longitudinal branching ratios, $L_{K^*\phi}$ etc.) corresponding to each randomly generated value for the list of input variables. We thus end up with $10^5$ numerical points for the observables providing a sample of their corresponding distributions. We then interpret the median of the distribution from this list of $10^5$ points as the central value of the observable. The uncertainties are inferred from the 1~$\sigma$ quantiles of the corresponding distribution.  

The SM values for the hadronic amplitudes $T$, $P$ and $ \Delta$ for longitudinally polarised final states within the QCD framework are given in Table~\ref{tab:T_P_delta}. As expected for penguin-mediated decays, $T$ and $P$ are strongly correlated, as indicated by the smallness of $\Delta=T-P$ compared to each of the two amplitudes. The corresponding results for the longitudinal branching ratios are given in Table~\ref{tab:BR_est}. The deviation between SM and experiment for the longitudinal $\mathcal{B}(B_d\to\bar{K}^{*0}\phi)$ and $\mathcal{B}(B_s\to K^{*0}\phi)$ is $0.35~\sigma$ and $1.26~\sigma$ respectively. These deviations are computed using the full distribution and the Maximum Likelihood method (see next subsection for details).

One may notice that experiment and theory agree well for $\bar{B}_d\rightarrow\bar{K}^{0*}\phi$ ($b\to s$ transition), whereas the central values are rather different for $\bar{B}_s\rightarrow K^{0*}\phi$ ($b\to d$ transition). This may
provide some clues on where potential NP effects may arise. However, we must bear in mind that there are large uncertainties attached to the branching ratios from both theoretical and experimental sides (with the exception of the experimental measurement for $\mathcal{B}(\bar{B}_d\rightarrow\bar{K}^{0*}\phi)$, which has an uncertainty of $\sim 6\%$).

\subsection{Results for $L_{K^*\phi}$}

The SM prediction for $L_{K^*\phi}$ can be obtained using Eqs.~(\ref{eq:Lgeneraldiscussion}),  Eq.~(\ref{eq:QCDFstructure})
and Eq.~(\ref{eq:QCDFstructure2}), the numerical inputs in Eq.~(\ref{eq:CKM}) and Table~\ref{tab:inputs}, leading to: 
\begin{equation}\label{eq:LKstphi_th}
L_{K^*\phi}^{\rm th}  = 22.04^{+7.06}_{-4.88}.
\end{equation}
The corresponding experimental measurement for $L_{K^*\phi}$ obtained from Eq.~(\ref{eq:LKstarphi}) using the measured branching ratios and longitudinal polarisation fractions (see Table~\ref{tab:BR_est}) is:
\begin{eqnarray}\label{eq:LKstphi_exp}
L_{K^*\phi}^{\rm exp} &=& 8.80^{+6.07}_{-2.97}. 
\end{eqnarray}
%Fig. \ref{fig:LKstphihist} displays the experimental and theoretical distributions for the $L_{K^*\phi}$ observable.

\begin{figure} 
\begin{center}
\includegraphics[width=7cm]{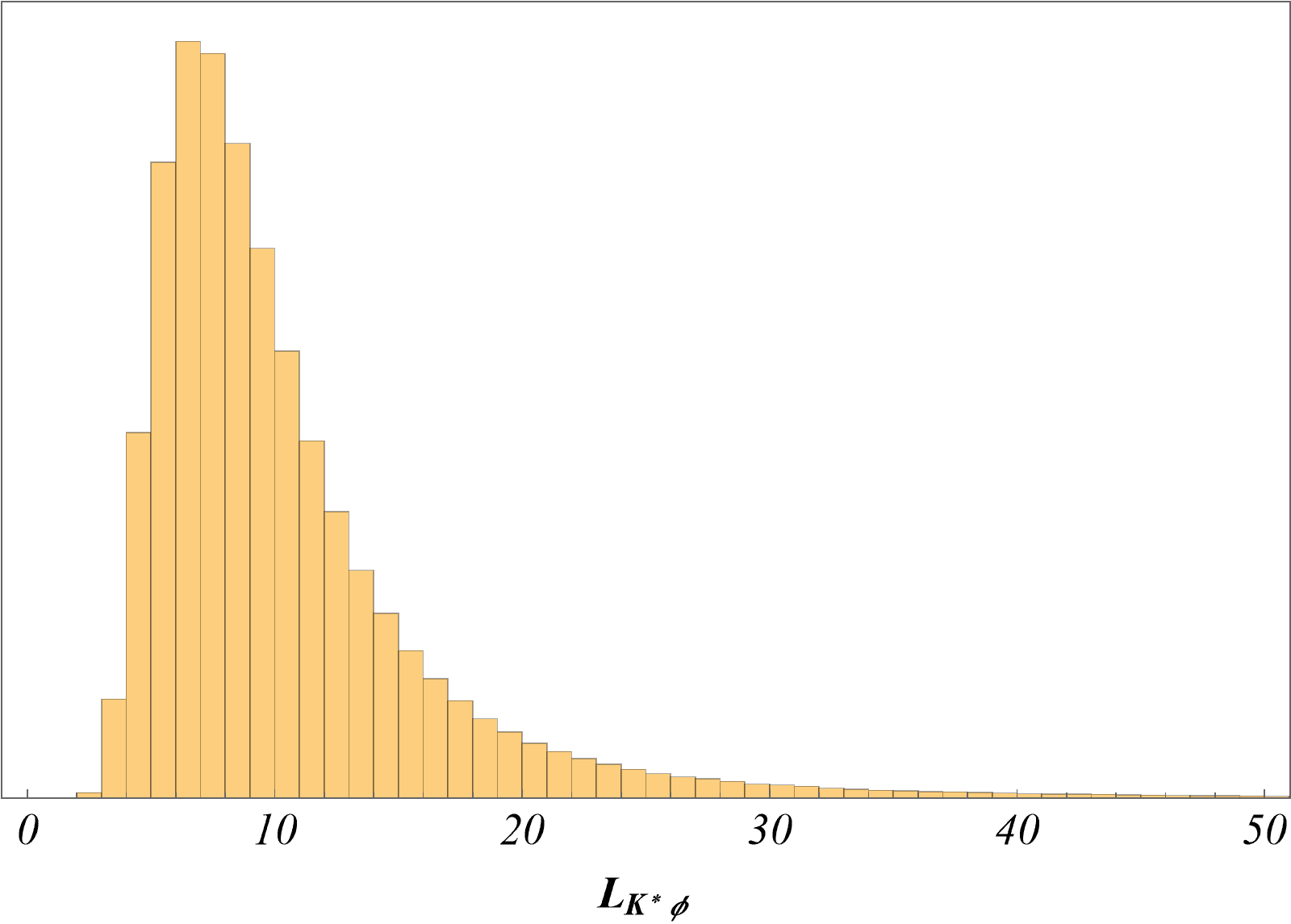}
\qquad \includegraphics[width=7cm]{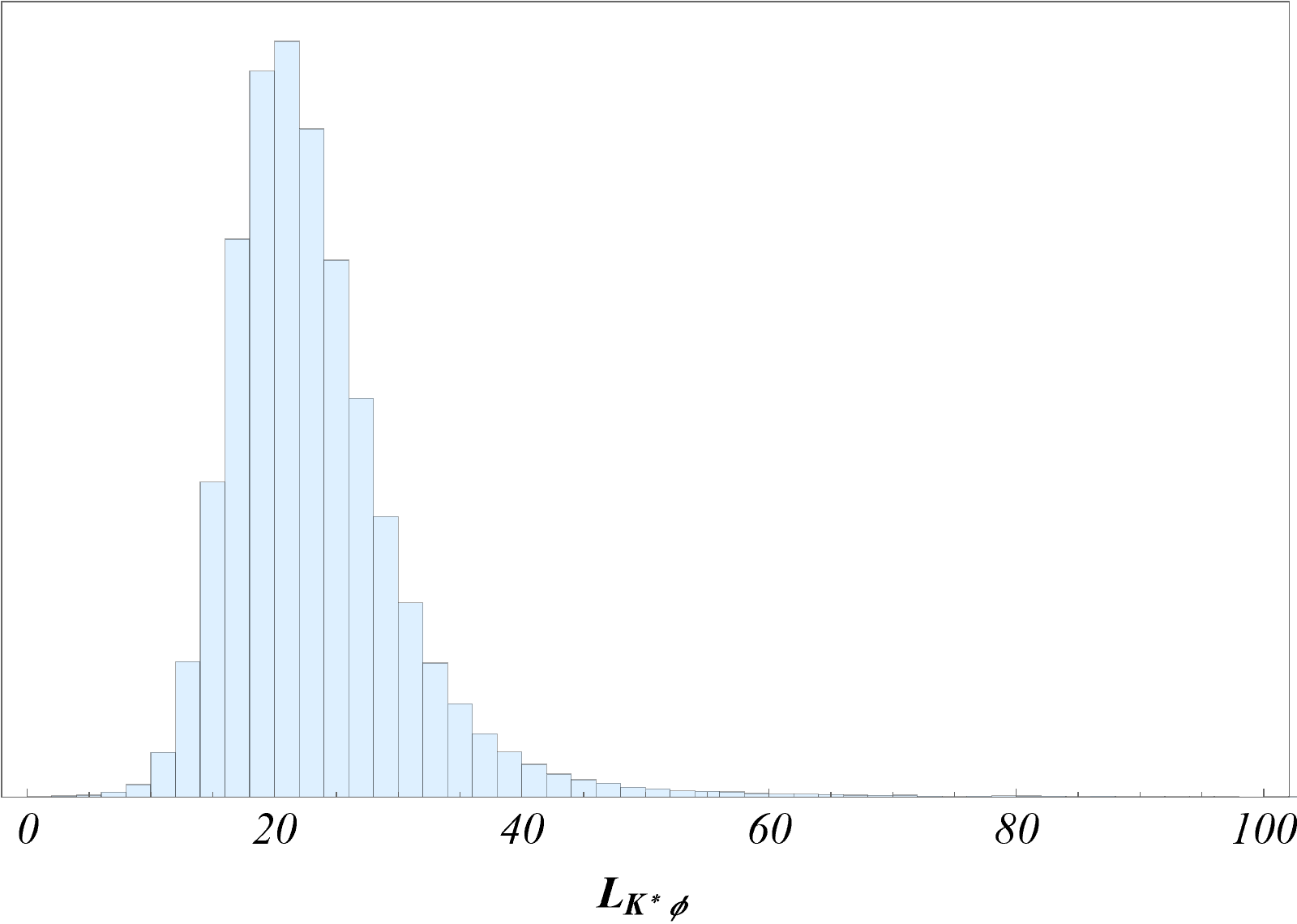}
\end{center}
\caption{The histograms corresponding to the distribution of the experimental measurement (left, yellow) and theoretical prediction (right, blue) for the $L_{K^*\phi}$ observable.}
\label{fig:LKstphihist}
\end{figure}

In Fig.~\ref{fig:LKstphihist}, both experimental and SM theoretical distributions  are displayed and are found to be asymmetric~\footnote{This is also reflected in the corresponding numbers quoted in Eqs.~(\ref{eq:LKstphi_th}) and (\ref{eq:LKstphi_exp}), where the central values are the medians and the uncertainties are estimated from the $1~\sigma$ quantiles of the respective distributions.}. The corresponding deviation between data and SM prediction is $1.48\ \sigma$. A few words are in order regarding the method implemented for the calculation of this deviation (as well as all other deviations quoted in our current and previous article~\cite{Biswas:2023pyw}). 
We have to deal with rather asymmetric distributions here, and it may prove naive and overconservative to symmetrize the uncertainties (e.g. keeping the larger uncertainty) in order to be able to calculate the pull in the usual way (difference of the theoretical and experimental central values, divided by the sum in quadrature of the theoretical and experimental errors). In our case, we determine the distribution of the random variable $X$ defined as the difference between the SM prediction and the experimental measurement.
We can then determine in which confidence interval the value $X=0$ lies according to this p.d.f.,  and convert this piece of information into a value in $\sigma$ units quoted as our pull~\footnote{We should mention here that the $2.6\sigma$ deviation in Ref.~\cite{Biswas:2023pyw} was obtained for $1/L_{K^*\bar{K}^*}$ whose p.d.f. for the  difference between theory and experiment is more symmetric as compared to $L_{K^*\bar{K}^*}$. When computed w.r.t to $L_{K^*\bar{K}^*}$, this deviation comes down to $\sim 2.3\sigma$.}.

A deviation between the SM determination and the experimental measurement of the $L_{K^{*}\phi}$ observable may originate from individual tensions in the 
longitudinal $b\to s$ or  $b\to d$ branching ratios used for its calculation. As discussed in the previous section, the $b\to s$ ($\bar{B_d}\rightarrow \bar{K}^{*0}\phi$) decay rate is consistent with the corresponding experimental measurement albeit with large uncertainties, whereas there is a small discrepancy in the $b\to d$ ($\bar{B_s}\rightarrow K^{*}\phi$) transition above $1~\sigma$. Consequently, the tension of $1.48~\sigma$ in $L_{K^*\phi}$ is driven by the $b\to d$ decay ratio ($\bar{B_s}\rightarrow K^{*}\phi$) appearing in the denominator.

The individual branching ratios defining $L_{K^*\phi}$ suffer from larger theoretical uncertainties. One may wonder whether $L_{K^*\phi}$ can be  predicted with smaller uncertainties than the branching ratios, like in the cases of $L_{K^{(*)}\bar{K}^{(*)}}$~\cite{Alguero:2020xca,Biswas:2023pyw}.
The decay ratios for $\bar{B_d}\rightarrow \bar{K}^{*0}\phi$ and $\bar{B_s}\rightarrow K^{*}\phi$ are not related through $U$-spin, in clear contrast with our previous optimized observables $L_{K^{(*)}\bar{K}^{(*)}}$. It is thus not obvious that the cancellation of hadronic uncertainties takes place for $L_{K^*\phi}$ in the same way as other observables.

However, the correlated hadronic uncertainties in hard-spectator scattering and in annihilation can still be correlated among the two modes based on the following arguments.
Firstly, the amplitudes of $\bar{B_d}\rightarrow \bar{K}^{*0}\phi$ and $\bar{B_s}\rightarrow K^{*0}\phi$ given in Eq.~(\ref{eq:QCDFstructure}) are dominated by the coefficients $a^c_4$. Both decay channels are given in terms of the same set of coefficients, i.e. $a_3, a^p_4, a_5, a^p_7, a^p_{10}, b_3, b^{\rm EW}_3$ (for $p=u,c$), although their evaluation is not identical: for $\bar{B_s}\rightarrow K^{*0}\phi$, they are arranged in two different blocks including the prefactors $A_{\phi K^*}$ and $A_{K^*\phi}$, due to the two different topologies occurring for this decay. This  difference affects specially $a^p_4$ and $a^p_{10}$ where the roles of the emitted meson and the meson receiving the spectator quark are swapped between the $B_d$ and $B_s$ modes. It
is conceptually important but numerically small
as this swapping only changes the Gegenbauer moments entering both amplitudes, with only small numerical differences, and the normalisation factors $A_{\phi K^*}$ and $A_{K^*\phi}$ differing by about $10\%$.
The annihilation topologies are identical besides the fact that the initial state for $\bar{B_s}\rightarrow K^{*0}\phi$ involves the decay constant $f_{B_s}$ whereas $\bar{B_d}\rightarrow \bar{K}^{*0}\phi$
depends on $f_{B_d}$, where the numerical difference is also about $10\%$. In view of these facts, we expect that the correlation  between the hadronic uncertainties holds also for the branching ratios involved in $L_{K^*\phi}$, so that this observable should be fairly protected from hadronic uncertainties.

\subsection{Comparison with $L_{K^*\bar{K}^*}$ and $L_{K\bar{K}}$}

As mentioned above, the deviation between the theoretical and the experimental measurements for $L_{K^*\phi}$ is $\sim 1.5 ~\sigma$ which is less than the corresponding deviations for the $L_{K^*\bar{K}^*}$ ($2.6\sigma$) and $L_{K\bar{K}}$ ($2.4~\sigma$) observables reported in Refs.~\cite{Alguero:2020xca,Biswas:2023pyw}. For all three modes, the central values of the experimental measurements and theoretical values obtained from QCDF are different. But the significance of the discrepancy depends on the uncertainties in each case, or in other words the spread of the distributions. We will now discuss and interpret the differences among the three cases.

Concerning the theoretical predictions, one should notice that the central values for $L_{K^*\bar{K}^*}$ and $L_{K^*\phi}$ are quite close to each other and both have comparable uncertainties. $L_{K\bar{K}}$ has a central value that is somewhat higher with manifestly lesser and almost symmetric uncertainties. Besides the difference in mass for $\phi$ and $K^*$ mesons, this is predominantly due to the form factors used as inputs for the QCDF predictions. $L_{K\bar{K}}$ involves the $B\to K$ form factors which have very recently been calculated by lattice~\cite{Parrott:2022rgu}, leading to estimates much more precise than $B_{d(s)}\to K^*,\phi$ form factors. 
For the contributing topologies shown in Fig.~\ref{fig:topologies}, one can see that the $B_s\to\phi$ form factor only affects the $b\to d$ transition in the denominator of $L_{K^*\phi}$ (Fig.~\ref{fig:topologies}c). Although this is not the only form factor present in the denominator, it enters the dominant topology through the combination $a_4^p-\frac{1}{2}a_{10}^p$. The other form factors ($B_{d,s}\to K^*$) needed for $L_{K^*\phi}$ are also used to predict $L_{K^*\bar{K}^*}$. 
It is thus not very surprising that that the central value and uncertainties of the theoretical predictions of $L_{K^*\phi}$ and $L_{K^*\bar{K}^*}$ are comparable. 

Another effect might explain further why the theoretical prediction for $L_{K^*\phi}$ is actually (slightly) more accurate than for $L_{K^*\bar{K}^*}$. In our earlier work on $L_{K\bar{K}}$ and $L_{K^*\bar{K}^*}$~\cite{Biswas:2023pyw}, we used form factor estimates at maximum recoil from Light-Cone Sum Rules (LCSR) taken from Ref.~\cite{Bharucha:2015bzk}. In this article, we use the corresponding form factors at $q^2=m_{V_2}^2$ \cite{Bartsch:2008ps} (where $V_2$ is the final state meson that does not pick up the spectator quark), taking the results from the combined fit of LCSR and Lattice results provided by the authors of Ref.~\cite{Bharucha:2015bzk}, with reduced uncertainties. This might account for the slightly smaller uncertainties on the theoretical prediction for $L_{K^*\phi}$ as compared to $L_{K^*\bar{K}^*}$.

On the experimental side, the two measured observables for vector-vector modes are found to be quite different. Indeed $L_{K^*\bar{K}^*}$ has a lower central value, which is almost half of that of $L_{K^*\phi}$. More strikingly, the uncertainties for the latter are much larger and more asymmetric than the former. 
The reason for this difference in the size of the uncertainties stem from the fact that in the experimental measurement for $L_{K^*\bar{K}^*}$ carried out in Ref.~\cite{Alguero:2020xca}, the authors used the ratio of branching ratios 
\begin{equation}\label{eq:BRratioKstarKstar}
\frac{\mathcal B_{B_d\to K^{*0}\bar{K}^{*0}}}{\mathcal B_{B_s\to K^{*0}\bar{K}^{*0}}} = 0.0758 \pm 0.0057 \stat \pm 0.0025 \syst \pm 0.0016 \, \left(\frac{f_s}{f_d} \right),
\end{equation}
measured by the LHCb collaboration~\cite{LHCb:2019bnl}. The correlation between these two modes was thus known experimentally and could be used in the prediction for the $U$-spin-based ratio $L_{K^*\bar{K}^*}$. Such a ratio/correlated measurement is currently unknown for the $K^*\phi$ final states. Hopefully a more precise experimental measurement for $L_{K^*\phi}$ could be obtained in the future once a correlated measurement of the corresponding branching ratios is available. The ratio of branching ratios needed could be particularly easy to measure at LHCb, with many systematics cancelling due to the same final state involved. A more precise measurement of the longitudinal polarisation fraction for $\bar{B}_s\to K^{*0}\phi$ will also help in reducing the experimental uncertainty.

Moreover, the increase in the central value for $L_{K^*\phi}$ as compared to $L_{K^*\bar{K}^*}$ is driven mainly by an increase in the numerator, due to a larger central value of $f_L(\bar{B}^0\to \bar{K}^{0*}\phi)$ with respect to $f_L(\bar{B}_s \to K^{*0}\bar{K}^{*0})$. Notice that 
 the experimental measurements for the full $b\to s,d$ branching ratios with $K^*\phi$ final states (provided in Table~\ref{tab:exp_inputs_Kstphi}) and the corresponding ones with $K^{*0}\bar{K}^{*0}$ final states (Table 2 of Ref.~\cite{Biswas:2023pyw}) are of similar sizes. 

As a conclusion, we see that differences in inputs for the form factors 
%(for the theoretical part) 
and in the knowledge of experimental correlations %(for the experimental part) 
explain why the $L$ observables exhibit different accuracies in the $K^*\bar{K}^*$, $K\bar{K}$ and $K^*(\bar{K}^*)\phi$ cases and deviations between SM and data.

%, leading to deviations of various significances between the SM and data.

\subsection{Penguin weak annihilation and universality}\label{sec:ann}
Before turning to the interpretation of these deviations in terms of NP, we would like to discuss further the possible impact of hadronic uncertainties on our optimised observables.

QCDF proceeds through a $1/m_B$ expansion where the leading-order term for longitudinally polarised vector-vector modes can be expressed in terms of  (universal nonperturbative) form factors and light-cone distribution amplitudes together with (pertubatively computable) hard-scattering kernels. However, subleading terms (spectator scattering and weak annihilation) can be numerically important: they exhibit infrared divergences which encode long-distance contributions that must be modelled and contribute significantly to the uncertainties of QCDF predictions. It is thus useful to check whether the $L$-observables are affected by the modelling of these contributions, in particular weak annihilation which was shown to be important for $K^*\bar{K}^*$ longitudinal branching ratios in Ref.~\cite{Alguero:2020xca}.

In Ref.~\cite{Beneke:2006hg}, the negative helicity  QCD  penguin annihilation amplitude defined by\\
$\beta^{c-}_3=B_{K^*\phi}/A_{K^*\phi}b^{c-}_3$ is assumed to be independent of the final state considered, and it is described as convolutions of hard-scattering kernels and light-cone distribution amplitudes which are divergent at the endpoints.
~This signals the breakdown of factorisation in this (power-suppressed) contribution, and these endpoint divergences corresponding to long-distance contributions are treated by replacing the endpoint divergences with the following model:
\begin{equation}\label{eq:XAmodel}
X_A=\left(1+\rho_A e^{i\varphi_A}\right) \ln{\frac{m_B}{\Lambda_h}}
\end{equation}
where $\Lambda_h=0.5$ GeV. In Ref.~\cite{Beneke:2006hg},
this endpoint divergence in the penguin annihilation term is assumed to be universal for the various 
polarisations. 

In order to explain the longitudinal polarisation fraction of $K^*\phi$ close to $50\%$ within the SM, one needs to assume a large negative helicity amplitude from unexpectedly large hadronic contributions. A large spectator scattering is not favoured in the case of the $B \to \rho \rho$ mode, as it has a polarisation fraction measured close to one~\cite{Belle:2006xxx}  and its transverse amplitude is dominantly affected by spectator scattering. Ref.~\cite{Beneke:2006hg} analysed both $B\to K^*\phi$ and $B\to K^*\rho$ data available at that time to check if they were in agreement with the universality assumption and to determine the parameters in Eq.~(\ref{eq:XAmodel}). All data showed a good agreement with a universal $X_A$ with the following ranges of parameters
\begin{equation}
\rho_A=0.5 \pm 0.2\,, \qquad \varphi_A=(-43\pm 19)^\circ\,.
\end{equation}

For our determination of $L$ observables, we actually need a much weaker version of universality,
which consists of uniform distributions for the parameters $\rho_A$ and $\varphi_A$ in the ranges
\begin{eqnarray}
0 \leq \rho_A\leq 1, \quad 0 \leq \varphi_A \leq 2 \pi,
\label{eq:annihilation_main}
\end{eqnarray}

We assume Eq.~(\ref{eq:annihilation_main})  to hold only among decays with the same final states (not all decays) and for the longitudinal polarisation only (and not all transverse and longitudinal polarisations).

For comparison we collect here the SM predictions in our more conservative set-up as in Eq.~(\ref{eq:annihilation_main}):
\begin{equation}
L_{K^*K^*}=19.54^{+9.09}_{-6.64}
\quad
L_{K^*\phi}=22.01^{+6.91}_{-4.81} \, 
\label{eq:Annihilation_scheme1}
\end{equation}
and using the stronger version of universality as in Ref.~\cite{Beneke:2006hg} but with our updated inputs:
\begin{equation}
L_{K^*K^*}=19.54^{+9.18}_{-6.45}
\quad
L_{K^*\phi}=21.89_{-4.38}^{+5.64
}
\label{eq:Annihilation_scheme2}
\end{equation}

We observe that the different models of weak annihilation affect neither the central values nor the uncertainties for the $L_{K^{*}\bar{K}^{*}}$ observables. This shows the robustness of this observable with respect to hadronic uncertainties, as encoded in the weak annihilation term $X_A$.  However, the observable $L_{K^*\phi}$ is slightly more sensitive to the annihilation scheme chosen, as the intervals of the asymmetric uncertainties are marginally reduced in Eq.~(\ref{eq:Annihilation_scheme2}) compared to Eq.~(\ref{eq:Annihilation_scheme1}). The impact remains however very limited and suggests that the modelling of the annihilation contribution has a limited impact on the hadronic uncertainties for the optimised observables discussed here.

\section{EFT analysis of NP in $B_{d,s}$-decays into $K^*\phi$, $K^*\bar{K}^*$, $K\bar{K}$ final states }\label{sec:LKstphi_dis}

We will try now to identify the sources of NP that could be responsible for the pattern of deviations observed in the three optimised $L$ observables corresponding to the $K^{(*)}\bar{K}^{(*)}, K^*(\bar{K}^*)\phi$ final states. Our analysis of longitudinal branching ratios for $K^*\phi$ final states in Sec.~\ref{sec:longpolandbrKstphi} suggests that $b\to d$ transitions might be more affected than $b\to s$ transitions. However, we prefer to keep the discussion general, assuming that NP may enter both $b\to d$ and $b \to s$ transitions and using  the $L$ observables as a guideline to determine the most promising NP scenarios.

The Hamiltonian that describes a general $b\to s,d$ decay is provided in Eq.~(\ref{eq:wet}) in appendix~\ref{app:WET}. For simplicity, we consider the following limiting assumptions: 
\begin{itemize}
    \item NP can only enter the operators already present in the weak effective theory (WET) for the SM at the $m_b$ scale (e.g., no right-handed currents or scalar operators).
    \item NP contributions to the short-distance Wilson coefficients at this scale are real (i.e., no new sources of CP violation).
    \item NP affects only the same operator for both $b\to s,d$ transitions in a predominant way.
\end{itemize}

Since the SM value for the $L_{K^*\phi}$ observable is rather close to its experimental measurement, an NP shift in many different pairs of Wilson coefficients ($C_{id}^{\rm NP}$, $C_{is}^{\rm NP}$) is certainly able to account for the (small) $1.48~\sigma$ deviation in $L_{K^*\phi}$. The situation is less open once we try to explain this result in the same framework with $L_{K^*\bar{K}^*}$ and $L_{K\bar{K}}$, which can be explained only with $C_{4f}^{\rm NP}$ or $C_{8gf}^{\rm NP}$ (with $f=d,s$) following Refs.~\cite{Alguero:2020xca,Biswas:2023pyw}.

\begin{figure}[t]\centering
\subfloat[]{\label{fig:c4_LKstKst}\includegraphics[width=0.5\textwidth,height=0.40\textwidth]{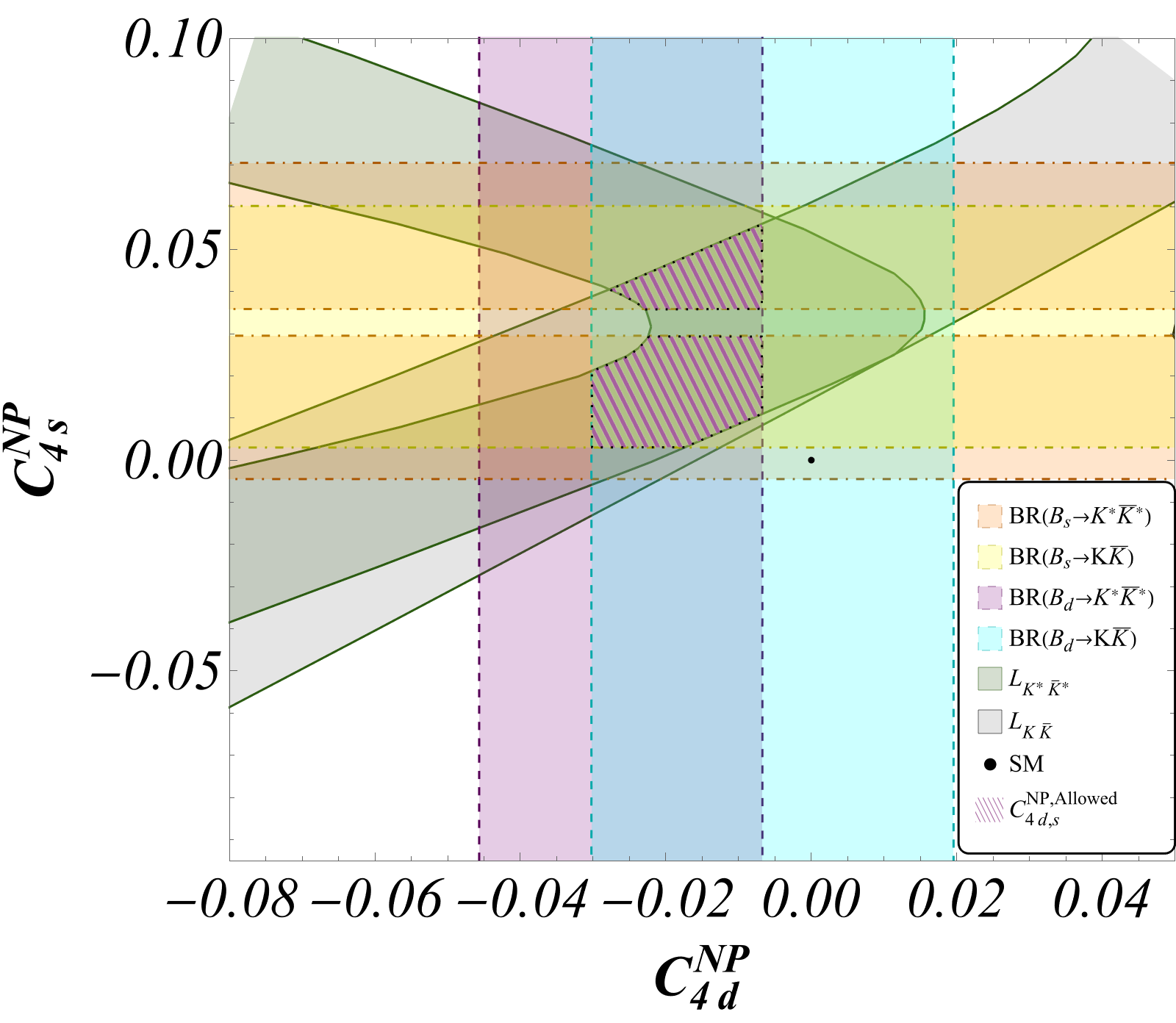}}~~~~~~
\subfloat[]{\label{fig:c4_LKstphi}\includegraphics[width=0.5\textwidth,height=0.40\textwidth]{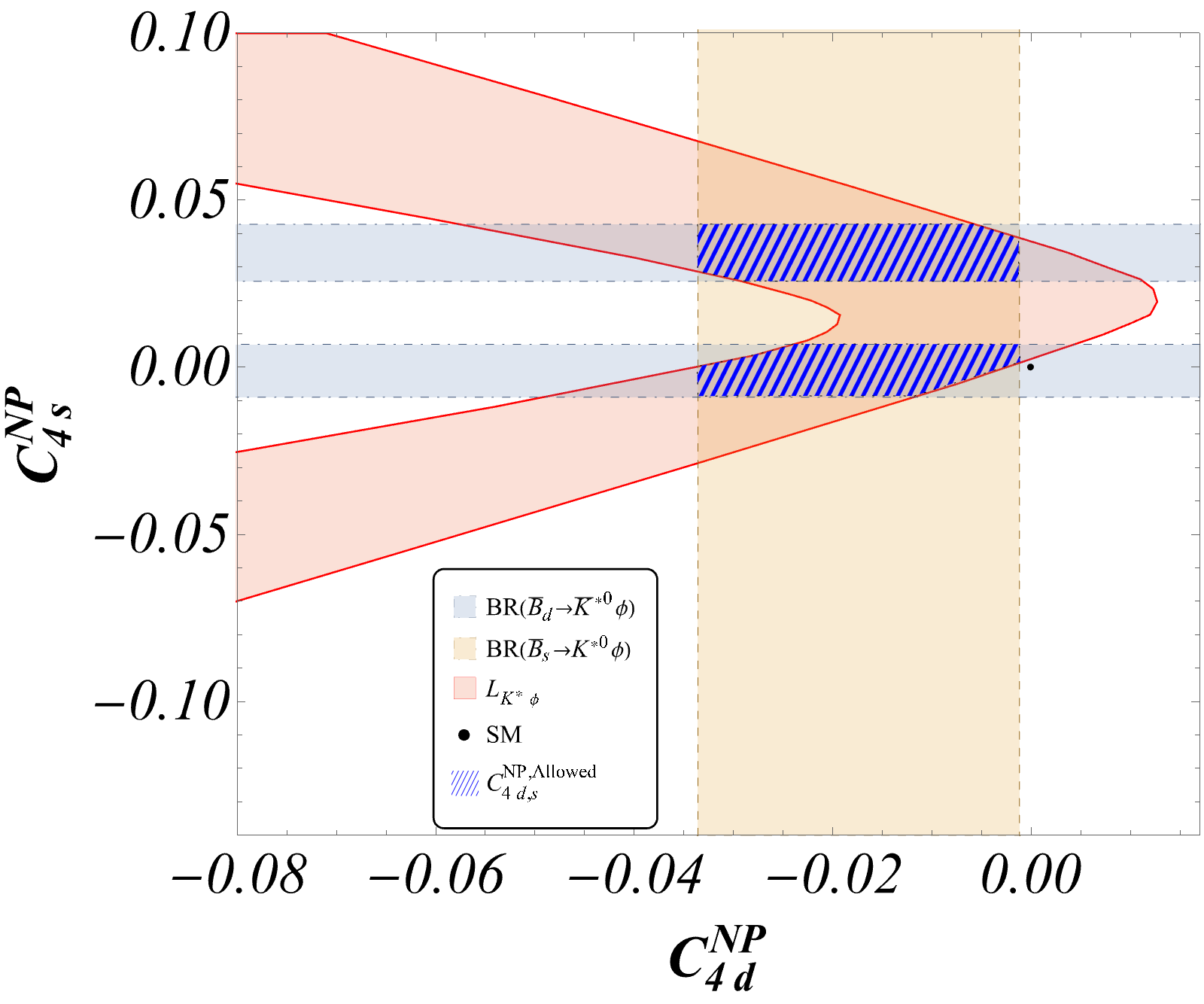}}\par 
\subfloat[]{\label{fig:c4_all}\includegraphics[width=0.5\textwidth,height=0.40\textwidth]{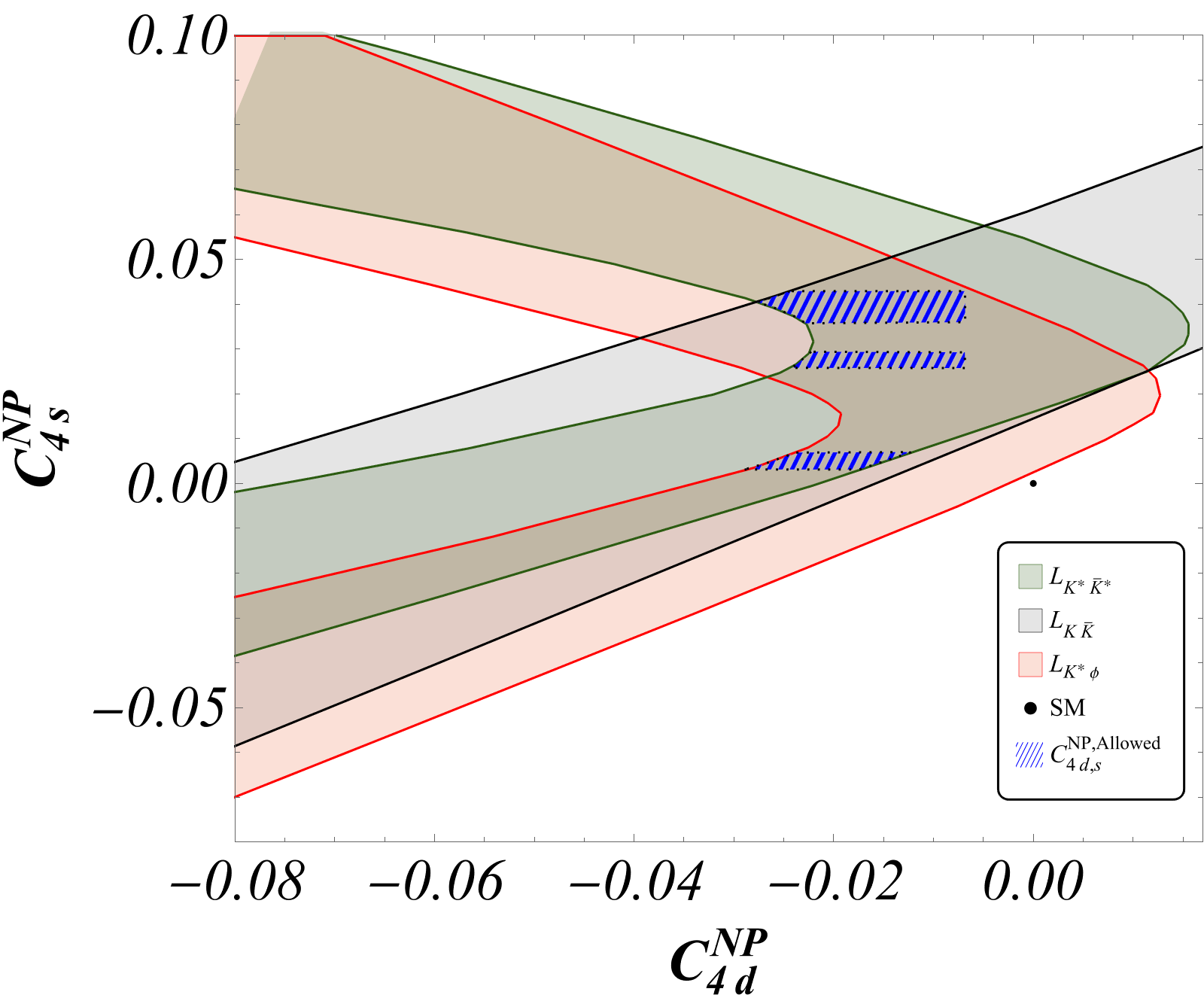}}
\caption{%We display the common region in the $C_{4d}^{\rm NP}-C_{4s}^{\rm NP}$ plane 
%that offers a simultaneous explanation for $L_{K^{*)}\bar{K}^{(*)}}$ and $L_{K^*\phi}$ along with all the corresponding $b\to d,s$ branching ratios. 
Fig.~\ref{fig:c4_LKstKst}: allowed (magenta, right-hatched) region in the $C_{4d}^{\rm NP}-C_{4s}^{\rm NP}$ plane considering the constraints from  $L_{K^{(*)}\bar{K}^{(*)}}$ along with the corresponding $b\to d,s$ branching ratios. Fig.~\ref{fig:c4_LKstphi}: same (blue, left-hatched) region for the $K^{(*)}\phi$ final states. Fig.~\ref{fig:c4_all}: common (blue, left-hatched) region satisfying the constraints from the $L$ observables and the branching ratios (not displayed) for all the three modes. The black dot corresponds to the SM point.}
\label{fig:c4comb}
\end{figure}
\begin{figure}[t]\centering
\subfloat[]{\label{fig:c8_LKstKst}\includegraphics[width=0.5\textwidth,height=0.40\textwidth]{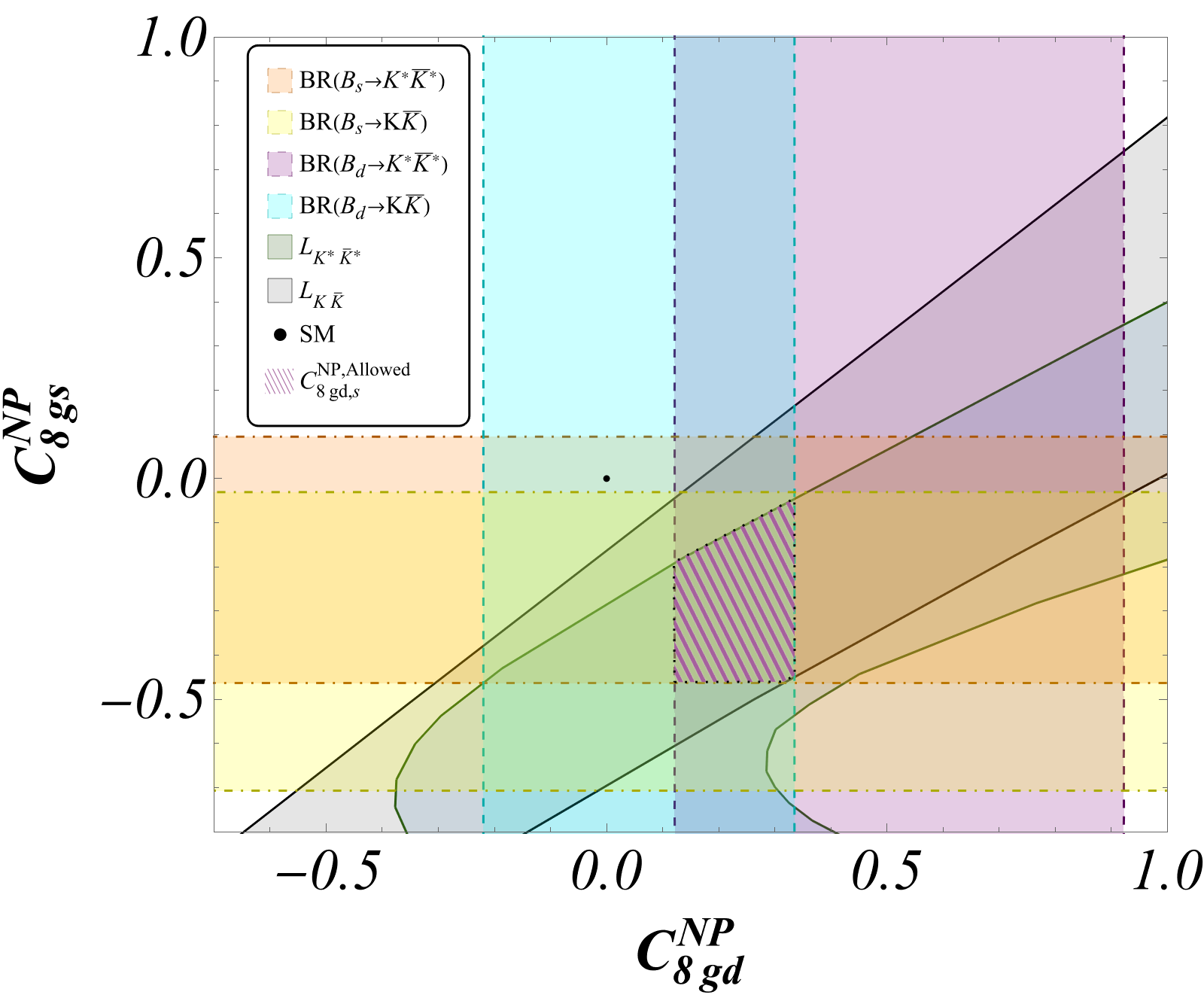}}~~~~~~
\subfloat[]{\label{fig:c8_LKstphi}\includegraphics[width=0.5\textwidth,height=0.40\textwidth]{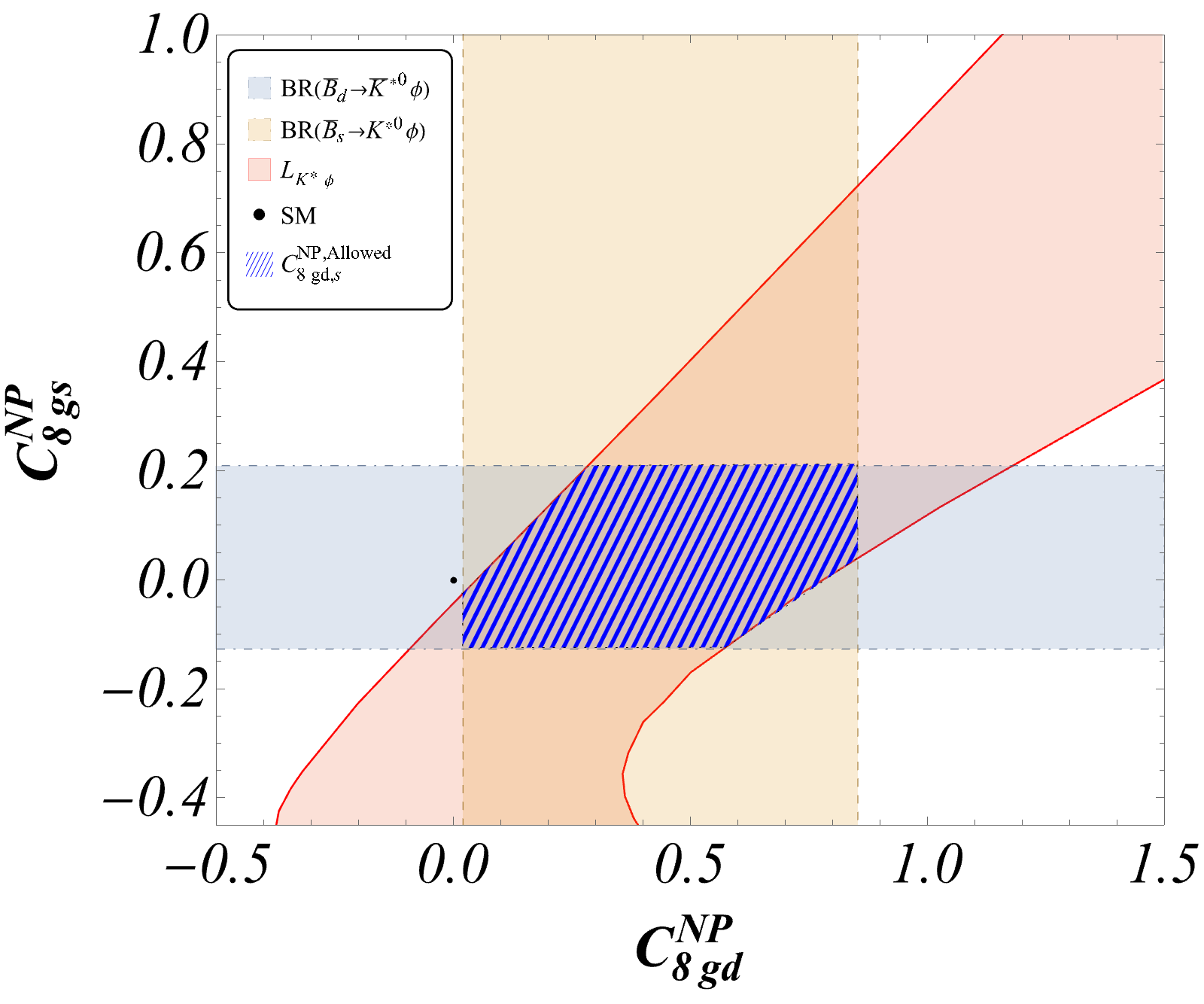}}\par 
\subfloat[]{\label{fig:c8_all}\includegraphics[width=0.5\textwidth,height=0.40\textwidth]{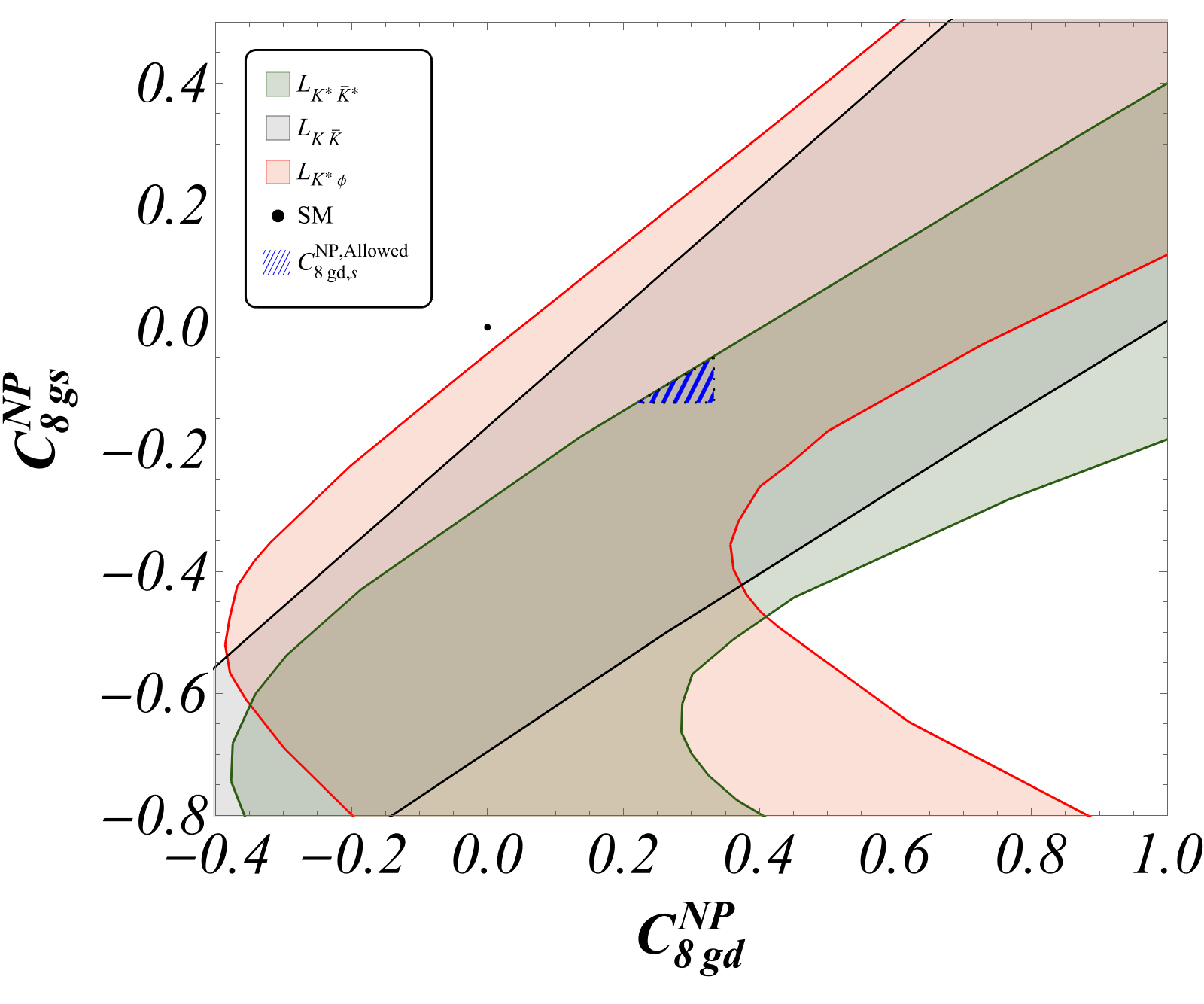}}
\caption{
Fig.~\ref{fig:c8_LKstKst}: allowed (blue) region in the $C_{8gd}^{\rm NP}-C_{8gs}^{\rm NP}$ plane
considering the constraints from  
$L_{K^{(*)}\bar{K}^{(*)}}$ along with the corresponding $b\to d,s$ branching ratios. Fig.~\ref{fig:c8_LKstphi}: same (blue) region for the $K^{(*)}\phi$ final states. 
Fig.~\ref{fig:c8_all}: common (blue) region satisfying the constraints from the $L$ observables and the branching ratios (not displayed) for all the three modes. The dot corresponds to the SM point.
}
\label{fig:c8comb}
\end{figure}

The region in the $C_{4d}^{\rm NP}-C_{4s}^{\rm NP}$ plane (or $C_4^{\rm NP}$ plane in the following) explaining the three $L$ observables along with all the corresponding longitudinal branching ratios is displayed in Fig.~\ref{fig:c4comb}. The bands in this figure (and the following) are obtained by drawing a contour for the values of Wilson coefficients for which there is an overlap of the 1~$\sigma$ intervals for the theoretical prediction and the experimental measurement of the observable of interest 
(branching ratio or $L$).
This approach is chosen for its simplicity, in order to determine which scenarios could be of interest to explain the pattern of deviations observed among these penguin-mediated modes. A more thorough statistical treatment (including a more elaborate treatment of the asymmetric distributions under consideration) would be required if we wanted to provide a precise statistical meaning to the (updated) common regions shown in Figs.~\ref{fig:c4comb} and~\ref{fig:c8comb}. Such ambitious statistical analysis is beyond the scope of this paper and will be presented elsewhere.

Fig.~\ref{fig:c4_LKstKst} displays the allowed region in the $C_4^{\rm NP}$ plane that offers a simultaneous explanation of both $L_{K^{(*)}K^{(*)}}$ along with the corresponding (longitudinal for $K^*\bar{K}^*$) branching ratios~\footnote{This plot corresponds to Fig.~20 of Ref.~\cite{Biswas:2023pyw} and it is reproduced here for comparison.}. Fig.~\ref{fig:c4_LKstphi} shows the region that can simultaneously explain $L_{K^*\phi}$ and the related longitudinal branching ratios involved in the construction of this observable. In Fig.~\ref{fig:c4_all}, we depict the allowed region in the $C_4^{\rm NP}$ plane that takes into account the constraints from all the three $L$ ratios and the corresponding longitudinal branching ratios~\footnote{We refrain from showing the horizontal and vertical bands associated with
$b\to s$ and $b\to d$ branching ratios in Fig.~\ref{fig:c4_all} to avoid overcrowding the plot.}. Similarly Fig.~\ref{fig:c8comb} depicts the allowed region in the $C_{8gd}^{\rm NP}-C_{8gs}^{\rm NP}$ plane (or $C_{8g}^{\rm NP}$ plane) which offers a simultaneous explanation of the $L_{K^{(*)}\bar{K}^{(*)}}$ and the corresponding longitudinal branching ratios (Fig.~\ref{fig:c8_LKstKst}), the allowed region consistent with constraints from $L_{K^*\phi}$ and the related longitudinal branching ratios (Fig.~\ref{fig:c8_LKstphi}) and the allowed region obtained after taking into account all the three $L's$ and the corresponding longitudinal branching ratios (Fig.~\ref{fig:c8_all}).

A few comments are in order regarding Figs.~\ref{fig:c4comb} and~\ref{fig:c8comb}:
\begin{itemize}
\item In Figs.~\ref{fig:c4_LKstKst} and ~\ref{fig:c4_LKstphi}  one can see two horizontal bands of the same colour. Indeed the  branching ratios have a quadratic dependence on the NP Wilson coefficients, allowing for two solutions for the Wilson coefficients for a given value of the branching ratios. In Fig.~\ref{fig:c8_LKstKst} and~\ref{fig:c8_LKstphi}
only one band is displayed because the second one requires a large contribution from NP (more than 3 times the magnitude of the SM value).
\item The two common regions in Fig.~\ref{fig:c4_LKstKst} from the $K^{(*)}\bar{K}^{(*)}$ modes get further reduced and split in three bands when the constraints from the $K^*(\bar{K}^*)\phi$ modes (Fig.~\ref{fig:c4_LKstphi}) are taken into account, leading to Fig.~\ref{fig:c4_all}.  It can be seen that there is an overlapping area for which $C_{4s}^{\rm NP}$ is very close to $0$ while keeping $C_{4d}^{\rm NP}$ around -0.02.
For the two remaining NP bands, both Wilson coefficients stay away from zero values with $C_{4s}^{\rm NP}$ ($C_{4d}^{\rm NP}$ ) preferring positive (negative) values.
In each region the range of allowed values for $C_{4s}^{\rm NP}$ is narrower than for $C_{4d}^{\rm NP}$, due to the fact that the SM value for the longitudinal branching ratio of the $\bar{B}_d\to\bar{K}^{*0}\phi$ decay is consistent with the data at the $0.35~\sigma$ level, while  $\bar{B}_s\to K^{*0}\phi$ is consistent at $1.26~\sigma$. 
\item Unlike Fig.~\ref{fig:c4comb}, only one common region is shown in the $C_{8g}^{\rm NP}$ plane for Fig.~\ref{fig:c8comb}. As already mentioned, the second region would require too large a contribution from NP. Similarly to the $C_4^{\rm NP}$ case, the range of allowed values for $C_{8gd}^{\rm NP}$ is (slightly) larger than that for $C_{8gs}^{\rm NP}$.
\item We can see a substantial reduction of the allowed range of values of NP Wilson coefficients when one includes the observables for the $K^*(\bar{K}^*)\phi$ final state, compared to what is allowed for the $K^{(*)}\bar{K}^{(*)}$ modes. The SM predictions for $L_{K^*\phi}$ and the corresponding longitudinal branching ratios (as computed within the framework of QCDF) are much closer to their experimental counterparts than in the case of $L_{K^{(*)}\bar{K}^{(*)}}$ observables and their corresponding branching ratios. As will be discussed later on, it is very important that LHCb provides an updated measurement for the $\bar{B}_d \to  \bar{K}^{*0}\phi$ mode, that so far has only been measured by Babar, Belle and CLEO.
\end{itemize}

In the near future one can foresee some improvements in the results presented above. There are  correlations that cannot be accounted for at present, which might change the allowed regions. For instance, the experimental measurements for the total branching ratios and longitudinal polarisations presented in Table~\ref{tab:BR_est} for the $K^*\phi$ final state are completely uncorrelated between the $B_d$ and $B_s$ modes, as they are obtained as an average over the measurements of different experimental collaborations. A measurement of these observables including their correlations would
change the accuracy of the measured value for $L_{K^*\phi}$, leading to a change in  
the width of the bands in the planes of NP Wilson coefficients and thus the range of the allowed values in the different scenarios.

\section{Including pseudoscalar-vector $B_{d,s}$ modes}\label{sec:PV_VP}

In order to constrain the NP Wilson coefficients $C_{4d,s}^{\rm NP}$ and $C_{8gd,s}^{\rm NP}$ one should take into account as many non-leptonic modes as possible which depends on these Wilson coefficients in a significant and controlled manner. Figs.~\ref{fig:c4_all} and \ref{fig:c8_all} are thus incomplete since they consider the branching ratios and $L$-observables corresponding only to the $K^{(*)0}\bar{K}^{(*)0}$ and $K^{*0}\phi$ modes, but:
\begin{itemize}
\item The mode $\bar{B_d}\to\bar{K}^0\phi$ is not considered. A PDG average~\cite{Workman:2022ynf} of 2012 Babar~\cite{BaBar:2012iuj} and 2003 Belle~\cite{Belle:2003ike} measurements 
exists for $\mathcal{B}(\bar{B_d}\to\bar{K}^0\phi)$:
\begin{equation}
\mathcal{B}(\bar{B_d}\to\bar{K}^0\phi)^{\rm exp}=(7.3\pm 0.7)\times 10^{-6}.
\label{eq:Kphi_exp_neutral}
\end{equation}
\item The untagged $\mathcal{B}({B_s}\to K^{*0}\bar{K}^0$) transition measured by LHCb~\cite{LHCb:2019vww} is not included:
\begin{equation}\label{mixedexp}
\mathcal{B}({B_s}\to K^{*0}\bar{K}^0)^{\rm exp}
+ \mathcal{B}({B_s}\to \bar{K}^{*0}{K}^0)^{\rm exp}
=(1.98\pm 0.28\pm 0.50)\times 10^{-5}.
\end{equation}
\end{itemize}
In the following subsections, we will look at the impact of these pseudoscalar-vector modes one after the other.

\subsection{Constraints on $C_{4s,8gs}^{\rm NP}$ from $\mathcal{B}(\bar{B}_d\to\bar{K}^{0}\phi)$} \label{subsec:BRBdKphi}
 We did not include the constraint from $\mathcal{B}(\bar{B}_d\to\bar{K}^{0}\phi)$ in Sec.~\ref{sec:LKstphi_dis} since the corresponding branching ratio for the $\bar{B_s}\to K^0\phi$ mode remains unmeasured till date, thus hindering us from quoting an experimental measurement for a corresponding optimised $L_{K\phi}$ observable. However, the measured branching ratio for $\mathcal{B}(\bar{B_d}\to\bar{K}^0\phi)$ itself sets additional constraints on $C_{4s,8gs}^{\rm NP}$ on top of the ones displayed in Figs. \ref{fig:c4_all} and $\ref{fig:c8_all}$.
 
\begin{figure}[t]\centering
\subfloat[]{\label{fig:c4_LKstphi_BdKphi}\includegraphics[width=0.5\textwidth,height=0.40\textwidth]{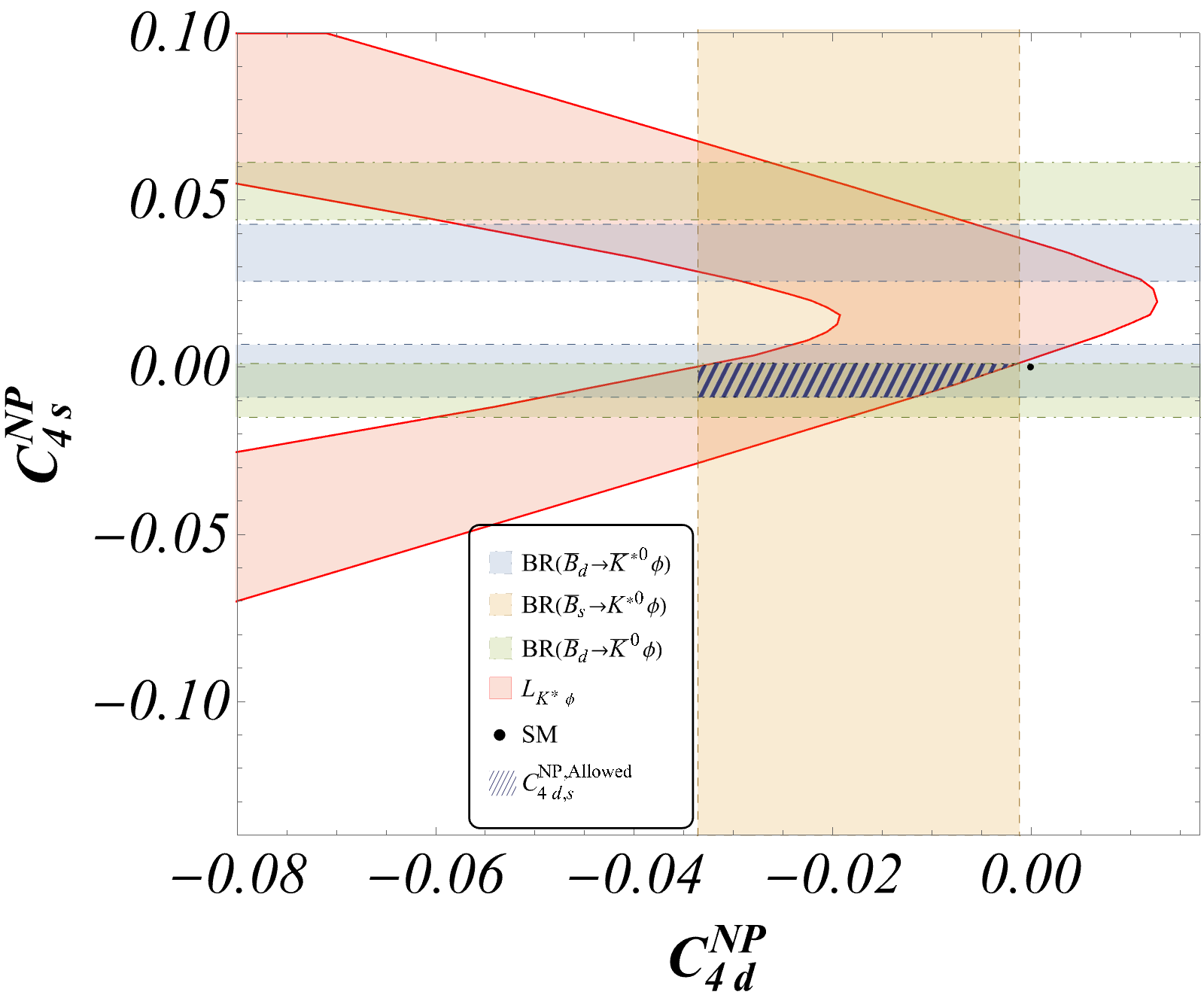}}~~~~~~
\subfloat[]{\label{fig:c4_all_BdKphi}\includegraphics[width=0.5\textwidth,height=0.40\textwidth]{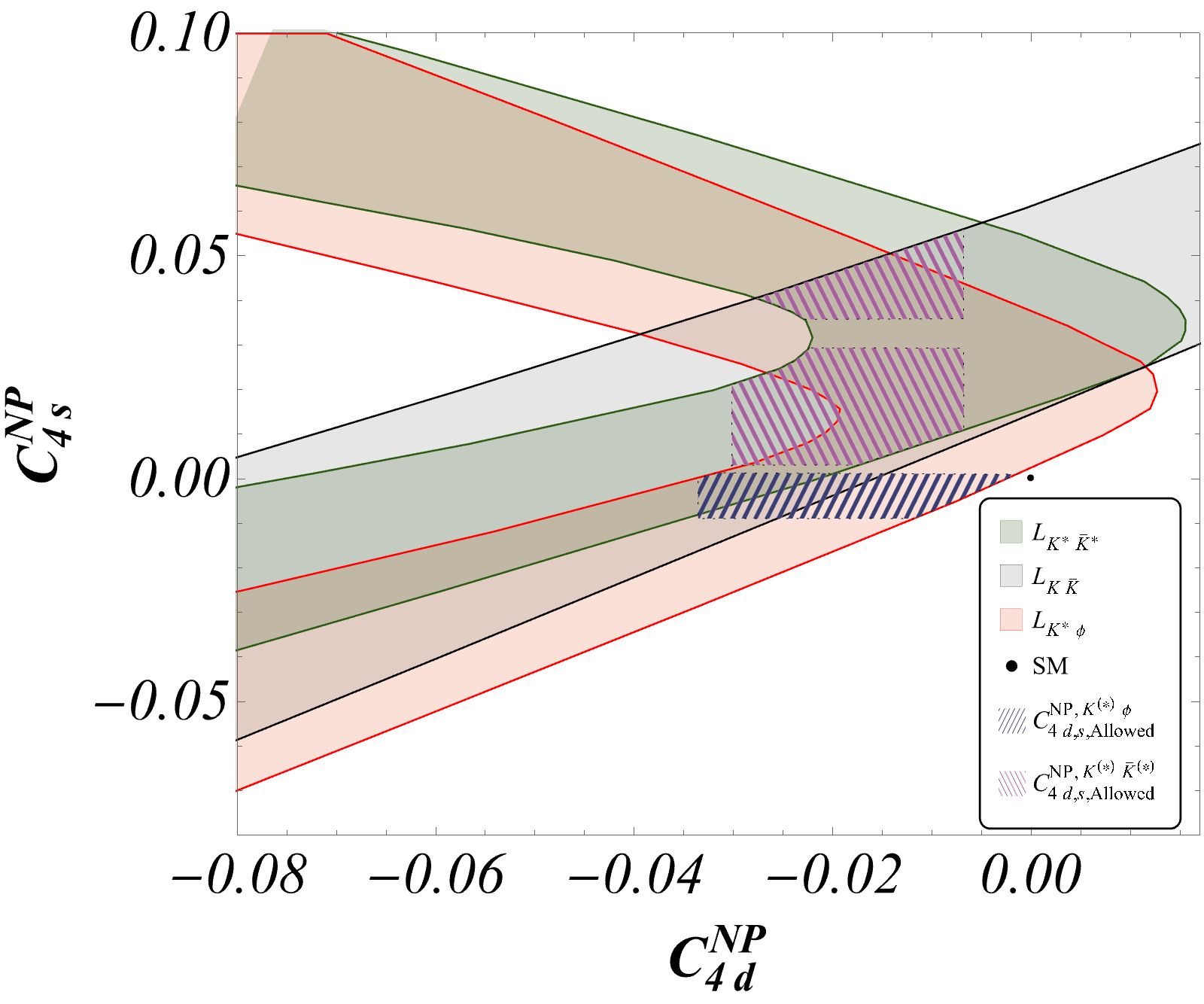}}\par 
\subfloat[]{\label{fig:c8_LKstphi_BdKphi}\includegraphics[width=0.5\textwidth,height=0.40\textwidth]{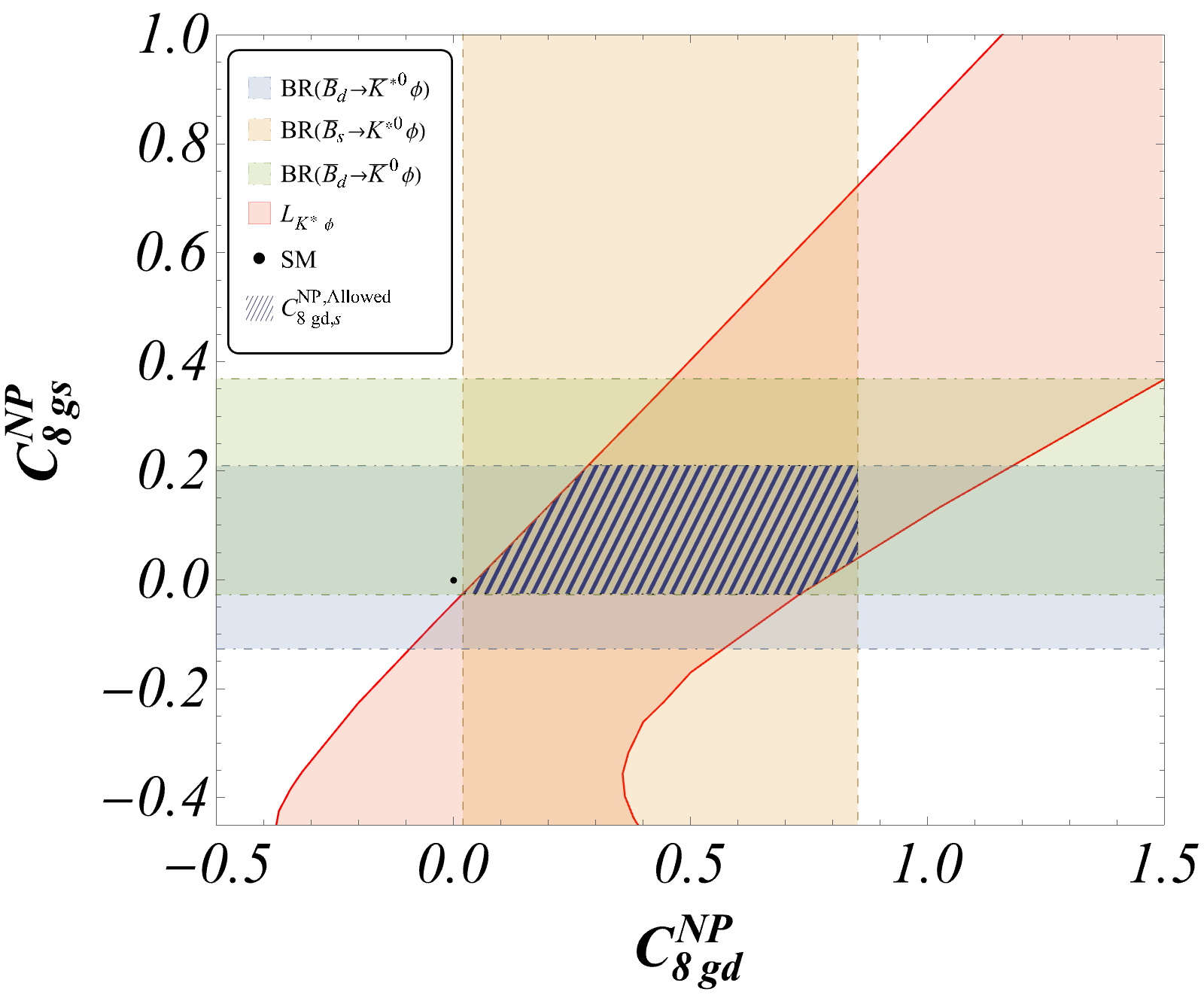}}~~~~~~
\subfloat[]
{\label{fig:c8_all_BdKphi}\includegraphics[width=0.5\textwidth,height=0.40\textwidth]{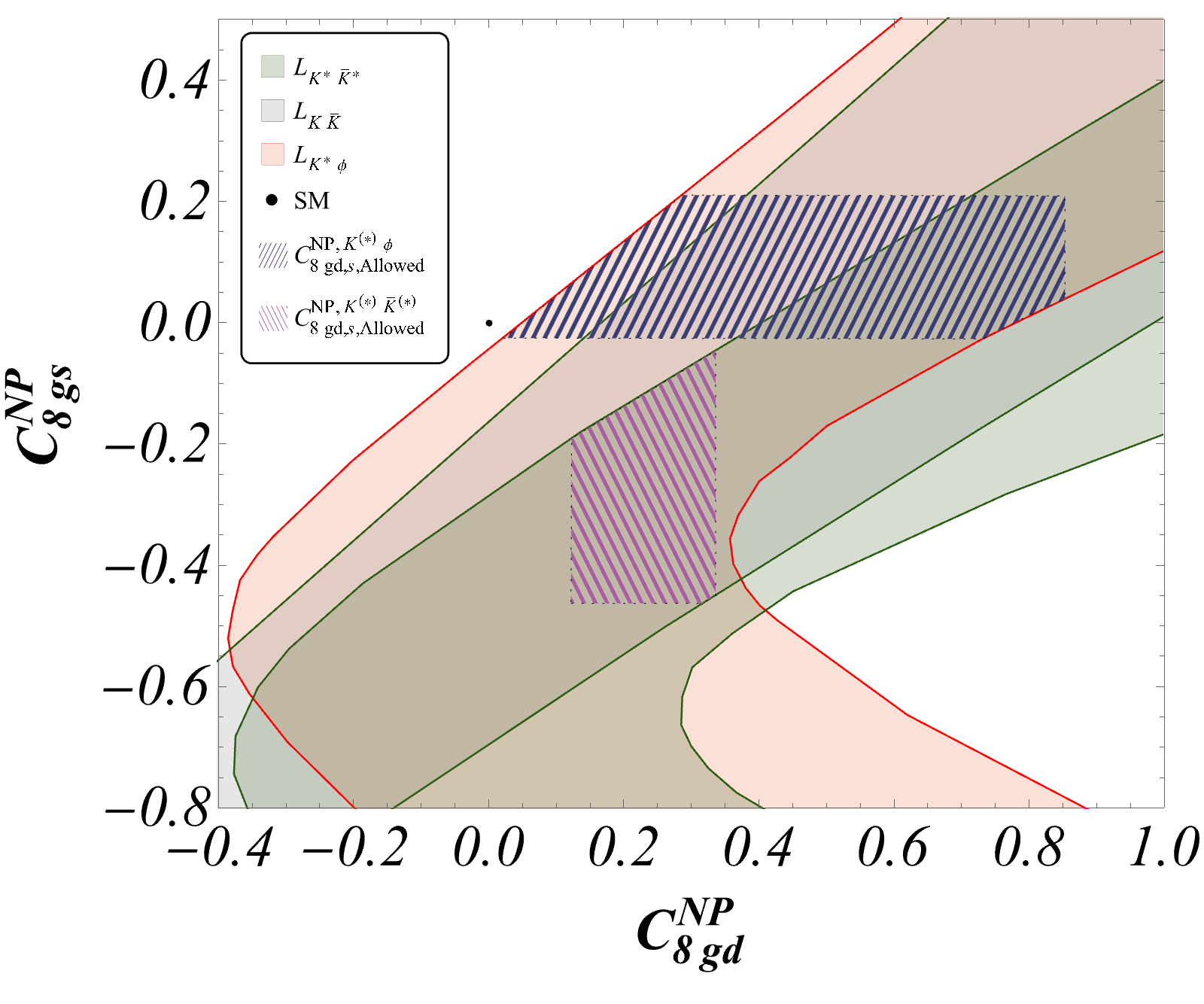}}
\caption{
Impact of including the constraint from $\mathcal{B}(\bar{B}_d\to\bar{K}^0\phi)$ on Figs.~\ref{fig:c4_LKstphi},~\ref{fig:c4_all},~\ref{fig:c8_LKstphi} and~\ref{fig:c8_all}. We find that the common blue hatched regions in Figs.~\ref{fig:c4_LKstphi} and~\ref{fig:c8_LKstphi} are reduced to the dark-blue hatched regions in Figs.~\ref{fig:c4_LKstphi_BdKphi} and~\ref{fig:c8_LKstphi_BdKphi} respectively. However, the blue hatched common regions in Figs.~\ref{fig:c4_all} and~\ref{fig:c8_all} completely vanish on including this constraint. It can be clearly seen from Figs.~\ref{fig:c4_all_BdKphi} and~\ref{fig:c8_all_BdKphi} that the region common to the observables $\mathcal{B}(\bar{B}_{d(s)}\to\bar{K}^{(*)0}(K^{*0})\phi)$ and $L_{K^*\phi}$ (dark-blue hatched) have no overlap with the region common to the $\mathcal{B}(\bar{B}_{d(s)}\to K^{(*)0}\bar{K}^{(*)0})$ and $L_{K^{(*)}\bar{K}^{(*)}}$ observables (magenta hatched).}
\label{fig:BdKphi_effect}
\end{figure}

We can predict this branching ratio within the SM in QCDF:
\begin{equation}
\mathcal{B}(\bar{B_d}\to\bar{K}^0\phi)^{\rm th}=(4.28^{+2.71}_{-1.50})\times 10^{-6},
\label{eq:Kphi_SM_neutral}
\end{equation}
using the same inputs and approach as for the other decay modes. The deviation of the SM prediction from the current experimental measurement is $1.31~\sigma$.  It is worth noticing that the corresponding deviation between the theoretical and experimental numbers for $\mathcal{B}(\bar{B_d}\to\bar{K}^{*0}\phi)$ is less significant than for $\mathcal{B}(\bar{B_d}\to\bar{K}^{0}\phi)$.

The resulting constraints in the $C_{4,8g}^{\rm NP}$ planes are shown in Fig.~\ref{fig:BdKphi_effect}. Figs.~\ref{fig:c4_LKstphi} and \ref{fig:c4_all} are superseded by Figs.~\ref{fig:c4_LKstphi_BdKphi} and \ref{fig:c4_all_BdKphi} while 
Figs.~\ref{fig:c8_LKstphi} and \ref{fig:c8_all} change to Figs.~\ref{fig:c8_LKstphi_BdKphi} and \ref{fig:c8_all_BdKphi}. The main impact of adding the constraint from this measurement is the following:
\begin{itemize}
\item In both $C_4^{\rm NP}$ and $C_{8g}^{\rm NP}$ planes, there exist regions of common overlap between the $K^*\phi$ observables ($\mathcal{B}(\bar{B}_{s(d)}\to K^{*0}(\bar{K}^{*0})\phi)$, $L_{K^*\phi}$) and $\mathcal{B}(\bar{B_d}\to\bar{K}^{0}\phi)$ as can be seen from Figs.~\ref{fig:c4_LKstphi_BdKphi} and~\ref{fig:c8_LKstphi_BdKphi} respectively.
\item However, once the constraints from the $K^{(*)0}\bar{K}^{*(0)}$ observables ($\mathcal{B}(\bar{B}_{s(d)}\to K^{(*)0}\bar{K}^{(*)0})$, $L_{K^{(*)}\bar{K}^{(*)}}$) are added, we find that no common region of overlap exists anymore. This is displayed in Figs.~\ref{fig:c4_all_BdKphi} and~\ref{fig:c8_all_BdKphi} for the  $C_4^{\rm NP}$ and $C_{8g}^{\rm NP}$ planes respectively, where the magenta hatched region represents the region of common intersection between the $K^{(*)0}\bar{K}^{*(0)}$ observables while the dark-blue hatched area stands for the NP region of overlap between the $K^{(*)}\phi$ observables.
\end{itemize}

In order to understand the tensions between the constraints with and without $\mathcal{B}(\bar{B}_d\to\bar{K}^{0}\phi)$, we consider the functional dependencies of the observables $\mathcal{B}(\bar{B}_d\to\bar{K}^{(*)0}\phi)$ on NP, taking only central values for the input parameters:
\begin{eqnarray}
10^6\times\mathcal{B}(\bar{B_d}\to\bar{K}^0\phi)^{\rm th,cv}=&4.89+6955.76 (C_{4s}^{\text{\rm NP}})^2&+13.08 (C_{8\text{gs}}^{\text{\rm NP}})^2-339.04 C_{4s}^{\text{\rm NP}}+15.30C_{8\text{gs}}^{\text{\rm NP}}\nonumber\\&-599.74 C_{4s}^{\text{\rm NP}} C_{8\text{gs}}^{\text{\rm NP}}&\label{eq:BdKphi_NP}\\
10^6\times\mathcal{B}(\bar{B_d}\to\bar{K}^{*0}\phi)^{\rm th,cv}=&4.65+8923.66 (C_{4s}^{\text{\rm NP}})^2&+17.03 (C_{8\text{gs}}^{\text{\rm NP}})^2-314.43 C_{4s}^{\text{\rm NP}}+14.94C_{8\text{gs}}^{\text{\rm NP}}\nonumber\\&-774.68 C_{4s}^{\text{\rm NP}} C_{8\text{gs}}^{\text{\rm NP}}&\label{eq:BdKstphi_NP}
\end{eqnarray}
The functional dependencies of the vector-vector and vector-pseudoscalar  branching ratios on the NP Wilson coefficients are very similar to each other. However, the experimental values deviate in a different way from their SM expectations. The values of $C_{4s,8gs}^{\rm NP}$ in Eq.~(\ref{eq:BdKphi_NP}) must account for a $1.31~\sigma$ deviation between theory and experiment while those in Eq.~(\ref{eq:BdKstphi_NP}) have no (or tiny) deviation ($0.35\sigma$) to explain. It is thus not surprising that the constraints on these NP coefficients turn out to be incompatible with each other at the $1~\sigma$ level when one takes both branching ratios into account. 

If one assumes a $2~\sigma$ uncertainty on all the experimental branching ratios, then the inconsistency understandably reduces, and one should find a common region in the $C_{4,8g}^{\rm NP}$ planes. However, this will also result in a decrease in the discrepancy for $L_{K^*\phi}$ from $1.48~\sigma$ to $0.47~\sigma$ (i.e. the discrepancy is reduced to about a third of its previous value). Naturally, a global fit with a proper statistical treatment of the various constraints would also help to build a more accurate picture of the discrepancies, but the naive exercise performed here allows us to identify potential difficulties in a common explanation of these observables when exploring NP in the $C_{4,8g}^{\rm NP}$ planes

In such a context, an updated measurement of the $\bar{B}_{d}\to\bar{K}^{0}\phi$ branching ratio by the LHCb collaboration would be highly desirable 
to shed new light on this situation, given that the incompatibility between the various constraints 
is marginal and could be confirmed or dispelled with a more accurate measurement of this branching ratio. The most recent measurements of this branching ratio from Babar (2012)~\cite{BaBar:2012iuj} and Belle (2003)~\cite{Belle:2003ike}  are one and two decades old respectively, and neither of them have been updated by LHCb. 

\subsection{Constraints on $C_{4s,8gs}^{\rm NP}$ from $\mathcal{B}({B_s}\to K^{*0}\bar{K}^0)+\mathcal{B}({B_s}\to \bar{K}^{*0}{K}^0)$} \label{subsec:BRBsKstK}

Similarly to the $\bar{B}_s \to K^0\phi$ decay, the $\bar{B}_d \to K^{*0}\bar{K}^0$ branching ratio (due to a $b\to d$ transition) 
remains unmeasured yet~\footnote{The LHCb collaboration has provided an upper limit for $\mathcal{B}({B_d}\to K^{*0}\bar{K}^0+c.c.)$~\cite{LHCb:2019vww}, where c.c denotes the conjugate mode ${B_d}\to \bar{K}^{*0}\bar{K}^0$.}. 

In a previous article devoted to  $K^{(*)0}\bar{K}^{(*)0}$ modes~\cite{Biswas:2023pyw}, we  
discussed the measurement for $\mathcal{B}({B_s}\to K^{*0}\bar{K}^0+c.c.)$~\cite{LHCb:2015oyu} and devoted a short discussion involving this measurement and its effect on the corresponding $L_{\rm Total}$ observable. However, due to the lack of an experimental number for $L_{\rm Total}$ we did not use this mode while constraining the NP Wilson coefficients $C_{4s,8gs}^{\rm NP}$ in Ref.\cite{Biswas:2023pyw}. 
Since we perform a more thorough analysis of all available measurements in the present article, we can take the opportunity to explore here the implications of the measurement of $\mathcal{B}({B_s}\to K^{*0}\bar{K}^0+c.c.)$

\begin{figure}[t]\centering
\subfloat[]{\label{fig:c4_KstK}\includegraphics[width=0.5\textwidth,height=0.40\textwidth]{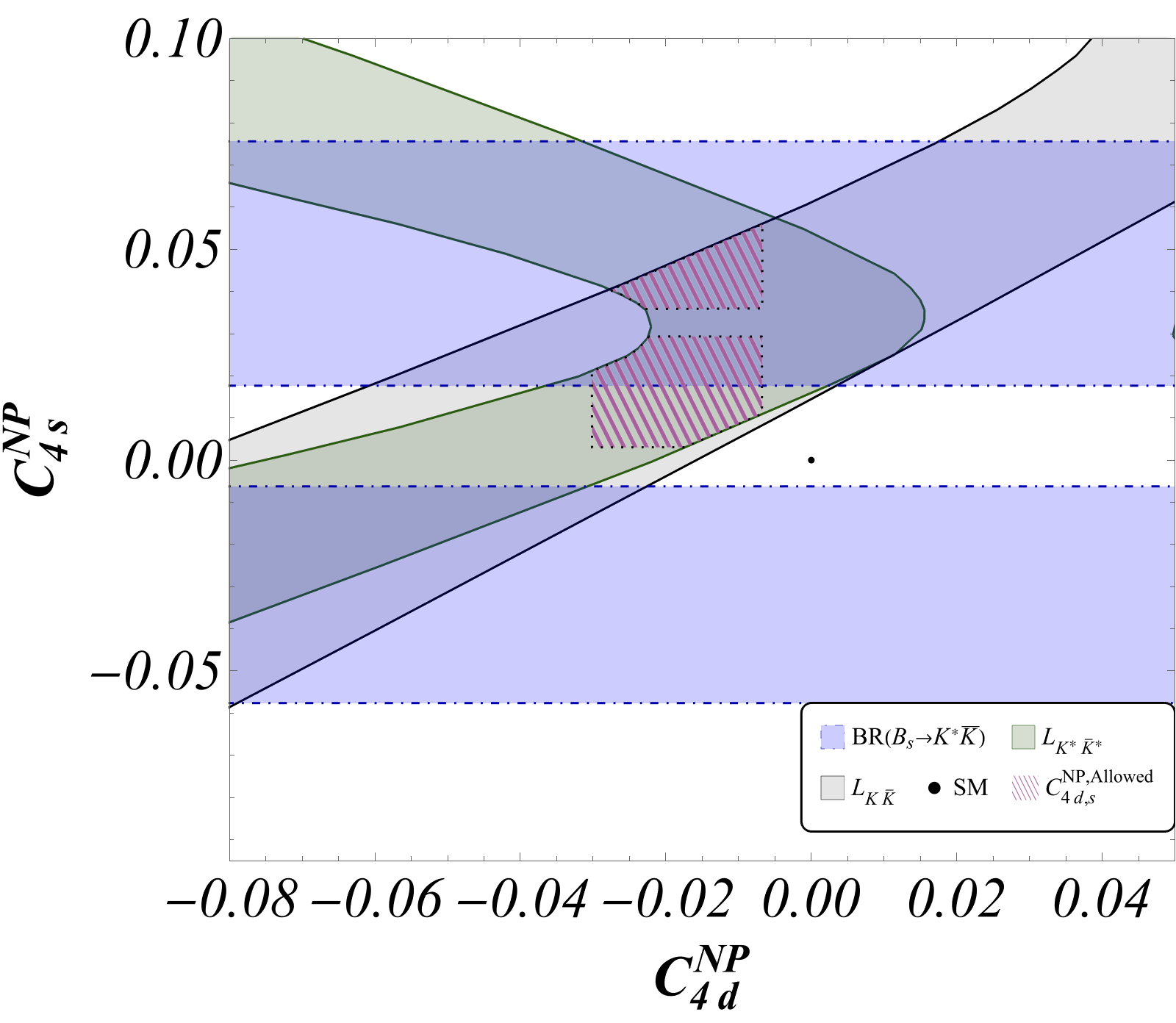}}~~~~~~
\subfloat[]{\label{fig:c8_KstK}\includegraphics[width=0.5\textwidth,height=0.40\textwidth]{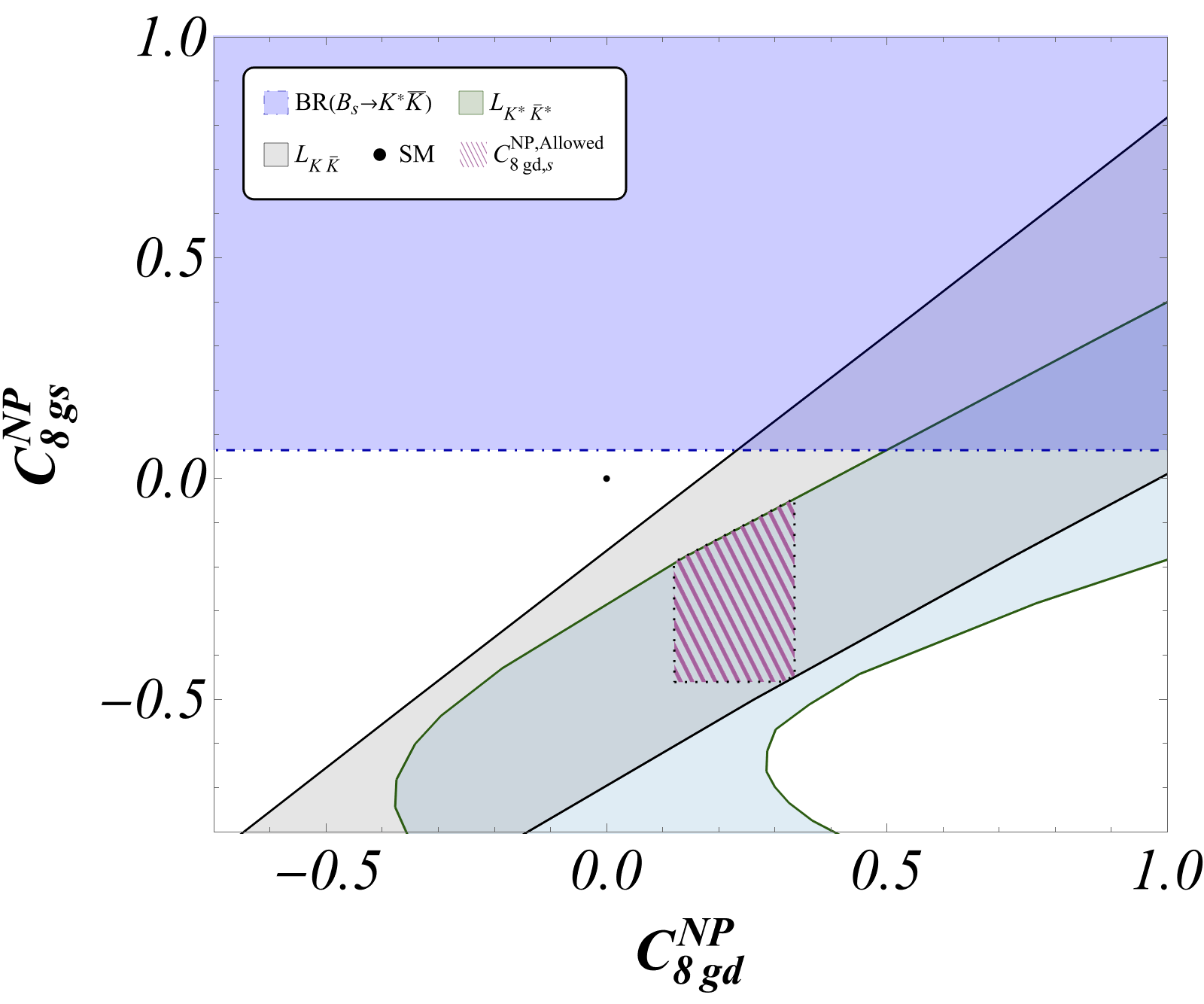}}
\caption{Impact of including the constraint from $\mathcal{B}(\bar{B}_s\to K^{*0}\bar{K}^0)$ along with those from the $B_{d,s}\to K^{(*0)}\bar{K}^{(*0)}$ modes. Fig.~\ref{fig:c4_KstK} (Fig.~\ref{fig:c8_KstK}) shows the impact of this mode on Fig.~\ref{fig:c4_LKstKst} (Fig.~\ref{fig:c8_LKstKst}). In both figures, the overlap of $\mathcal{B}(\bar{B}_{d,s}\to K^{(*)0}\bar{K}^{(*)0})$ and $L_{K^{(*)}\bar{K}^{(*)}}$ yields the magenta region, whereas $\mathcal{B}(\bar{B}_s\to K^{*0}\bar{K}^0)$ yields the horizontal blue band. We do not display the $\mathcal{B}(\bar{B}_s\to K^{(*)0}\bar{K}^{(*)0})$ bands in order to avoid overcrowding and confusion. 
In the $C_{4}^{\rm NP}$ plane (Fig.~\ref{fig:c4_KstK}), including $\mathcal{B}(\bar{B}_s\to K^{*0}\bar{K}^0)$ shrinks the common region close to the SM in Fig.~\ref{fig:c4_LKstKst}, whereas the common region away from the SM remains intact. In the $C_{8g}^{\rm NP}$ plane (Fig.~\ref{fig:c8_KstK}), there is no overlap for $\mathcal{B}(\bar{B}_s\to K^{*0}\bar{K}^0)$ and the $K^{(*)}K^{(*)}$ observables.}
\label{fig:BsKstK_effect}
\end{figure}

Fig.~\ref{fig:BsKstK_effect} shows the impact of the $\mathcal{B}(\bar{B}_s\to K^{*0}\bar{K}^0)$ constraint in the $C_4^{\rm NP}$ (\ref{fig:c4_KstK}) and the $C_{8g}^{\rm NP}$ (\ref{fig:c8_KstK}) plane. Including $\mathcal{B}(\bar{B}_s\to K^{*0}\bar{K}^0)$ together with $K^{(*)0}\bar{K}^{(*)0}$ observables 
results in a shrinkage of the magenta region near the SM in the $C_4^{\rm NP}$ plane while preserving the one away from the SM completely, as can be seen 
in Fig.~\ref{fig:c4_LKstKst}. In Fig.~\ref{fig:c8_KstK}, we can see
that there is no overlap among the constraints in the $C_{8g}^{\rm NP}$ plane. In both Figs.~\ref{fig:c4_KstK} and~\ref{fig:c8_KstK}, the magenta region represents the overlap of $\mathcal{B}(\bar{B}_{d,s}\to K^{(*)0}\bar{K}^{(*)0})$ and $L_{K^{(*)}\bar{K}^{(*)}}$, while $\mathcal{B}(\bar{B}_s\to K^{*0}\bar{K}^0)$ is represented by the horizontal blue band.

Since there is an overlap between the $K^{(*)}\bar{K}^{(*)}$ observables and $\mathcal{B}(\bar{B}_s\to K^{*0}\bar{K}^0+c.c.)$ in the $C_4^{\rm NP}$ plane, 
we would like to look for a common intersection of $\mathcal{B}(\bar{B}_s\to K^{*0}\bar{K}^0+c.c.)$ with all the other observables discussed in this article: $\mathcal{B}(\bar{B}_{d(s)}\to \bar{K}^{*0}(K^{*0})\phi)$, $L_{K^*\phi}$ and $\mathcal{B}(\bar{B}_d\to \bar{K}^{0}\phi)$. However, it is clear from the discussion in Sec.~\ref{subsec:BRBdKphi} that there is no common NP explanation in the $C_4^{\rm NP}$ plane. In fact, the effect of including constraints from $\mathcal{B}(\bar{B}_s\to K^{*0}\bar{K}^0+c.c.)$ to 
Fig.~\ref{fig:c4_all_BdKphi} will result in a reduction of the magenta hatched region, as can be understood from Fig.~\ref{fig:c4_KstK}.

Turning to the $C_{8g}^{\rm NP}$ plane, we may explain why the $K^{(*)}\bar{K}^{(*)}$ observables and $\mathcal{B}(\bar{B}_s\to K^{*0}\bar{K}^0+c.c.)$ do not yield an overlap region. The SM prediction for the latter branching ratio using our updated inputs is $(8.35_{-2.51}^{+5.02})\times 10^{-6}$~\cite{Beneke:2003zv}~\footnote{Note that this SM prediction corresponds to $\mathcal{B}(\bar{B}_s\to K^{*0}\bar{K}^0+\bar{B}_s\to\bar{K}^{*0} K^0)$, i.e. the sum and not the average, which is what the LHCb measurement reported in Ref.~\cite{LHCb:2019vww}.} and
the deviation between theory and experiment in Eq.~(\ref{mixedexp}) is $1.44~\sigma$. We can determine the functional dependence of this observable on the NP Wilson coefficients $C_{4s,8gs}^{\rm NP}$, fixing all the relevant input parameters to their central values:
\begin{eqnarray}
10^5\times\mathcal{B}(\bar{B}_s\to K^{*0}\bar{K}^0+c.c.)^{\rm th,cv}=&&+0.87+547.57 (C_{4s}^{\text{\rm NP}})^2+1.12 (C_{8\text{gs}}^{\text{\rm NP}})^2-8.23 C_{4s}^{\text{\rm NP}} \hfill\cr&&+0.95 C_{8\text{gs}}^{\text{\rm NP}} -47.10 C_{4s}^{\text{\rm NP}} C_{8\text{gs}}^{\text{\rm NP}}\label{eq:BsKstK_NP}.
\end{eqnarray}
This could be compared with the functional dependencies for the branching ratios $\mathcal{B}(\bar{B}_s\to K^{(*)0}\bar{K}^{(*)0})$ which we reproduce from Eq.(6.1) in Ref.~\cite{Biswas:2023pyw}):
\begin{eqnarray}
10^6\times\mathcal{B}(\bar{B}_s\to K^{*0}\bar{K}^{*0})^{\rm th,cv}=&&+4.36+3257.92 (C_{4s}^{\text{\rm NP}})^2+8.94 (C_{8\text{gs}}^{\text{\rm NP}})^2-212.07 C_{4s}^{\text{\rm NP}}\nonumber\cr&&+11.42 C_{8\text{gs}}^{\text{\rm NP}} -340.21 C_{4s}^{\text{\rm NP}} C_{8\text{gs}}^{\text{\rm NP}}\label{eq:BsKstKst_NP},\\
10^5\times\mathcal{B}(\bar{B}_s\to K^{0}\bar{K}^{0})^{\rm th,cv}=&&+2.86+181.75 (C_{4s}^{\text{\rm NP}})^2+1.14 (C_{8\text{gs}}^{\text{\rm NP}})^2-42.04 C_{4s}^{\text{\rm NP}}\cr&&+3.52 C_{8\text{gs}}^{\text{\rm NP}} -28.47 C_{4s}^{\text{\rm NP}} C_{8\text{gs}}^{\text{\rm NP}}\label{eq:BsKK_NP}.
\end{eqnarray}

Since the NP Wilson coefficients are expected to be small compared to one, we can focus on the linear terms. All branching ratios depend on a given NP Wilson coefficient ($C_{4s}^{\rm NP}$ or $C_{8gs}^{\rm NP}$) with the same sign but not the same weight, with deviations of different sizes to account for between theory and experiment.
 The NP contribution in the $K^*\bar{K}^*$ branching ratio has to explain a small deviation of $0.9~\sigma$, whereas NP for the $K\bar{K}$ final state has to account for a (somewhat) larger deviation of $1.6~\sigma$ (see Eqns.~(B.3)-(B.5) in Ref.~\cite{Biswas:2023pyw}). The deviation for $K^*\bar{K}+c.c.$ is close to that for $K\bar{K}$, while that for $K^*\bar{K}^*$ is somewhat smaller. However, we see that the magnitude of the linear term for $C_{8gs}^{\rm NP}$ for $K^*\bar{K}^*$ and $K\bar{K}$ is approximately 12 and 4 times (respectively) that for  $K^*\bar{K}+c.c.$. This smaller weight explains the lack of overlap in the $C_{8g}^{\rm NP}$ plane, since it cannot accommodate a deviation of $\sim1.4~\sigma$, which is of the same order as the deviation for $K\bar{K}$ and greater than the deviation for $K^*\bar{K}^*$.

\section{Considering all modes together} \label{sec:allmodes}

\subsection{Two-operator scenarios} \label{sec:two-operator}

Sections~\ref{subsec:BRBdKphi} and~\ref{subsec:BRBsKstK} show that the simplified assumption of NP affecting only one type of $b\to s,d$ operator at a time ($C_{4d,4s}$ or $C_{8gd, 8gs}$) is unable to explain all the vector-vector, pseudoscalar-pseudoscalar and vector-pseudoscalar $B_{d,s}$-meson branching ratios along with the $L_{K^{(*)}\bar{K}^{(*)}}$ and $L_{K^*\phi}$ observables:
\begin{itemize}
\item $\mathcal{B}(\bar{B}_{d,s}\to\bar{K}^{*0}\phi)$, $L_{K^*\phi}$ and $\mathcal{B}(\bar{B_d}\to\bar{K}^{0}\phi)$ can be explained simultaneously by assuming that NP affects either $C_{4f}^{\rm NP}$ or $C_{8gf}^{\rm NP}$ ($f=s, d$). However, there is no overlap found after adding the branching ratios and $L$ optimised observables corresponding to the $K^{(*)0}\bar{K}^{(*)0}$ final states.
\item $\mathcal{B}(\bar{B}_s\to K^{*0}\bar{K}^{0}+c.c.)$ has a substantial overlap with $K^{(*)0}\bar{K}^{(*)0}$ observables in the $C_{4}^{\rm NP}$ plane, but none in the $C_{8g}^{\rm NP}$ plane.
On the inclusion of the $K^{(*)0}\phi$ observables, neither $C_{4f}^{\rm NP}$ nor $C_{8gf}^{\rm NP}$ hypotheses can provide a simultaneous explanation of all the involved modes. 
\end{itemize}

\begin{figure}\centering
\subfloat[]{\label{fig:b2d_46}\includegraphics[width=0.42\textwidth,height=0.32\textwidth]{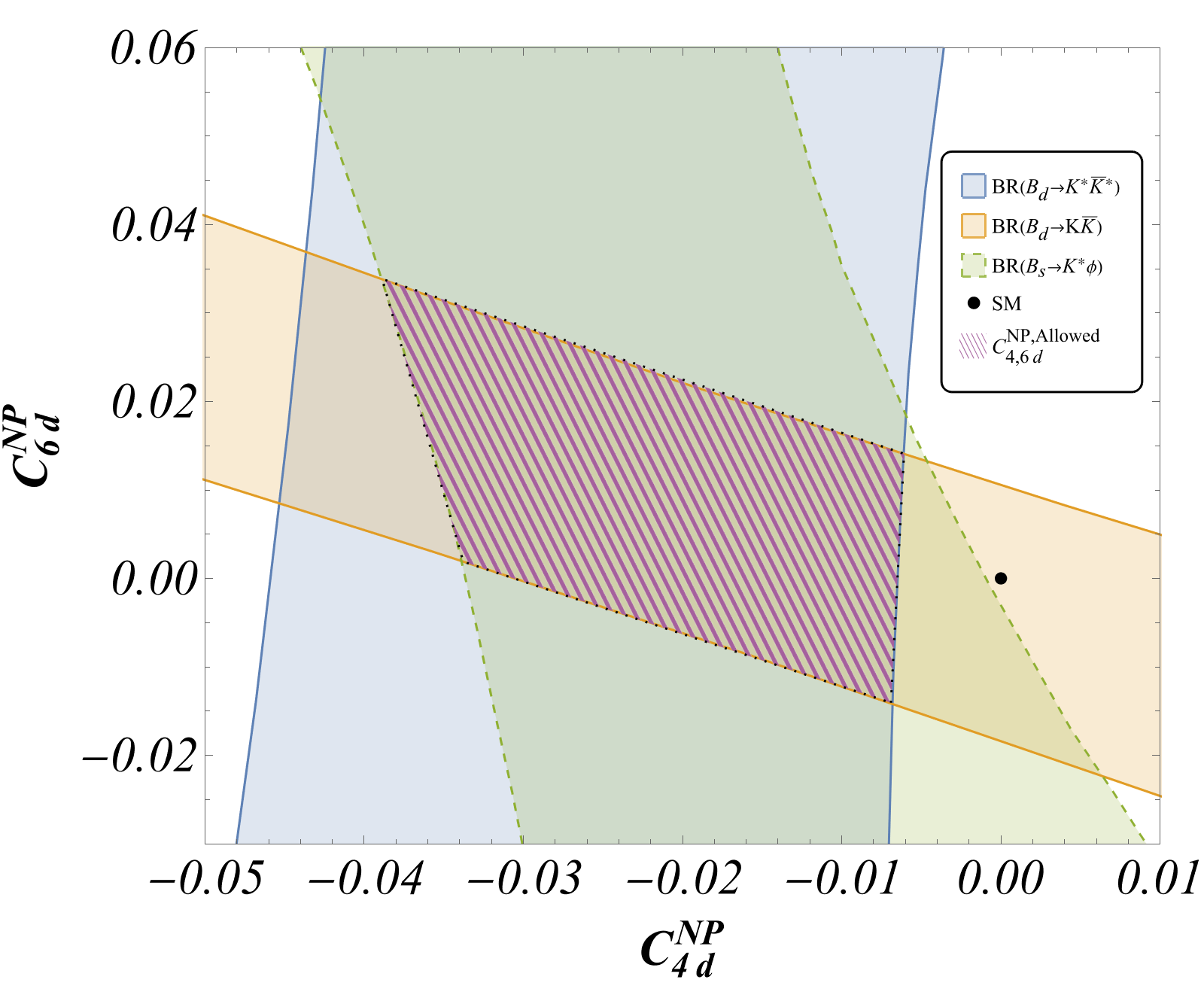}}~~~~
\subfloat[]{\label{fig:b2d_68}\includegraphics[width=0.42\textwidth,height=0.32\textwidth]{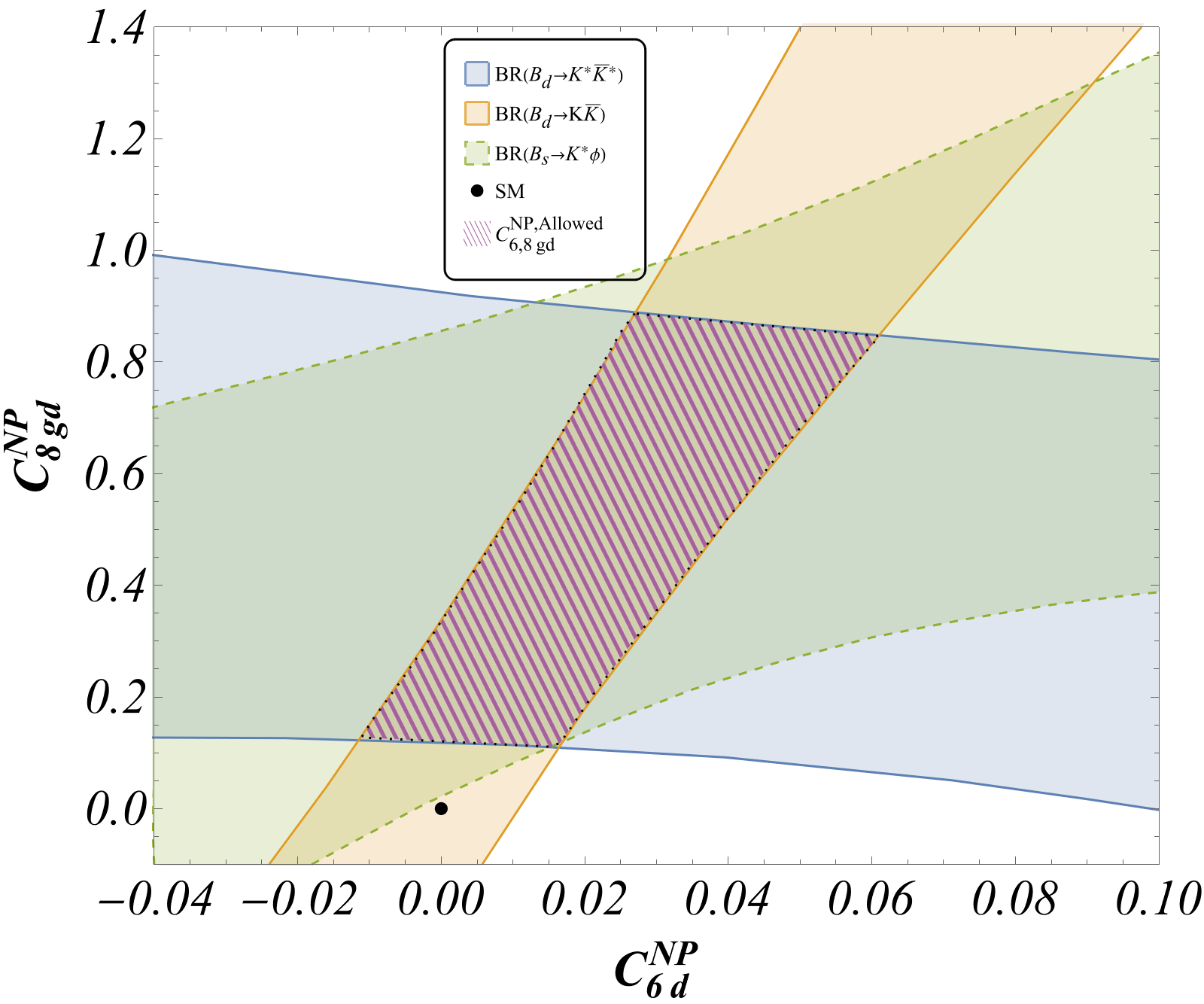}}\par
\subfloat[]{\label{fig:b2d_48}\includegraphics[width=0.42\textwidth,height=0.32\textwidth]{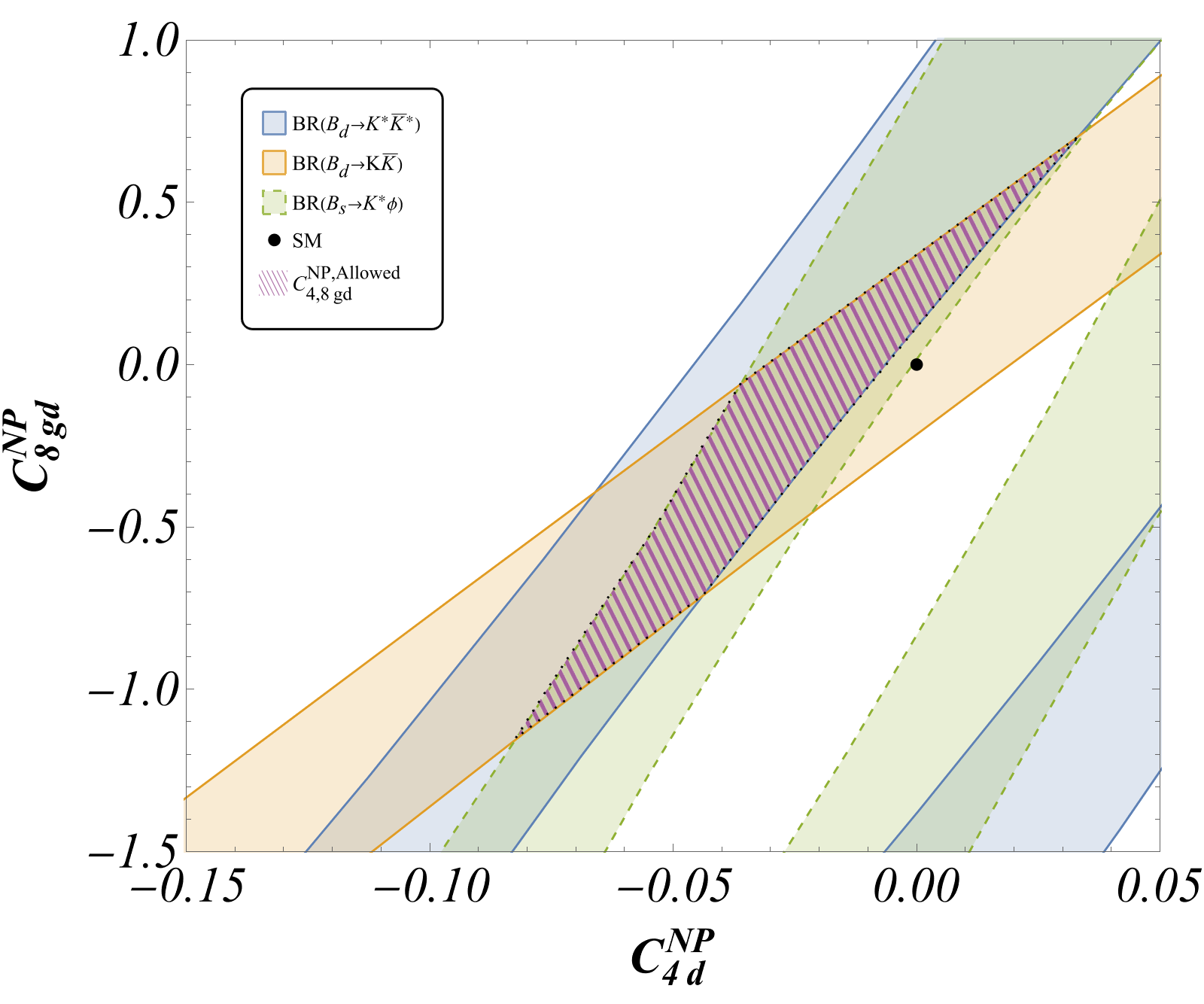}}~~~~
\subfloat[]
{\label{fig:b2s_46}\includegraphics[width=0.42\textwidth,height=0.32\textwidth]
{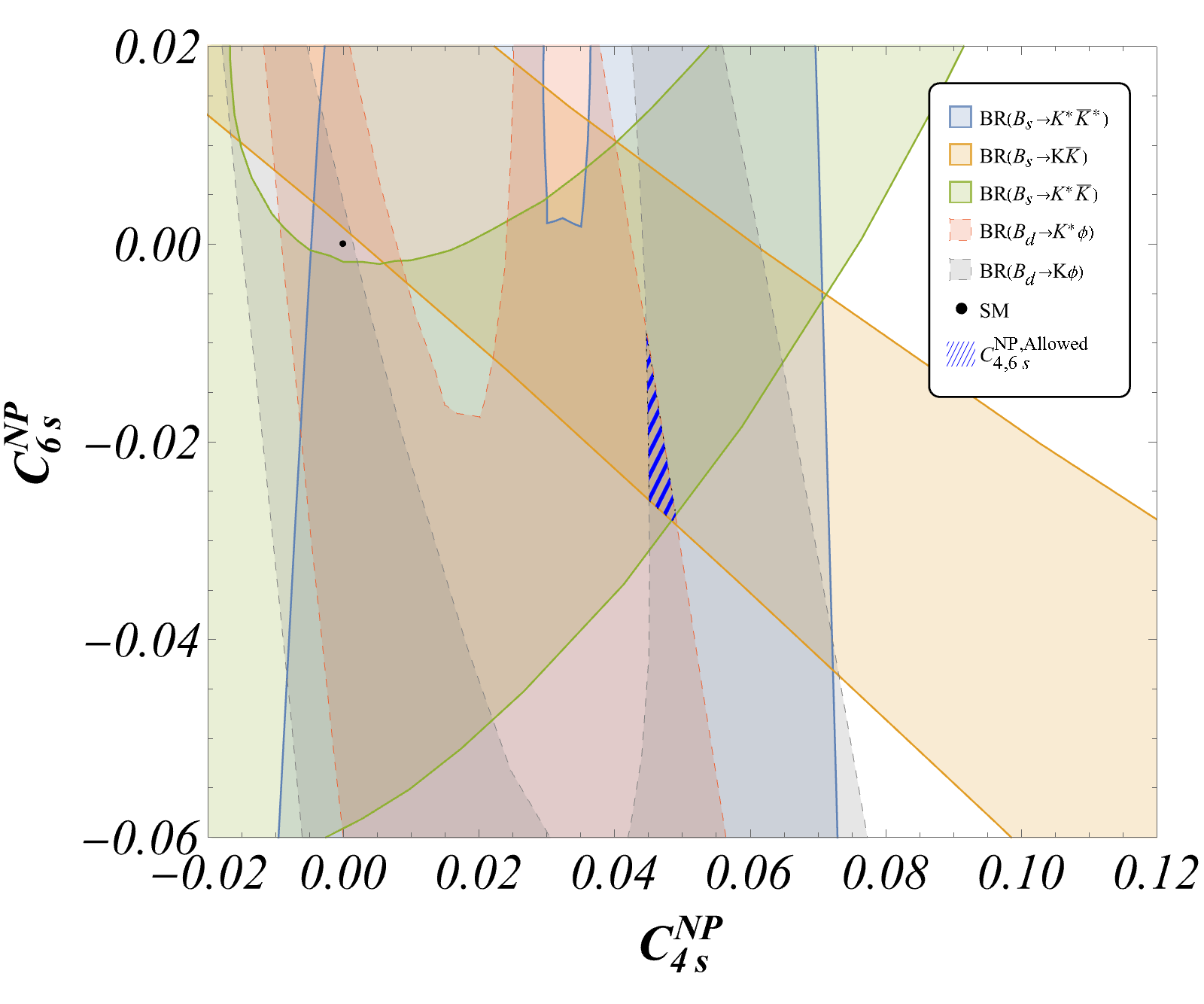}}\par
\subfloat[]
{\label{fig:b2s_68}\includegraphics[width=0.42\textwidth,height=0.32\textwidth]
{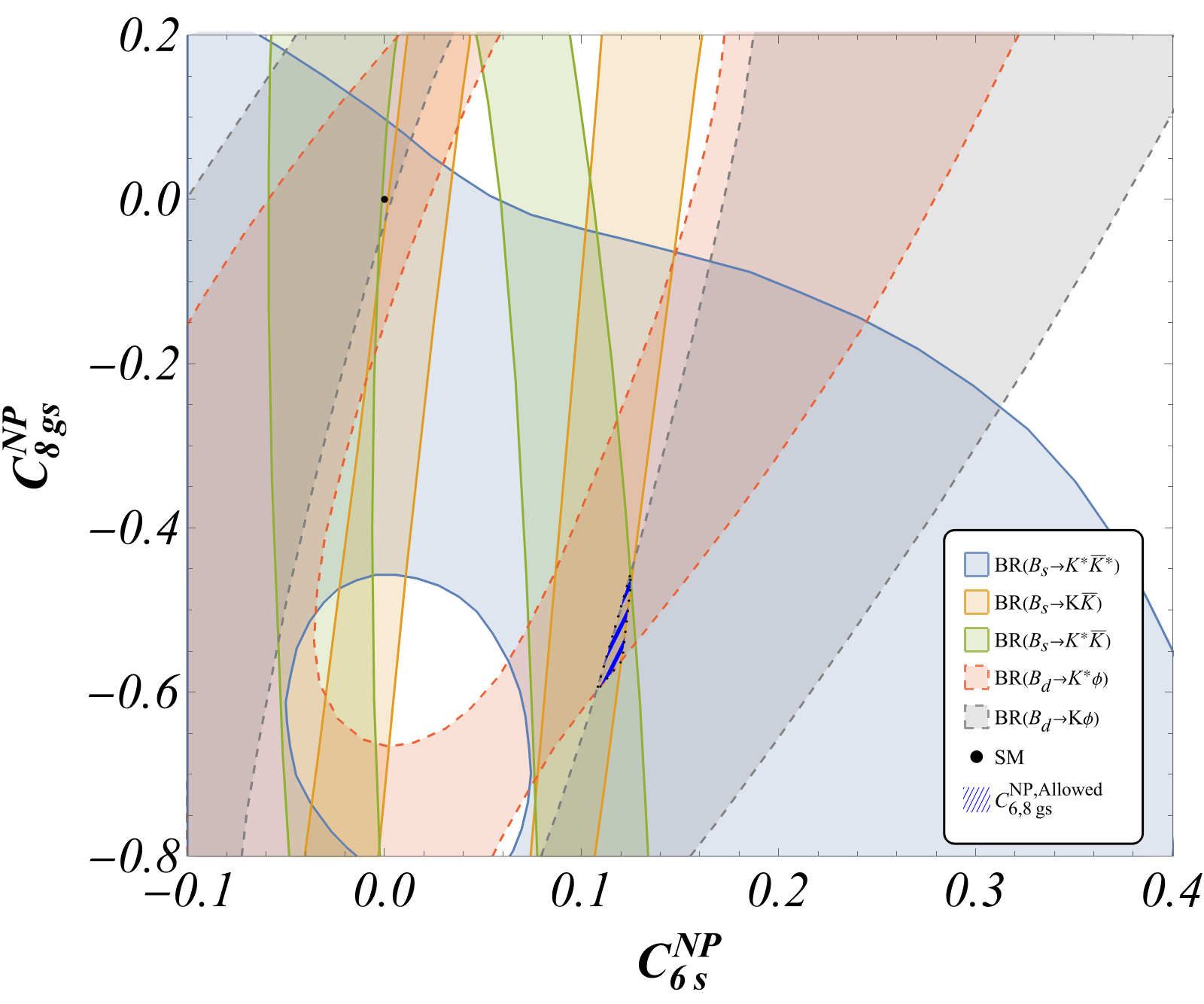}}~~~~
\subfloat[]
{\label{fig:b2s_48}\includegraphics[width=0.42\textwidth,height=0.32\textwidth]
{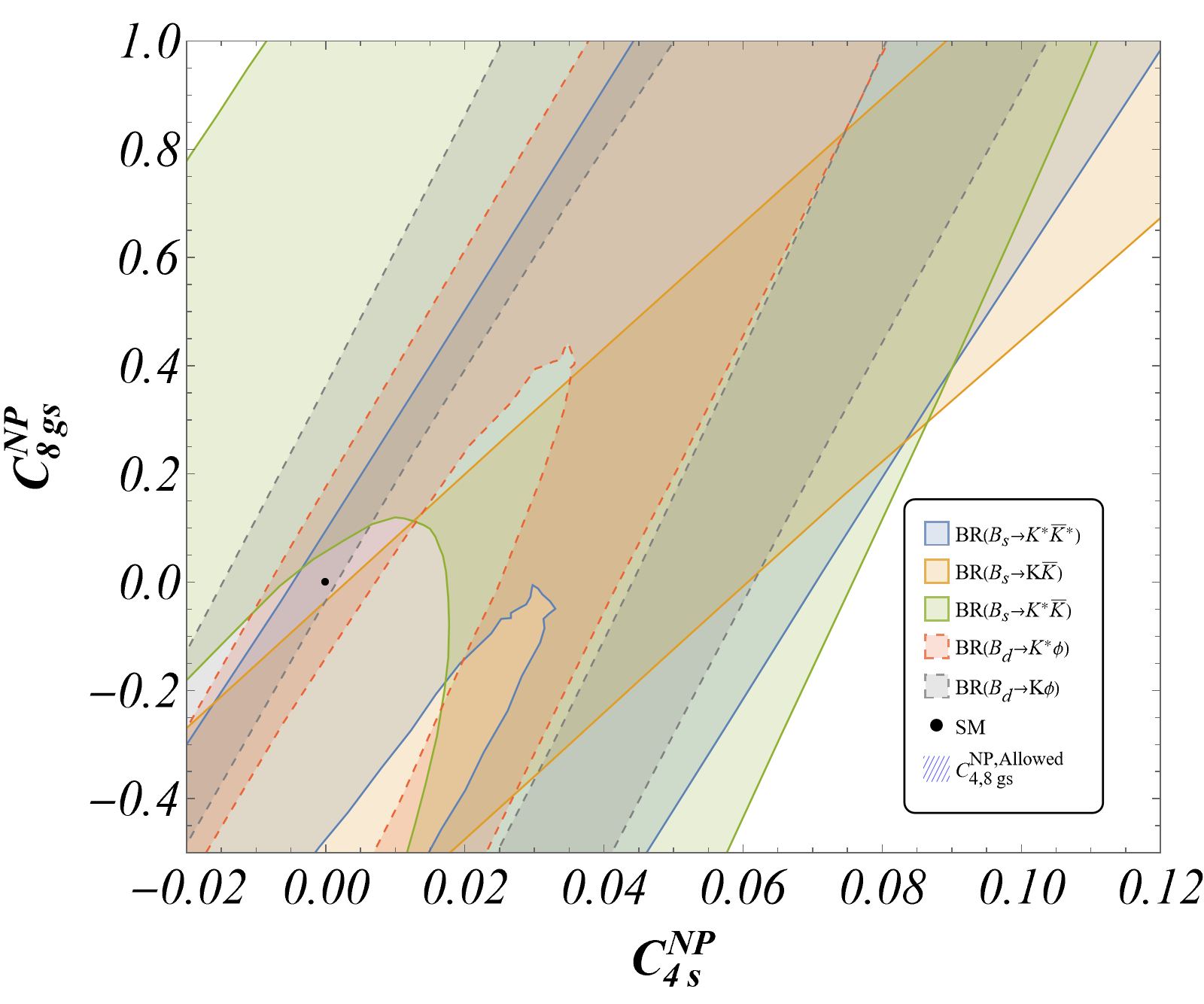}}
\caption{Two-operator scenarios. Figs.~\ref{fig:b2d_46},~\ref{fig:b2d_68} and~\ref{fig:b2d_48} correspond to the scenarios $(C_{4d}^{\rm NP},C_{6d}^{\rm NP})$, $(C_{6d}^{\rm NP},C_{8gd}^{\rm NP})$ and $(C_{4d}^{\rm NP},C_{8gd}^{\rm NP})$ respectively, where the magenta hatched area is the overlap for three relevant $b\to d$ branching ratios. Figs.~\ref{fig:b2s_46},~\ref{fig:b2s_68} and~\ref{fig:b2s_48} represent scenarios $(C_{4s}^{\rm NP},C_{6s}^{\rm NP})$, $(C_{6s}^{\rm NP},C_{8gs}^{\rm NP})$ and $(C_{4s}^{\rm NP},C_{8gs}^{\rm NP})$ respectively, where the blue hatched regions is the overlap for all the five relevant $b\to s$ branching ratios together. The SM is represented by the black dot. According to Fig.~\ref{fig:b2s_48}, the scenario $(C_{4s}^{\rm NP},C_{8gs}^{\rm NP})$ cannot provide a simultaneous explanation of all five $b\to s$ branching ratios.}
\label{fig:2_param_sc}
\end{figure}

In this section we will relax this assumption and allow for a non-zero NP contribution to two $b\to s$ operators and the corresponding two $b\to d$ operators.
In order to get a handle on which other operator should be switched on, we first recall the discussion in appendix A.2 of Ref.~\cite{Biswas:2023pyw}. In short, an NP contribution to $C_{6f}$ by itself is unable to simultaneously explain all the $K^{(*)}\bar{K}^{(*)}$ observables, but can explain the $K\bar{K}$ modes.
Indeed, $(C_{6q} + C_{5q}/3)$ $(q=d,s)$  appears in the chirally enhanced part $a^c_6$ of the numerically dominant combination $\alpha^c_{4}=a^c_4 + r_{\chi} a^c_6$ contribution to the $K\bar{K}$ mode (and pseudoscalar-pseudoscalar final states in general). Although  $a^c_4$ is also numerically dominant for the $K^*\bar{K}^*$ final states, it receives no similar contribution at the leading order. This induces a stronger dependence on the NP Wilson coefficient $C_{6q}^{\rm NP}$ for $\bar{B_q}\to K^0\bar{K}^0$ compared to $\bar{B_q}\to K^{*0}\bar{K}^{*0}$ $(q=d,s)$. This explains why $C_{6q}^{NP}$ can explain $L_{K\bar{K}}$ along with the associated $K^0\bar{K}^0$ branching ratios but not their vector-vector counterparts.

However, we have found that a two-operator scenario involving $C_{6f}$ together with either $C_{4f}$ or $C_{8gf}$ could provide a simultaneous NP explanation for all the $K^{(*)}\bar{K}^{(*)}$ observables combined, and by extension, for all the modes considered in this article. We will thus consider three scenarios:
$C_{4f}-C_{6f}$, $C_{6f}-C_{8gf}$ and $C_{4f}-C_{8gf}$. We will call them ``two-operator'' scenarios, bearing in mind that we allow an NP contribution for both $f=s,d$ for each of the two operators (hence 4 NP contributions).
%For each two-operator scenario,
%Individual branching ratios involve either  a $b\to d$ or a $b\to s$ transition and will either depend on $(C_{id}^{\rm NP}$, $C_{jd}^{\rm NP})$ or $(C_{is}^{\rm NP}$, $C_{js}^{\rm NP})$ in each two-operator scenario ($i\neq j$ fixed among $\{4,6,8g\}$). 

For each of these above mentioned two-operator scenarios, the related branching ratios (being either a $b\to d$ or a $b\to s$ transition) will either depend on $C_{id}^{\rm NP}$ \& $C_{jd}^{\rm NP}$ or $C_{is}^{\rm NP}$ \& $C_{js}^{\rm NP}$ ($i, j = 4,6,8g$, $i\neq j$). In other words, it is possible to obtain a common region of overlap in the $b\to d(s)$ 2-dimensional plane where all the $b\to d(s)$ branching ratios are simultaneously satisfied. However, the optimized $L$ observables cannot be represented in the same two-dimensional spaces since they depend on four NP Wilson coefficients (two from the $b\to s$ branching ratio in the numerator, two from the $b\to d$ branching ratio in the denominator). 

However, we can test whether particular combinations of points that separately belong to the common regions of overlap in the $b\to d$ and the $b\to s$ planes can also simultaneously explain $L_{K^{(*)}\bar{K}^{(*)}}$ and $L_{K^*\phi}$. This test can be performed using the same approach of overlaps as done for the one-operator scenarios in our current and previous articles. We identify the combinations $(C_{id}^{\rm NP},C_{jd}^{\rm NP})$ and $(C_{is}^{\rm NP},C_{js}^{\rm NP})$ that explain the $b\to d$ and $b\to s$ branching ratios separately, and then identify the combinations $(C_{id}^{\rm NP}, C_{jd}^{\rm NP}, C_{is}^{\rm NP}, C_{js}^{\rm NP})$, if any, that can also simultaneously explain the three optimised observables $L$. A set of common 4d-points obtained by inspection is provided as an ancilliary file.

Fig.~\ref{fig:2_param_sc} displays all the relevant two-operator scenarios $(C_{if}^{\rm NP},C_{jf}^{\rm NP})$ ($i, j = 4,6,8g$, $i\neq j$, $f=d,s$) that provide simultaneous NP explanations for the $b\to d$ and $b\to s$ branching ratios separately. A few comments and observations are in order:
\begin{itemize}
\item It is easier to find points providing simultaneous explanations for the $b\to d$ branching ratios combined than for $b\to s$ ones. This can be attributed to mainly two reasons: the number of constraints is different (five $b\to s$ branching ratios as compared to three for $b\to d$) and the level of agreement between data and SM predictions is different (only one of the three $b\to d$ branching ratios ($\bar{B}_{d}\to K^0\bar{K}^{0}$) is compatible with the SM at the $1~\sigma$ level, which is the case for three of the five $b\to s$ branching ratios ($\bar{B}_{d}\to K^{(*)0}\phi$ and $\bar{B}_{s}\to K^0\bar{K}^0$)).
\item The plots obtained for the two-operator scenarios can be used to guess how one-operator hypotheses will fare.
For instance, in Fig.~\ref{fig:b2d_46}, the $C_{6d}=0$ line intersects the magenta region at $C_{4d}^{\rm NP}\approx -0.03$ and $-0.008$ respectively. Fig.~\ref{fig:b2d_48} shows that this range for $C_{4d}^{\rm NP}$ remains the same along the $C_{8gd}^{\rm NP}=0$ line. One can compare this with Fig.~\ref{fig:c4_all} and check that the same range in $C_{4d}^{\rm NP}$ explains all the $b\to d$ branching ratios. On the other hand, the $C_{6s}^{\rm NP} = 0$ line in Fig.~\ref{fig:b2s_46} can provide a common explanation for $\mathcal{B}(\bar{B}_s\to K^{(*)0}\bar{K}^{(*)0})$, $\mathcal{B}(\bar{B}_s\to K^{*0}\bar{K}^0+c.c.)$ and $\mathcal{B}(\bar{B}_d\to\bar{K}^{*0}\phi)$ (but not $\mathcal{B}(\bar{B}_d\to\bar{K}^{0}\phi)$) for a range of $C_{4s}^{\rm NP}$ values $\sim[0.025,0.04]$ (the $C_{8gs}^{\rm NP}=0$ line in Fig.~\ref{fig:b2s_48} exhibits the same feature). This  points towards the fact that $C_{4s}$ can provide a simultaneous explanation for all the $b\to s$ branching ratios apart from $\mathcal{B}(\bar{B}_d\to\bar{K}^0\phi)$. Hence, we do not find a common region of overlap in Fig.~\ref{fig:c4_all_BdKphi} but find overlaps in Fig.~\ref{fig:c4_KstK}.
\item It can be seen from Fig.~\ref{fig:b2s_48} that the scenario $(C_{4s}^{\rm NP},C_{8gs}^{\rm NP})$ cannot explain all the $b\to s$ branching ratios simultaneously. This is in agreement with the difficulties encountered by the one-operator scenarios $C_{4d,s}^{\rm NP}$ and $C_{8gd,s}^{\rm NP}$ to account for all the observables ($L$ optimised observables and branching ratios). In Figs.~\ref{fig:b2s_48} and $\ref{fig:b2s_68}$, following the $C_{4s}^{\rm NP}=0$ and the $C_{6s}^{\rm NP}=0$ line respectively shows that all the $K^{(*)}\bar{K}^{(*)}$ and the $K^{(*)}\phi$ modes cannot be simultaneously explained. The same conclusion is reached from Figs.~\ref{fig:b2s_46} and \ref{fig:b2s_48} following the $C_{6s}^{\rm NP}=0$ and $C_{8gs}^{\rm NP}=0$ lines. A one-operator explanation for all these observables together is thus not possible.
\end{itemize}
From Fig.~\ref{fig:2_param_sc}, the potential candidates for two-parameter models that can simultaneously explain all three $b\to d$ and five $b\to s$ branching ratios (in the sense of an overlap of $1~\sigma$ intervals between theory and experiment for each observable) are $(C_{4f},C_{6f})$ and $(C_{6f},C_{8gf})$. 

\subsection{Adding the charged modes}\label{subsec:charge_modes}

\begin{figure}[t]\centering
\subfloat[]{\label{fig:c46_charge}\includegraphics[width=0.5\textwidth,height=0.40\textwidth]{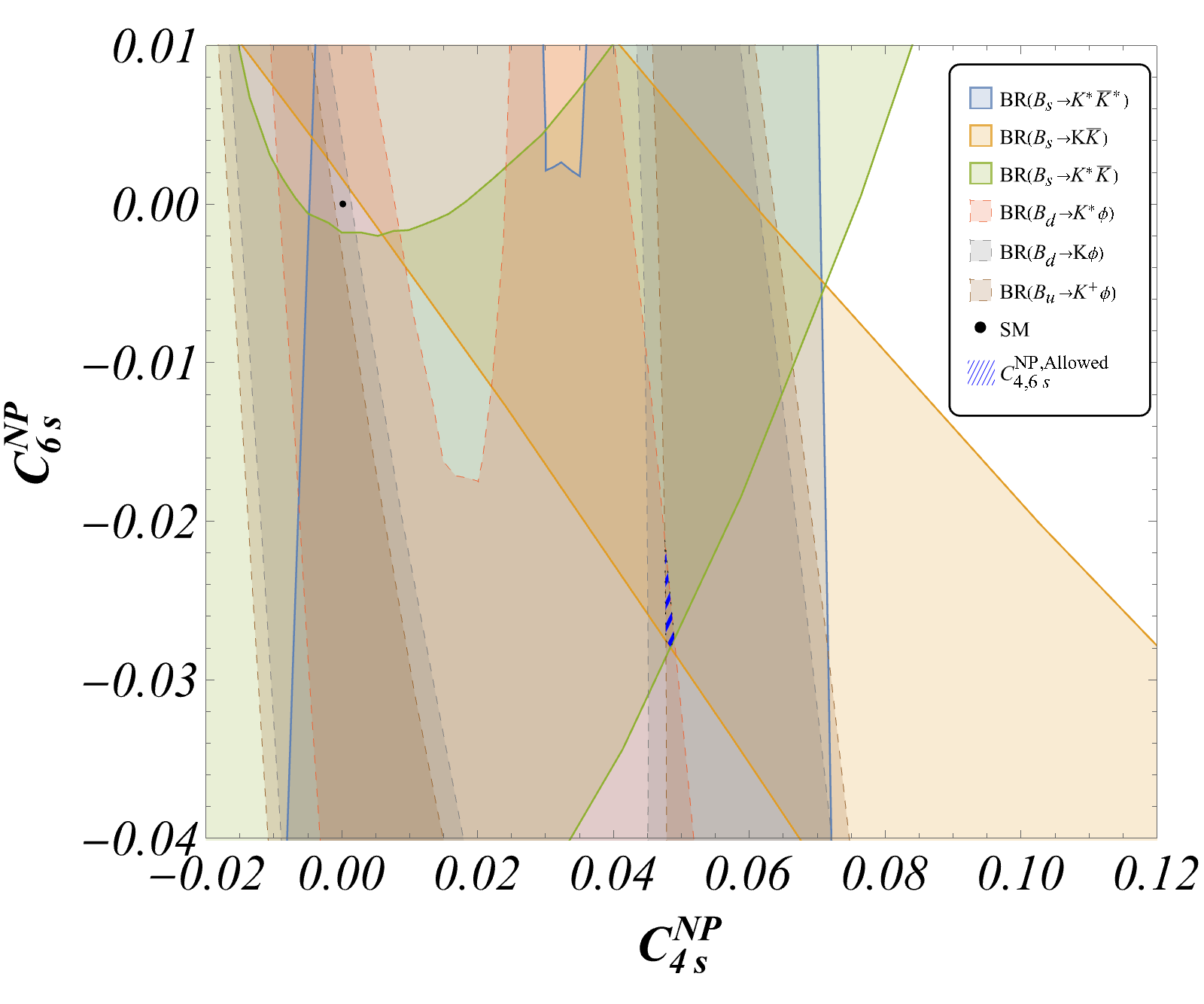}}~~~~~~
\subfloat[]{\label{fig:c46_charge_zoom}\includegraphics[width=0.5\textwidth,height=0.40\textwidth]{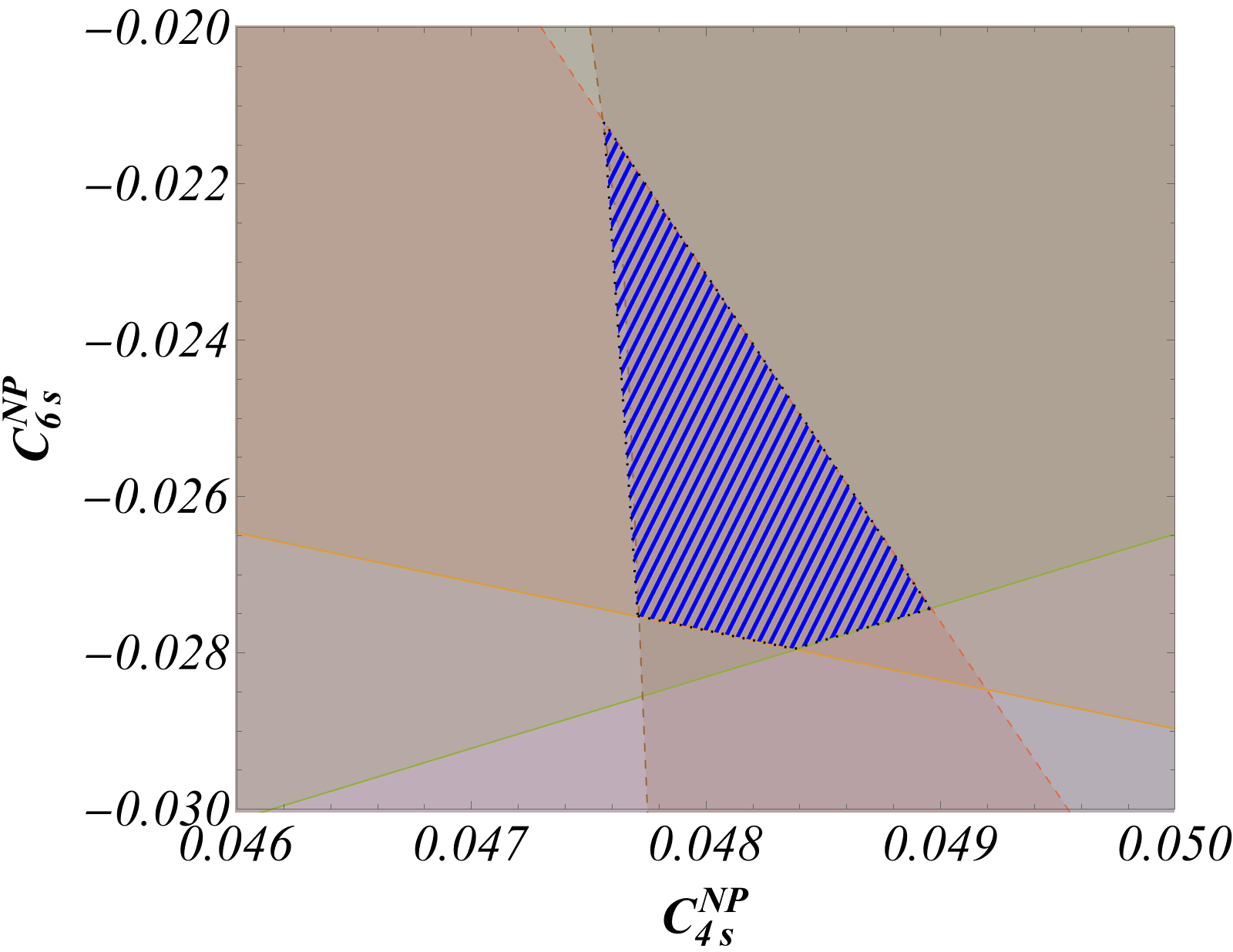}}\par 
\subfloat[]{\label{fig:c68_charge}\includegraphics[width=0.5\textwidth,height=0.40\textwidth]{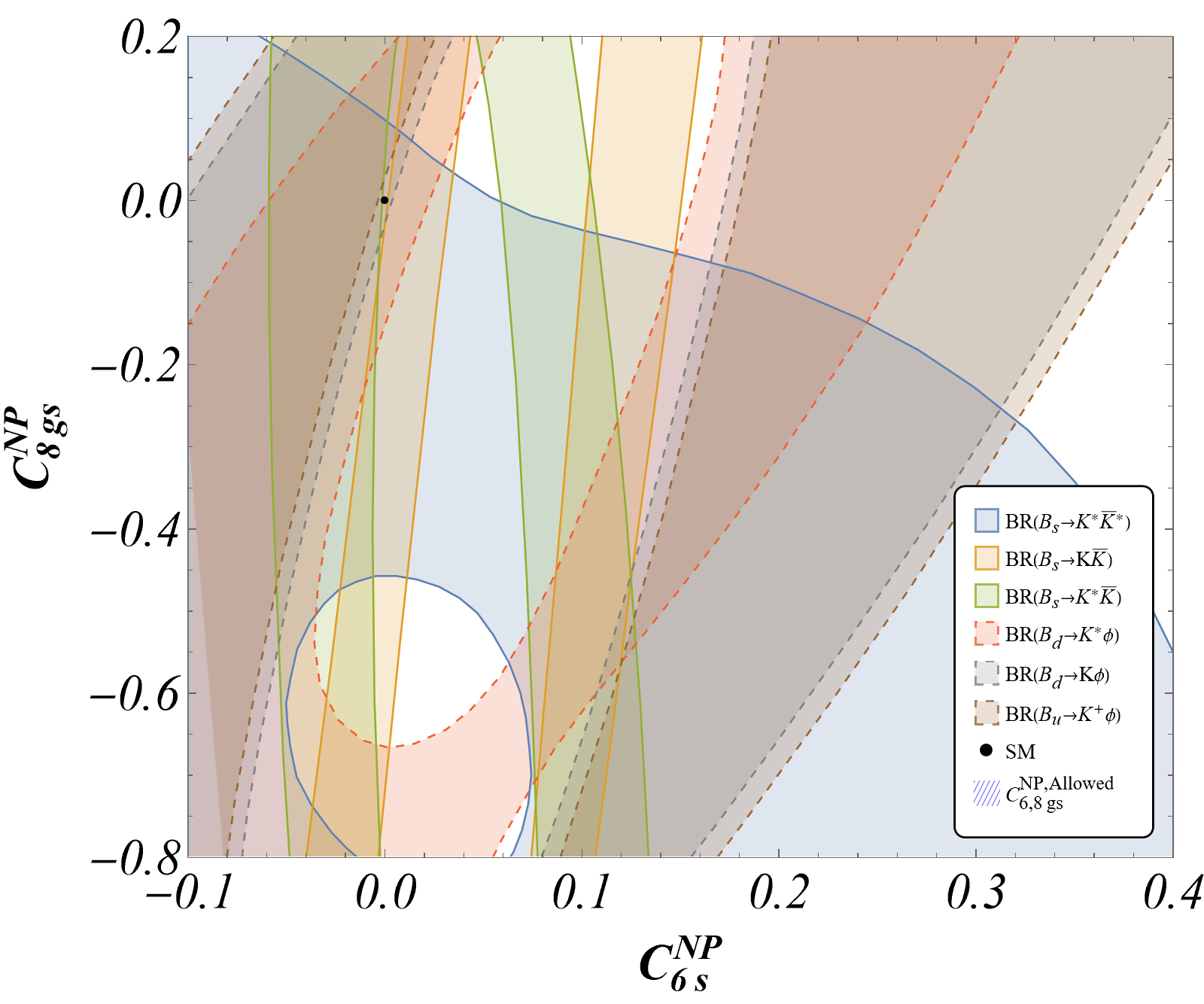}}
%\subfloat[]{\label{fig:c68_charge_zoom}\includegraphics[width=0.5\textwidth,height=0.40\textwidth]{btos_BR_c68_wBuKphi_zoomed.png}}
\caption{Figs.~\ref{fig:c46_charge} and~\ref{fig:c68_charge} show the effect of including constraints from $\mathcal{B}(B^-\to K^-\phi)$ to Figs.~\ref{fig:b2s_46} and~\ref{fig:b2s_68} respectively. Fig.~\ref{fig:c46_charge_zoom} is the zoomed in version of Fig.~\ref{fig:c46_charge}, highlighting the common blue-hatched region that can explain $\mathcal{B}(B^-\to K^-\phi)$ along with the other $b\to s$ branching ratios. No such common region exists when constraints from $\mathcal{B}(B^-\to K^-\phi)$ is added to Fig.~\ref{fig:b2s_68}.}
\label{fig:charged_mode}
\end{figure}

In the previous section we have shown  that the 2-operator scenarios $(C_{4f},C_{6f})$ and $(C_{6f},C_{8gf})$ can successfully accommodate all the eight (three $b\to d$ and five $b\to s$) $B_{d,s}$-branching ratios discussed so far. We turn to the charged modes $B^-\to K^{(*)-}\phi$ now and explore their impact on these scenarios. Both modes entail a $b\to s$ transition at the quark level. The lowest-order (in $\alpha_s$) topologies contributing to these transitions are similar to Figs.~\ref{fig:Bd_CA} and~\ref{fig:Bd_CS} with the spectator $\bar{d}$ quark replaced by a $\bar{u}$ quark.  The complete theoretical expressions corresponding to these transitions under the framework of QCDF can be obtained from Ref.~\cite{Bartsch:2008ps} for $K^{*-}\phi$ and ref.~\cite{Beneke:2003zv} for $K^{-}\phi$ respectively. 

The experimental measurement and SM predictions for the longitudinal branching ratio for the charged vector-vector mode are provided in Table~\ref{tab:BR_est}. They are in perfect agreement with the caveat that the uncertainties, especially for the SM prediction, are rather large (as expected). 
Moreover, the theoretical predictions for the charged and neutral $K^*\phi$ modes are rather close and completely consistent with each other (well below one sigma)~\footnote{This is expected since these two modes are related by isospin symmetry.}.

The experimental measurement of the branching ratio for the pseudoscalar-vector mode is~\cite{Workman:2022ynf}
\begin{equation}
\mathcal{B}(B^-\to K^{-}\phi)^{\rm exp}=(8.8^{+0.7}_{-0.6})\times 10^{-6}\label{eq:Kphi_exp_charge}
\end{equation}
and its SM prediction calculated in QCDF is
\begin{equation}
\mathcal{B}(B^-\to K^{-}\phi)^{\rm th}=(4.67^{+2.98}_{-1.63})\times 10^{-6}\label{eq:Kphi_SM_charge}.
\end{equation}
The deviation between the experimental measurement and the SM prediction for the charged pseudoscalar-vector final state is $1.48\sigma$.

We now discuss the constraints induced on the $(C_{4f},C_{6f})$ and $(C_{6f},C_{8gf})$ scenarios. Since both charged modes are $b\to s$ transitions, they should affect the common regions shown in Figs.~\ref{fig:b2s_46} and~\ref{fig:b2s_68}.

If we look at the vector-vector mode in Table~\ref{tab:BR_est}, the uncertainties on the theoretical predictions are pretty comparable between neutral and charged modes, whereas the uncertainty on the experimental number for the branching ratio of the charged mode is more than three times the magnitude of the uncertainty for the neutral mode. The charged mode should thus not yield any further constraint given this large uncertainty, which we checked explicitly. 

We will thus show only the constraints coming from $\mathcal{B}(B^-\to\ K^{-}\phi)$ in Fig.~\ref{fig:charged_mode}. 
Figs.~\ref{fig:c46_charge} and~\ref{fig:c68_charge} show the effect of the inclusion of this constraint to Figs.~\ref{fig:b2s_46} and~\ref{fig:b2s_68} respectively. It can clearly be seen that its inclusion results in a substantial reduction of the common region in the $C_{4s}-C_{6s}$ plane, while leaving no common region in the $C_{6s}-C_{8gs}$ plane. This can be understood as follows. The SM predictions for both charged and neutral $K\phi$ modes are consistent (well below one $\sigma$) as can be verified from Eqs.~\ref{eq:Kphi_SM_charge} and~\ref{eq:Kphi_SM_neutral}. However, there is a discrepancy at the level of 1.5 $\sigma$ between the experimental measurements Eqs.~\ref{eq:Kphi_exp_charge} and~\ref{eq:Kphi_exp_neutral}. There is a $\sim 9\%$ (20\%) increase in the central value of the SM prediction (experimental measurement) for the charged vector-pseudoscalar mode compared to the neutral case.
This is in stark contrast to the vector-vector case, where the central values are exactly the same for the experimental measurements, and the corresponding increase of the SM predictions from the neutral mode to the charged one is again $\sim 9\%$. This tension in pseudoscalar-vector neutral and charged modes (not seen in the vector-vector case) results in 
%results in the band corresponding to the charged pseudoscalar-vector final state 
further constraining the common region obtained in Fig.~\ref{fig:b2s_46}, while completely discarding the one in Fig.~\ref{fig:b2s_68}. 

It is worth stressing at this point that this discrepancy in experimental values for charged and neutral modes is somewhat surprising since these two modes are related by isospin symmetry: one expects the experimental numbers for these modes to be much closer (as seen for the vector-vector final states $K^{*0,-}\phi$). This could stem from the fact that the PDG number for $\mathcal{B}(B^-\to K^{-}\phi$) is an average over four different measurements (Babar \cite{BaBar:2012iuj}, CDF \cite{CDF:2005apk}, Belle \cite{Belle:2004drb} and CLEO \cite{CLEO:2001ium}) over the years 2001 to 2012, with a scale factor of 1.1 showing a slight dispersion of the values. This suggests that the experimental average for the charged mode should be considered with some caution.
Indeed, if we assume an uncertainty of $2\sigma$ on the current experimental average for this mode, we find that both the $C_{4f}-C_{6f}$ and $C_{6f}-C_{8gf}$ scenarios can well accommodate this observable. As such, we feel that it is very desirable to have 
 an updated and precise measurement for this mode by LHCb/Belle II in the near future, to confirm or not the current PDG average.

Assuming that the above constraints are valid, it remains to be seen whether the points within the common regions can also explain the three optimised observables $L$ simultaneously within 1$\sigma$ of their experimental measurements.
\begin{figure}[ht]\centering
\subfloat[]{\label{fig:expplt}\includegraphics[width=0.5\textwidth,height=0.4\textwidth]{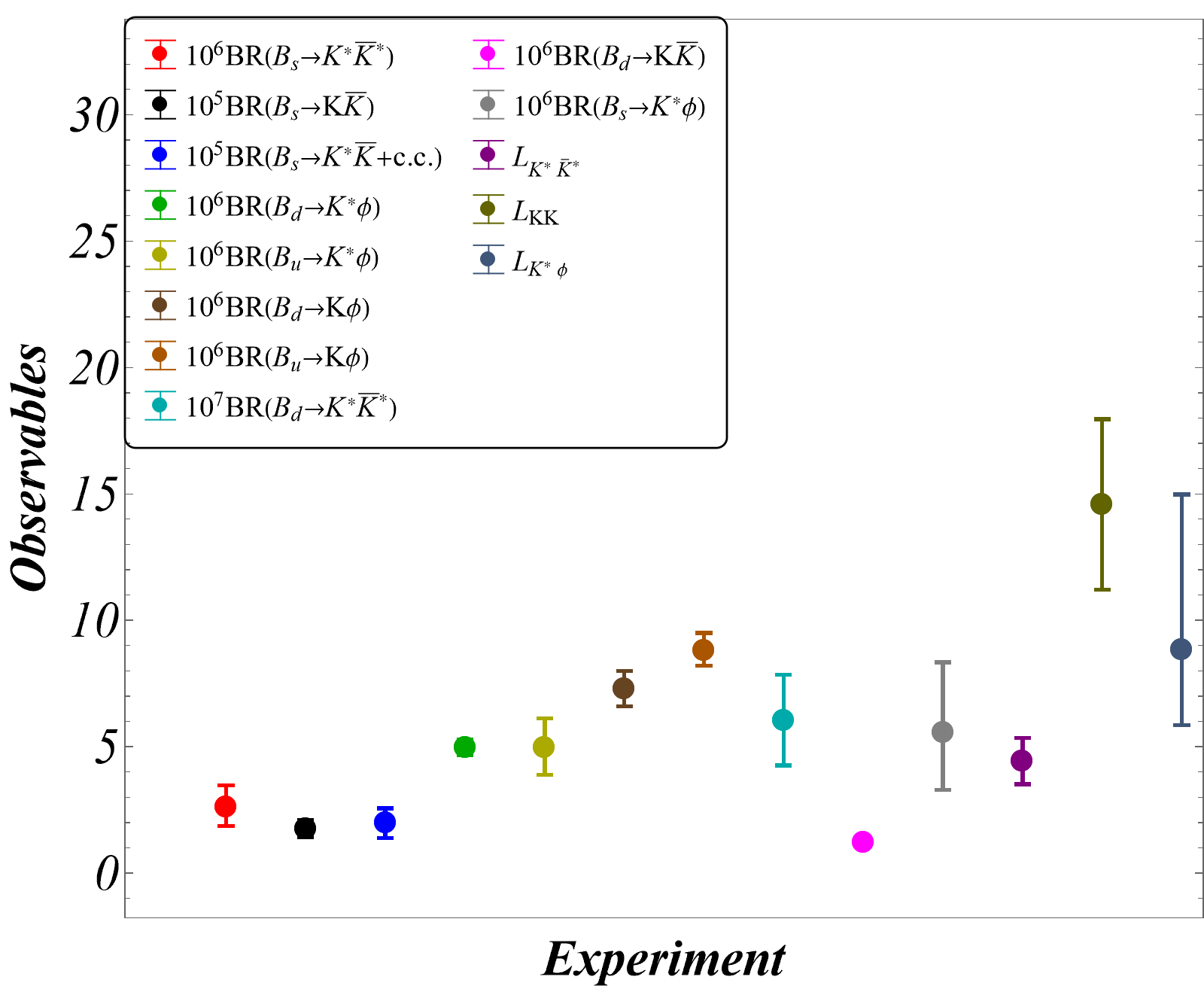}}~~~~
\subfloat[]{\label{fig:smplt}\includegraphics[width=0.5\textwidth,height=0.4\textwidth]{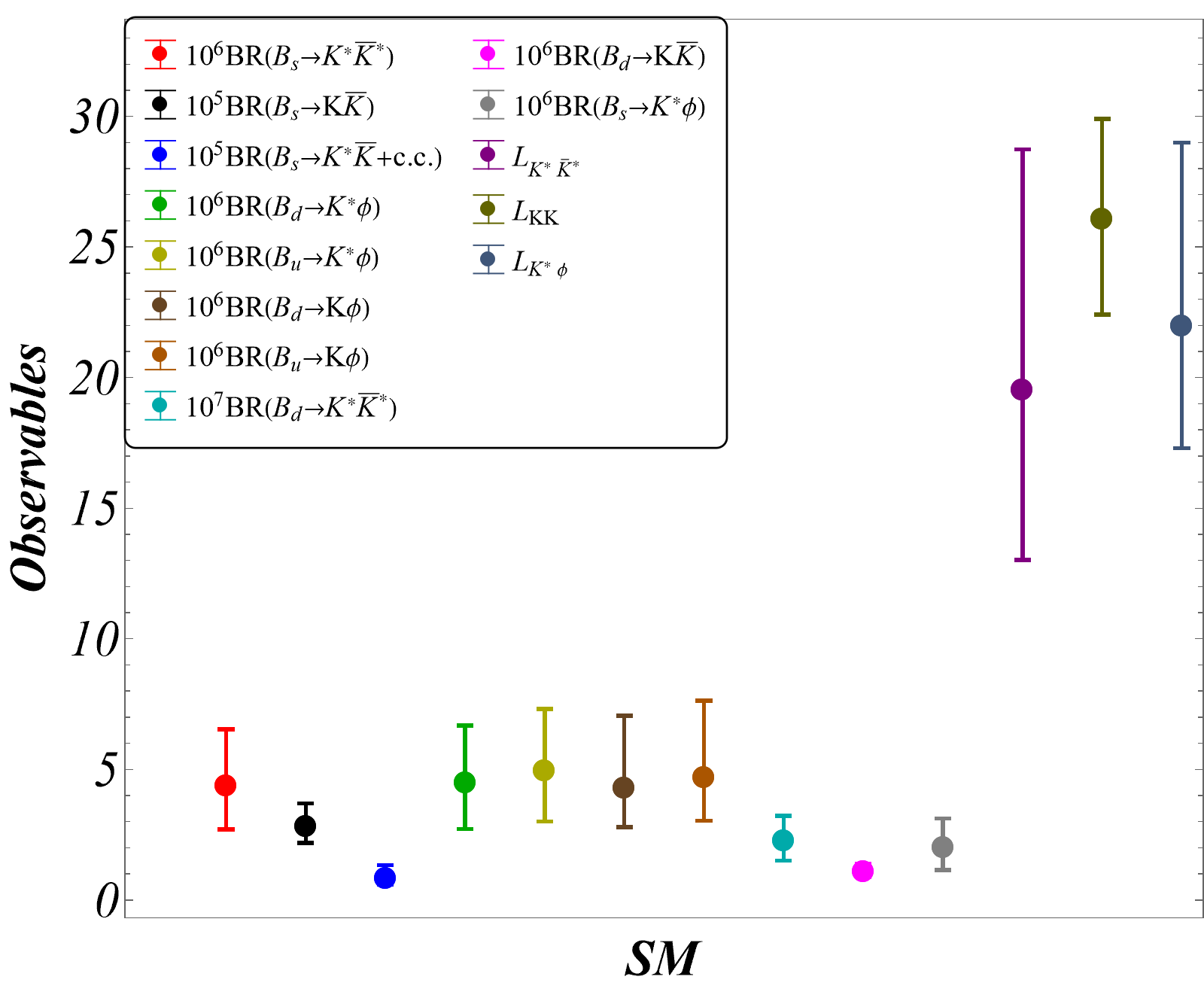}}\par
\subfloat[]{\label{fig:46plt}\includegraphics[width=0.5\textwidth,height=0.4\textwidth]{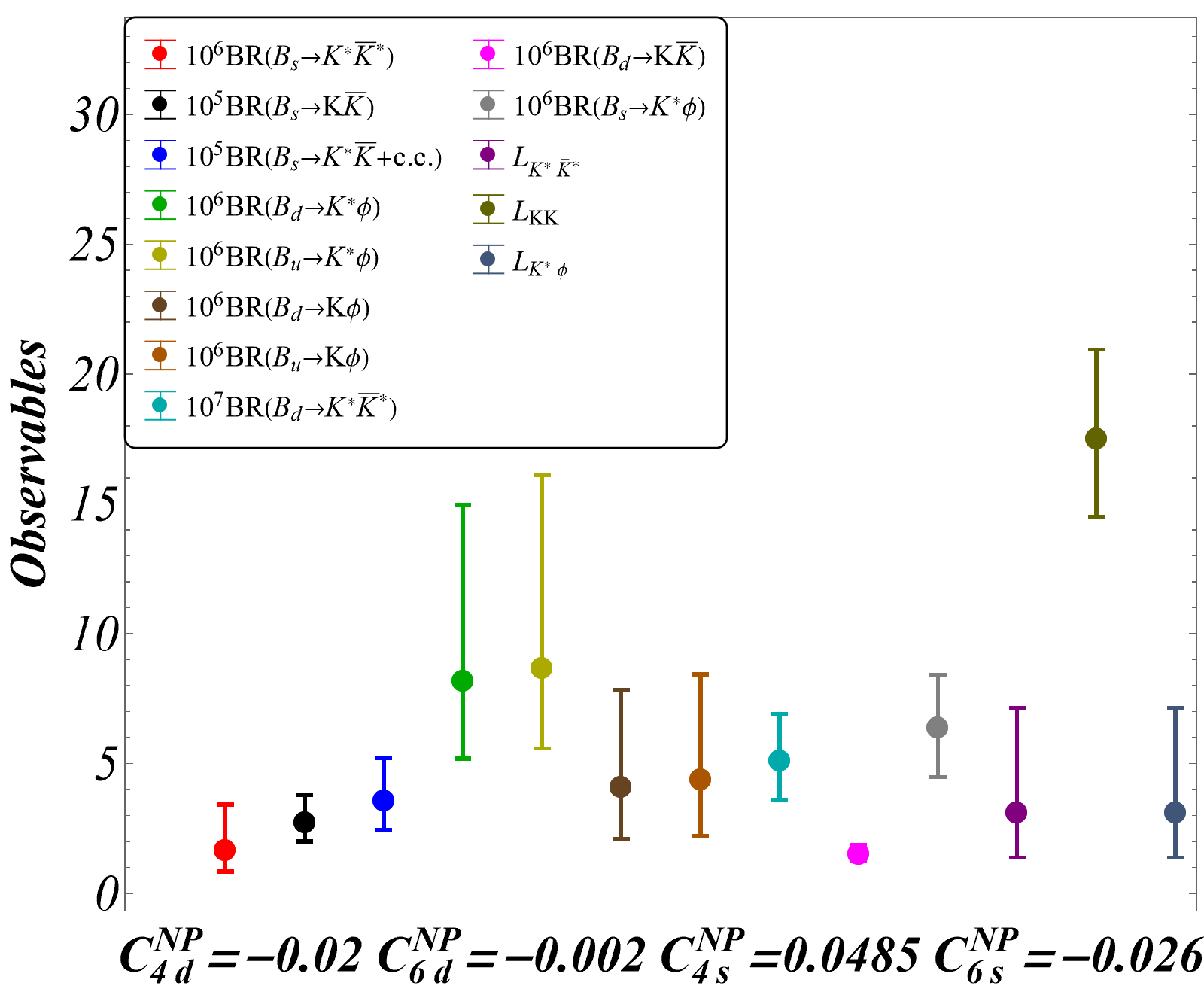}}
%\subfloat[]{\label{fig:68plt}\includegraphics[width=0.5\textwidth,height=0.4\textwidth]{c68plt.png}}
\caption{Experimental measurements and theoretical predictions of the thirteen observables (three $b\to d$ branching ratios, seven $b\to s$ branching ratios and three optimised observables $L$) in the SM and for benchmark values in the two-operator scenario $(C_{4f},C_{6f})$. Fig.~\ref{fig:46plt} is fully consistent with Fig.~\ref{fig:expplt} within $1\sigma$, contrary to Fig.~\ref{fig:smplt}.}
\label{fig:2_param_NP}
\end{figure}
In Fig.~\ref{fig:2_param_NP}, we display a benchmark scenario for the pair $(C_{4f},C_{6f})$ (Fig.~\ref{fig:46plt}) 
and compare it against the experimental measurements (Fig.~\ref{fig:expplt}) and the SM predictions (Fig.~\ref{fig:smplt}) for all the thirteen observables (three optimised observables $L$, three $b\to d$ and seven $b\to s$ branching ratios). This benchmark scenario satisfactorily explains all the observables within 1~$\sigma$ of their experimental determinations. 

In summary, it is not possible to explain the experimental data for the branching ratios available and optimized observables in $B_{d,s}\rightarrow K^{(*)} K^{(*)}$, $B_{d,s}\rightarrow K^{*} \phi$, $B_d\rightarrow K \phi$, $B^-\rightarrow K^{(*)} \phi$, $B_{s}\rightarrow K^{*} K$ 
if we consider scenarios with NP entering only one operator of the Weak effective Hamiltonian shown in Appendix  \ref{app:WET}. Our results show how the minimal scenarios that can address the measurements should include at least two operators, for instance for the pair $(C_{4f},C_{6f})$. 
Interestingly, although $C_{6f}$ cannot by itself explain the observables corresponding to the vector-vector final state $K^{*0}\bar{K}^{*0}$, it proves an essential ingredient for the two-operator scenarios able to explain the thirteen observables of interest.

Together with this manuscript we provide the  ancillary file ``Q4\_Q6\_ancilliary.dat". 
%and ``Q6\_Q8g\_ancilliary.dat". 
Each file is a list of numerical 4-plets. Fig.~\ref{fig:46plt} 
%and~\ref{fig:68plt} have
has been generated by evaluating the observables corresponding to one benchmark 4-plet (the numbers for which have been specified in the figure itself). In ``Q4\_Q6\_ancilliary.dat"  the arrangement is 
$\{C_{4d},C_{6d},C_{4s},C_{6s}\}$
%while in ``Q6\_Q8g\_ancilliary.dat" is $\{C_{6d},C_{8gd},C_{6s},C_{8gs}\}$
, where $C_{if}$ corresponds to the NP Wilson coefficient for the operator $Q_{if}$.

\section{Conclusions}\label{sec:conclusion}

Hints of NP have been reported in several neutral-current $b\to s\mu\mu$ and charged-current $b\to c\tau\nu$ decays over the past decade. If the ratios $R_{K^{(*)}}$  are now in agreement with lepton universality between the first and the second generations, the $R(D^{(*)})$ observables still point towards flavour  universality violation with respect to the third generation at the $\sim 3~\sigma$ level. Moreover, there are persistent indications of deviations in the $b\to s\mu\mu$ transitions, regarding the optimised angular observable $P_5^\prime$ for $B\to K^*\mu\mu$ as well as several branching ratios ($B\to K\mu\mu$, $B\to K^*\mu\mu$, $B_s\to \phi\mu\mu$).

If these deviations are really due to NP, the latter is expected to affect other processes, in particular rare processes sharing some features with $b\to s\mu\mu$ and/or $b\to c\tau\nu$. In this context it is particularly interesting to consider rare hadronic $B$ decays corresponding to $b\to sq\bar{q}$ transitions. Such neutral-current processes are mediated through loop diagrams in the SM and could thus have an enhanced sensitivity to NP, in parallel with $b\to s\mu\mu$. However, such transitions are expected to suffer from larger theoretical uncertainties which may reduce their sensitivity to NP.

In previous articles~\cite{Alguero:2020xca,Biswas:2023pyw}, we showed that $B_d$ and $B_s$ decays into $K^{(*)}\bar{K}^{(*)}$ final states are particularly well suited to define optimised observables $L_{K^{(*)}\bar{K}^{(*)}}$ as ratios of branching ratios, with a reduced sensitivity to the hadronic uncertainties 
that affect the branching ratios.
 Indeed, the hadronic uncertainties arising in these penguin-mediated processes are expected to be correlated by the $U$-spin symmetry, which is fully exploited in the optimised observables $L_{K^{(*)}\bar{K}^{(*)}}$. This allowed for an exploration of the NP hypotheses able to explain the breakdown of the $U$-spin symmetry observed by LHCb in the longitudinal polarisation of $K^*\bar{K}^*$ decays, reflected by the optimised observable $L_{K^*\bar{K}^*}$, as well as the rest of the optimised observables and branching ratios of interest for the $K^{(*)}\bar{K}^{(*)}$ modes.

In this article, we have looked at another pair of penguin-mediated decay modes involving the $\bar{B}_d(\bar{B}_s)\to\bar{K}^{*0}(K^{*0})\phi$ processes. They exhibit a few differences in comparison to those discussed in the previous articles; for instance, they receive the contribution from an additional penguin topology, the $B_{d(s)}$ decay is generated by a $b\to s(d)$ transition and, moreover, the modes are not related  through $U$-spin. However, a careful analysis of the amplitudes within QCD factorisation points towards correlations similar to those in $K^{(*)}\bar{K}^{(*)}$,  so that one can define an optimised observable $L_{K^*\phi}$ with a weaker sensitivity to the hadronic uncertainties than the individual branching ratios. It turns out that the experimental data  for the longitudinal branching ratios and optimised observables of these modes deviate less significantly from the corresponding QCDF predictions than for the $K^{(*)}\bar{K}^{(*)}$ modes.

It is then interesting to analyse these deviations in $K^*(\bar{K}^*)\phi$ modes together with those in $K^{(*)}\bar{K}^{(*)}$ modes in terms of the Effective Field Theory valid at the $m_b$ scale. 
Assuming NP to affect only the operators that already contribute to the SM effective field theory at the $m_b$ scale, all results can be accounted for within one standard deviation if there is an NP shift in the Wilson coefficients of either the QCD penguin operator $C_{4f}$ or the chromomagnetic dipole operator $C_{8gf}$, assuming that both $f=s,d$ operators are affected in each case. The deviation between the theoretical and experimental measurements for the $b\to s$ longitudinal branching ratio $\bar{B}_d\to\bar{K}^{*0}\phi$ ($0.35~\sigma$) is less than  that for the $b\to d$ transition $\bar{B}_s\to K^{*0}\phi$ mode ($1.26~\sigma$). This yields allowed ranges of values for $C_{4s,8gs}$ smaller in size than that for $C_{4d,8gd}$, i.e. NP is more important for $b\to d$ transitions. 

We also analyzed the vector-pseudoscalar final state with $\mathcal{B}(\bar{B}_d\to \bar{K}^0\phi)$. The experimental value of the associated optimised observable cannot be determined due to the lack of measurement for $\mathcal{B}(\bar{B}_s\to K^0\phi)$. Analysing the only available branching ratio $\mathcal{B}(\bar{B}_d\to \bar{K}^0\phi)$, we can see that the current value of this branching ratio is in tension with the rest of the observables if one considers NP in $C_{4f}$ or $C_{8gf}$ ($f=d,s$). It would thus be highly desirable to have new measurements of both branching ratios $\bar{B}_{d,(s)}\to\bar{K}^{0}(K^{0})\phi$, if possible with their experimental correlations. Both LHCb and Belle II could contribute to these improvements in a very useful way.

We then considered $\mathcal{B}(\bar{B}_s\to K^{*0}\bar{K}^0+c.c.)$. This observable could  have been included with the $K^{(*)}\bar{K}^{(*)}$ final states in Ref.~\cite{Biswas:2023pyw}, but it was discarded at that time since that article focused on optimized $L$ ratios. We considered it here and found that the experimental measurement for $\mathcal{B}(\bar{B}_s\to K^{*0}\bar{K}^0+c.c.)$ deviates from the SM by $1.44~\sigma$, and that the previous NP scenarios for $C_{4f}$ or $C_{8gf}$ ($f=d,s$) cannot accommodate this deviation easily.  As mentioned above we already encountered problems  
 when considering NP scenarios involving a single and identical type of operator for both $b\to d$ and $b\to s$ transitions.

Therefore we turned to more elaborated scenarios, and we found some that could accommodate all (currently) available (and relevant)
branching ratios involving the $K^{(*)}\bar{K}^{(*)}$, $K^*\bar{K}$ and neutral $K^*(\bar{K}^{(*)})\phi$ final states. More specifically, we considered two-operator scenarios $(Q_{if},Q_{jf})$ with $i\neq j$ denoting operators in the low-energy EFT at the scale $m_b$, and $f=d,s$. These scenarios allowing NP in 4 distinct Wilson coefficients can explain our set of data for $i=4,j=6$ and $i=6,j=8g$. Interestingly, 
although $C_{6f}$ alone is unable to explain observables corresponding to the $K^*\bar{K}^*$ final states, it is a mandatory ingredient in the two-operator scenarios that we identified.

After
 all the relevant penguin-dominated neutral branching ratios were analysed, we probed whether 
 the NP scenarios could also explain the companion charged modes entailing a $b\to s$ transition: i.e. $B^-\to K^{(*)-}\phi$. We found that the two-operator scenario $(C_{4f},C_{6f})$ could explain $\mathcal{B}(B^-\to K^{(*)-}\phi)$ along with all the other observables including the $L$'s, albeit with a (substantially) reduced parameter space. On the contrary, the scenario $(C_{6f},C_{8gf})$ cannot explain the pattern of measurements. 
 However, the data on $\mathcal{B}(B^-\to K^-\phi)$ are slightly suspicious, as they are in slight tension with the data on its neutral counterpart.
  This disparity is a bit surprising since these two modes are related by isospin symmetry, which is known to be very accurate (the experimental measurements for the corresponding vector-vector modes are, as expected, extremely consistent). Interestingly, neither $\mathcal{B}(\bar{B}_d\to\bar{K}^0\phi)$ nor $\mathcal{B}(B^-\to K^-\phi)$ have been measured by LHCb or Belle II till date. 
  A shift in these two measurements could have an important effect on our analysis. For instance, a lower value for $\mathcal{B}(\bar{B}_d\to\bar{K}^0\phi)$ would result in a consistent one-operator explanation involving all the observables from neutral modes. A shift in $\mathcal{B}(B^-\to K^-\phi)$ reducing its deviation with respect to the current measurement for $\mathcal{B}(\bar{B}_d\to\bar{K}^0\phi)$ would result in $(C_{6f},C_{8gf})$ being a viable two-operator scenario along with $(C_{4f},C_{6f})$.  
 In view of this, we consider that a precise and up-to-date measurement of $\mathcal{B}(B^-\to K^-\phi)$ is warranted at this point of time. We further suggest a correlated measurement of the two branching ratios $\mathcal{B}(\bar{B}_d(B^-)\to \bar{K}^0(K^-)\phi)$ to test the validity of the isospin symmetry and in turn confirm or dismiss the picture of NP scenarios found here with the current data. 

Finally, we would like to stress that our current and previous analyses are only an exploratory first step to identify ``common regions" and 
more sophisticated global fits would be needed in order to quantify these disagreements and difficulties and assess the scenarios considered in a statistically meaningful way. In this sense 
a more rigorous statistical analysis that includes the combination of asymmetric distributions to determine confidence intervals is left for a future work.
We have also only looked at the scope of real NP Wilson coefficients for a simultaneous explanation of the $L$'s and branching ratios in our current and early papers. The relevance of imaginary NP Wilson coefficients, along with the inclusion of data on CP-asymmetries is also an interesting direction that we would like to pursue in the near future.

In summary, larger sets of data should be considered within a more appropriate statistical apparatus, in order to perform global fits to the corresponding Wilson coefficients and determine if the data favour  the SM or invite us to consider NP contributions in rare hadronic $B$ decays. 

\section*{Acknowledgements}
The authors acknowledge Tim Gershon for valuable comments. This project has received support from the European Union’s Horizon 2020 research and innovation programme under the Marie Sklodowska-Curie grant agreement No 860881-HIDDeN [S.D.G.] and the Marie Sklodowska-Curie grant agreement No 945422 [G.T-X.]. This research has been supported by the Deutsche Forschungsgemeinschaft (DFG, German Research Foundation) under grant 396021762 - TRR 257 ``Particle Physics Phenomenology after the Higgs Discovery”. J.M. gratefully acknowledges the financial support from ICREA under the ICREA Academia programme. J.M. and A.B. also received financial support from Spanish Ministry of Science, Innovation and Universities (project PID2020-112965GB-I00) and from the Research Grant Agency of the Government of Catalonia (project SGR 1069). 

\newpage

\appendix
\section{Weak Effective Theory}\label{app:WET}

At the scale $m_b$, the relevant effective Hamiltonian separating small and large-distances for $b\to s,d$ transitions in non-leptonic $B$-decays is
\begin{equation}\label{eq:wet}
H_{\rm eff}=\frac{G_F}{\sqrt{2}}\sum_{p=c,u} \lambda_p^{(s,d)}
 \Big({\cal C}_{1s,d}^{p} Q_{1s,d}^p + {\cal C}_{2s,d}^{p} Q_{2s,d}^p+\sum_{i=3 \ldots 10} {\cal C}_{is,d} Q_{is,d} + {\cal C}_{7\gamma s,d} Q_{7\gamma s,d} + {\cal C}_{8gs,d} Q_{8gs,d}\Big) \,,
\end{equation}
where $\lambda_p^{(s,d)}=V_{pb}V^*_{ps,d}$. We follow the conventions and definitions of Ref.~\cite{Beneke:2001ev}:
\begin{align}
 Q_{1f}^p &= (\bar p b)_{V-A} (\bar f p)_{V-A} \,,  & Q_{7s} &= (\bar f b)_{V-A} \sum_q\,\frac{3}{2} e_q (\bar q q)_{V+A} \,, \nonumber \\[-2.2mm]
 Q_{2f}^p &= (\bar p_i b_j)_{V-A} (\bar f_j p_i)_{V-A} \,, & Q_{8f} &= (\bar f_i b_j)_{V-A} \sum_q\,\frac{3}{2} e_q (\bar q_j q_i)_{V+A} \,, \nonumber \\[-2.2mm]
 Q_{3f} &= (\bar f b)_{V-A} \sum_q\,(\bar q q)_{V-A} \,, &Q_{9f} &= (\bar f b)_{V-A} \sum_q\,\frac{3}{2} e_q (\bar q q)_{V-A} \,, \nonumber \\[-2.2mm]
 Q_{4f} &= (\bar f_i b_j)_{V-A} \sum_q\,(\bar q_j q_i)_{V-A} \,, & Q_{10f} &= (\bar f_i b_j)_{V-A} \sum_q\,\frac{3}{2} e_q (\bar q_j q_i)_{V-A} \,, \nonumber\\[-2.2mm]
 Q_{5f} &= (\bar f b)_{V-A} \sum_q\,(\bar q q)_{V+A} \,, &Q_{7\gamma f} &= \frac{-e}{8\pi^2}\,m_b\bar f\sigma_{\mu\nu}(1+\gamma_5) F^{\mu\nu} b \,,\nonumber \\[-2.2mm]
 Q_{6f} &= (\bar f_i b_j)_{V-A} \sum_q\,(\bar q_j q_i)_{V+A} \, , &Q_{8gf} &= \frac{-g_s}{8\pi^2}\,m_b\, \bar f\sigma_{\mu\nu}(1+\gamma_5) G^{\mu\nu} b \,, \nonumber
%\label{operators}
\end{align}

with $f=s,d$. In the above $(\bar q_1 q_2)_{V\pm A}=\bar q_1\gamma_\mu(1\pm\gamma_5)q_2$, 
$i,j$ are colour indices, $e_q$ are the electric charges of the quarks in units of $|e|$.
$ Q_{1f,2f}^p$ are the left-handed current-current operators,  $ Q_{3f\ldots 6f}$ and
$ Q_{7f\ldots 10f}$ are QCD and electroweak penguin operators, and $Q_{7\gamma f}$ and $Q_{8gf}$ are electromagnetic and chromomagnetic dipole operators.
A summation over $q=u,d,s,c,b$ is implied. 
In the SM, ${\cal C}_1^c$  is the largest coefficient and it corresponds to the colour-allowed tree-level contribution from the W exchange, whereas ${\cal C}_2^c$  is colour suppressed. QCD-penguin operators are numerically suppressed, and the electroweak operators even more so.

\section{Input parameters\label{app:inputs}}

We present the central values and errors of the parameters used in calculating the theoretical predictions for the $K^*(\bar{K}^{(*)})\phi$ observables ($L$'s and branching ratios) in Table~\ref{tab:inputs}.

\begin{table}
	\begin{center}
\tabcolsep=1.55cm\begin{tabular}{|c|c|c|}
\hline\multicolumn{3}{|c|}{$B_{d,s}$ Distribution Amplitudes (at $\mu=1$ GeV)~\cite{Khodjamirian:2020hob,Ball:2006nr} } \\ \hline
$\lambda_{B_d} $ [GeV]&$\lambda_{B_s}/\lambda_{B_d}$& $\sigma_B$\\
\hline
$0.383\pm0.153$&$1.19\pm0.14$&$1.4\pm0.4$\\ \hline
\end{tabular}

\vskip 1pt

\tabcolsep=0.930cm\begin{tabular}{|c|c|c|c|}
\hline\multicolumn{4}{|c|}{$K^*$ Distribution Amplitudes (at $\mu=2$ GeV)~\cite{Ball:2007rt}}  \\ \hline
$\alpha_1^{K^*}$&
$\alpha_{1,\perp}^{K^*}$&
$\alpha_2^{K^*}$&
$\alpha_{2,\perp}^{K^*}$\\
\hline
$0.02\pm0.02$&$0.03\pm0.03$&$0.08\pm0.06$&$0.08\pm0.06$\\ \hline
\end{tabular}

\vskip 1pt

\tabcolsep=1.245cm\begin{tabular}{|c|c|c|c|}
\hline\multicolumn{4}{|c|}{$\phi$ Distribution Amplitudes (at $\mu=2$ GeV)~\cite{Ball:2007rt}}  \\ \hline
$\alpha_1^{\phi}$&
$\alpha_{1,\perp}^{\phi}$&
$\alpha_2^{\phi}$&
$\alpha_{2,\perp}^{\phi}$\\
\hline
$0$&$0$&$0.13\pm0.06$&$0.11\pm0.05$\\ \hline
\end{tabular}

\vskip 1pt

\tabcolsep=1.27cm\begin{tabular}{|c|c|c|}
\hline\multicolumn{3}{|c|}{Decay Constants for $B$ mesons (at $\mu=2$ GeV)~\cite{FlavourLatticeAveragingGroup:2019iem} and $K$ meson~\cite{Workman:2022ynf}}  \\ \hline
$f_{B_d}$&$f_{B_s}/f_{B_d}$&$f_K$\\
\hline
$0.190\pm0.0013$&$1.209\pm0.005$&$0.1557\pm0.0003$\\ \hline
\end{tabular}

\vskip 1pt

\tabcolsep=0.120cm\begin{tabular}{|c|c|c|c|c|c|}
\hline\multicolumn{6}{|c|}{Decay Constants for $K^*, \phi, \rho, \omega$ (at $\mu=2$ GeV)~\cite{Bharucha:2015bzk,RBC-UKQCD:2008mhs}}  \\ \hline
$f_{K^*}$&$f^\perp_{K^*}/f_{K^*}$&$f_{\phi}$&$f^\perp_{\phi}/f_{\phi}$&$f_\rho$&$f_\omega$\\
\hline
$0.204\pm 0.007$&$0.712\pm0.012$&$0.233\pm0.004$&$0.750\pm0.008$&$0.213\pm0.005$&$0.197\pm0.008$\\ \hline
\end{tabular}
\vskip 1pt

\tabcolsep=0.11cm\begin{tabular}{|c|c|c|c|c|}
\hline\multicolumn{5}{|c|}{$B_{d,s}\to K^*,\phi$ form factors~\cite{Bharucha:2015bzk} and B-meson lifetimes (ps)~\cite{HeavyFlavorAveragingGroup:2022wzx}}  \\ \hline
$A_0^{B_s\to K^*}(q^2=m_\phi^2)$&$A_0^{B_d\to K^*}(q^2=m_\phi^2)$ &$A_0^{B_s\to\phi}(q^2=m_{K^*}^2)$& $\tau_{B_d}$ & $\tau_{B_s}$\\
\hline
$0.380 \pm 0.024$ &$ 0.393 \pm 0.039$ & $0.438\pm0.024$ & $1.519\pm0.004$ & $1.520\pm0.005$  \\ \hline
\end{tabular}

\vskip 1pt

\tabcolsep=1.325cm\begin{tabular}{|c|c|c|c|}
\hline
\multicolumn{4}{|c|}{Mass and decay widths for $\rho, \omega$ (GeV)~\cite{Workman:2022ynf}}  \\ \hline
$m_\rho$ & $\Gamma_\rho$ & $m_\omega$ &
$\Gamma_\omega$\\ \hline
$0.7745$ &$	0.1484$&$0.7827$&$0.0087$\\
\hline
\end{tabular}

\vskip 1pt

\tabcolsep=1.22cm\begin{tabular}{|c|c|c|}
\hline\multicolumn{3}{|c|}{$B_{d}\to K$~\cite{Parrott:2022rgu}, $B_s\to K$~\cite{Flynn:2023nhi} and $B_s\to\phi$ form factors}  \\ \hline
$f_0^{B_s}(q^2=m_\phi^2)$&$f_0^{B_d}(q^2=m_\phi^2)$& $A_0^{B_s\to\phi}(q^2=m_K^2)$\\
\hline
$0.336 \pm 0.023$ &$ 0.340 \pm 0.011$&$ 0.426 \pm 0.024$ \\ \hline
\end{tabular}

\vskip 1pt

\tabcolsep=0.737cm\begin{tabular}{|c|c|c|c|}
\hline
\multicolumn{4}{|c|}{Wolfenstein parameters~\cite{Charles:2004jd} }  \\ \hline
$A$ & $\lambda$ & $\bar\rho$ &
$\bar\eta$\\ \hline
$0.8132^{+0.0119}_{-0.0060}$ &$	0.22500^{+0.00024}_{-0.00022}$&$0.1566^{+0.0085}_{-0.0048}$&$0.3475^{+0.0118}_{-0.0054}$\\
\hline
\end{tabular}

\vskip 1pt

\tabcolsep=0.240cm\begin{tabular}{|c|c|c|c|c|c|c|c|}
	\hline
	\multicolumn{8}{|c|}{QCD scale and masses [GeV]~\cite{Workman:2022ynf}}\\
		\hline 
 $\bar{m}_b(\bar{m}_b)$  & $m_b/m_c$ & $m_{B_d} $& $m_{B_s} $& $m_{K^*} $& $m_{\phi} $ & $m_{K} $ &$\Lambda_{{\rm QCD}}$
      \\ \hline
      $4.18$  & $4.577\pm0.008$  & $5.27966$ & $5.36692$&$0.89555$&$1.01946$&$0.497611$&$0.225$
      \\ \hline
\end{tabular}
\vskip 1pt
\tabcolsep=0.646cm\begin{tabular}{|c|c|c|c|c|c|}
\hline
	\multicolumn{6}{|c|}{SM Wilson Coefficients (at $\mu=4.18$ GeV)}\\
		\hline 
${\cal C}_1$ &  ${\cal C}_2$ & ${\cal C}_3$ & ${\cal C}_4$ &  ${\cal C}_5$ & ${\cal C}_6$
      \\ \hline
 1.082 & -0.191 & 0.014 & -0.036 & 0.009 &  -0.042
      \\ \hline
 ${\cal C}_{7}/\alpha_{em}$ & ${\cal C}_{8}/\alpha_{em}$ & ${\cal C}_{9}/\alpha_{em}$ &  ${\cal C}_{10}/\alpha_{em}$ & ${\cal C}^{\rm eff}_{7\gamma}$ &  ${\cal C}^{\rm eff}_{8g}$
      \\ \hline
 -0.011 & 0.060 & -1.254 & 0.224 & -0.318 & -0.151 
      \\ \hline
		\end{tabular}
		\caption{Input parameters. For the form factors, we use the Lattice+LCSR fit results provided as .json files by the authors of Ref.~\cite{Bharucha:2015bzk}. For the mass and the decay width of the $\rho$ meson, we use a weighted average over the corresponding ``Neutral only: $e^+e^-$", ``Neutral only: Photoproduced" and ``Neutral only: Other reactions" measurements from \cite{Workman:2022ynf}.}
		\label{tab:inputs}
	\end{center}
\end{table}

\newpage

\bibliographystyle{JHEP}

\bibliography{main.bib}

\end{document}